\documentclass[11pt,prd,onecolumn,showpacs,amsmath,amssymb,aps,floats,floatfix,nofootinbib]{revtex4-1}
\usepackage[colorlinks=true,urlcolor=blue,anchorcolor=blue,citecolor=blue,filecolor=blue,linkcolor=blue,menucolor=blue,pagecolor=blue,linktocpage=true]{hyperref} 

\usepackage{graphicx}
\usepackage{verbatim}
\usepackage[multidot]{grffile}  
\usepackage{dcolumn}
\usepackage{amsmath}
\usepackage{amsfonts}
\usepackage{amssymb}
\usepackage{color}
\usepackage{footmisc}
\usepackage{latexsym}
\usepackage{slashed} 
\usepackage{pstricks}
\usepackage{indentfirst}
\usepackage{mathrsfs}
\usepackage{multirow}
\usepackage{epsfig,psfrag}
\usepackage{subfigure}
\usepackage{mathtools}
\usepackage{setspace} 
\usepackage[utf8]{inputenc} 
\usepackage[scientific-notation=true]{siunitx} 
\graphicspath{{figs/}}
\usepackage{float}
\usepackage{cleveref}

\newcommand{\MS}{\overline{\mathrm{MS}}}

\def\bac{\begin{array} {c}}

  \makeatother

  \allowdisplaybreaks 

\begin{document}

  \title{ Fate of false vacuum in singlet-doublet fermion extension model with RG improved effective action}
  \author{Yu Cheng}
  \email{chengyu@mail.ecust.edu.cn}
  \author{Wei Liao}
  \email{liaow@ecust.edu.cn}
  \affiliation{
  \vskip 0.5cm
   Institute of Modern Physics,  School of Sciences,\\
  East China University of Science and Technology, 130 Meilong Road, Shanghai 200237, P. R. China}

  \begin{abstract}
  We study the effective potential and the Renormalization Group(RG) improvement to the 
  effective potential of Higgs boson in a singlet-doublet fermion dark matter extension of the 
  Standard Model(SM), and in general singlet-doublet fermion extension models with several 
  copies of  doublet fermions or singlet fermions. We study the stability of the electroweak 
  vacuum with the RG improved effective potential in these models 
  beyond the SM.   We study the decay of the electroweak vacuum using 
  the RG improved effective potential in these models beyond the SM.  In this study we consider the 
  quantum correction to the kinetic term in the effective action and consider the RG improvement 
  of the kinetic term. Combining all these effects, we find that the decay rate of the false vacuum is slightly 
  changed when calculated using the RG improved effective action in the singlet-doublet fermion dark matter 
  model. In general singlet-doublet fermion extension models, we find that the presence of several copies of 
  doublet fermions can make the electroweak vacuum stable if the new Yukawa couplings are not large. 
  If the new Yukawa couplings are large, the electroweak vacuum can be turned into metastable or unstable 
  again by the presence of extra fermions.  
  \end{abstract}

  \maketitle
  \tableofcontents
  \clearpage

  \section{Introduction}
  
  Quantum contribution to the effective potential is important to understand
  the properties of the scalar field, e.g. the property of the ground-state and the behavior at large energy scale.
  For example, radiative correction can make a vacuum unstable and trigger spontaneous symmetry 
  breaking~\cite{Coleman:1973jx}.  The RG improved effective potential, which re-sums contributions of
  large logarithms, is important to understand the behavior of effective potential at large energy scale.
  For example, the RG-improved effective potential is crucial in reducing the dependence
  on the renormalization scale when calculating quantities related with physical parameters ~\cite{Ford:1992mv, Bando et.al., Casas:1994qy}.
   
   It is well known that a false vacuum can decay via tunneling ~\cite{Langer, Coleman:1977py, Callan:1977pt} and become
   unstable. In the SM,  the Higgs quartic self-coupling can become negative 
   at an energy scale around $ 10^{10}$ GeV. This makes the electroweak(EW) vacuum unstable. 
   The decay rate of the EW vacuum can be calculated using an approximate bounce solution of the Higgs potential with
   a negative Higgs quartic coupling ~\cite{Isidori:2001bm,Chigusa:2018uuj,Chigusa:2017dux}.
   Calculation of vacuum decay using RG improved effective potential in the SM does not give much difference, because the bounce solution
   is dominated by behavior at high energy scale and  the RG-improved effective potential in the SM at high energy scale is accidentally close
   to the Higgs potential with running quartic coupling~\cite{Degrassi:2012ry}.
   
   In extension of the SM, the situation can be quite different. The RG improved effective potential is possible to be very different
   from the potential using a running Higgs quartic coupling. As an example, we consider a singlet-doublet fermion dark matter(SDFDM) 
   extension of the SM. We show that the RG improved effective potential can be quite different from the tree-level potential aided with
   running Higgs quartic coupling. Then we study the vacuum stability in this extension of SM. We study the false vacuum decay
   using RG improved effective potential in this SDFDM model. We also study the quantum contribution to the kinetic term in the effective action
   in this model beyond the SM and consider the RG improvement of the kinetic term. 
   After taking all these effects into account we find that the false vacuum decay rate is just slightly changed using 
   the RG improved effective action in the SDFDM model, although the RG effective potential is significantly different 
   from the tree-level form of the Higgs potential with running quartic self-coupling.
   We also perform these analyses in general singlet-doublet fermion extension models in which several copies of singlet fermions or several copies
   of doublet fermions are considered. We find that the presence of several extra doublet fermions can make the EW vacuum stable
   if the new Yukawa couplings are not large.
  
  The article is organized as follows. In section \ref{sec:RGEandeffpotential}, we first briefly review the SDFDM model. 
  We study the threshold effect caused by
  the extra fermions in this model beyond the SM and study the running of the Higgs quartic coupling in this model. 
  Then we study the effective potential and the RG improved effective potential in this model. 
  In Section \ref{sec:vacuumstability} we study the vacuum stability in SDFDM model. 
  We calculate the renormalization of the kinetic term in the effective action in the SM and in SDFDM model. 
  We study the RG improvement of the kinetic term and calculate the decay rate of the false vacuum.
  In Section \ref{GSDM}, we do the analysis in general singlet-doublet fermion extension model.
  Details of calculation are summarized in Appendix \ref{sec:thersholdeffect}, \ref{betafunction}, \ref{derivativecorr}, and \ref{sec:thersholdeffectgeneral}.
  We summarize in conclusion.

  \section{Effective potential and RG improved effective potential in SDFDM model}
  \label{sec:RGEandeffpotential} 
  \subsection{The SDFDM model}
  
  In addition to the SM fields, the SDFDM model has $\mathrm{SU}(2)$ doublet fermions $\psi_{L, R}=\left(\psi_{L,R}^{0}, \psi_{L,R}^{-}\right) ^T$  with Y = -1/2 and  singlet fermions $S_{L, R}$. Here, $L$ ($R$) refers to the left(right) chirality.
  As a singlet,  $S$ can be either Dirac type or Majorana type fermion~\cite{Cohen:2011ec, Freitas:2015hsa}. 
  In this model, the neutral fermion can be a dark matter candidate.
  The vacuum properties of SDFDM model with a Majorana type mass have been discussed in \cite{Wang:2018lhk}.
  In this article we  work on the Dirac type mass \cite{Yaguna:2015mva}.

  The relevant terms of $\psi$ and $S$ in the Lagrangian are:
  \begin{equation}
	\begin{aligned} 
	& \mathcal{L}_{\mathrm{SDFDM}}= \overline{\psi} i  \slash \hspace{-0.25cm} D \psi
	+\overline{S} i \slash \hspace{-0.2cm} \partial S \\
	& -M_{D} \overline{\psi}_{L} \psi_{R}-M_{S} \overline{S}_{L} S_{R} -y_{1} \overline{\psi}_{L} \tilde{H} S_{R}-y_{2} \overline{\psi}_{R} \tilde{H} S_{L}+\mathrm{H.c.} \end{aligned} 
	\label{Lagrangian}
  \end{equation}
  where $M_{D,S}$  are the mass parameters, $y_{1,2}$ the new Yukawa couplings, 
  $H$ is the SM Higgs doublet with $Y=1/2$, and $\tilde H=i\sigma_2H^{*}$ .
   We impose a $Z_2$ symmetry in the Lagrangian with the new fermions $\psi$ and $S$ odd and SM fermions even under the $Z_2$ operation.
   This guarantees the lightest of these new fermions to be stable, and makes it a dark matter candidate if it is neutral.

  After the EW symmetry breaking, the mass matrix of $S_{L, R}$ and  the neutral component of  $\psi_{L,R}$($\psi_{L,R}^{0}$)
   is given as
  \begin{equation}
  M=\left(\begin{array}{cc}{M_{S}} & {\frac{y_{2} v}{\sqrt{2}}} \\ {\frac{y_{1} v}{\sqrt{2}}} & {M_{D}}\end{array}\right),
  \label{Mmatrix}
  \end{equation}
  where $v=246 \mathrm{~GeV}$. Mixings between these two neutral fermions are generated by this mass matrix.
  The mixing angles $\theta_{L,R}$ appear in the diagonalization of the mass matrix using two unitary mass matrices, that is
   \begin{equation}
  M^{d}=\left(\begin{array}{cc}{M_{\chi_{1}^{0}}} & {0} \\ {0} & {M_{\chi_{2}^{0}}}\end{array}\right)=U_{L}^{\dagger} M U_{R}
  \end{equation}
with
  \begin{equation}
  U_{L, R}=\left(\begin{array}{cc}{\cos \theta_{L, R}} & {\sin \theta_{L, R}} \\ {-\sin \theta_{L, R}} & {\cos \theta_{L, R}}\end{array}\right)
  \end{equation}
 and
        \begin{eqnarray}
        && M_{\chi_1^0}^2=\frac{1}{2}(T_M-\sqrt{T_M^2-4 ~D_M^2}),   \label{Mchi1} \\
        && M_{\chi_2^0}^2=\frac{1}{2}(T_M+\sqrt{T_M^2-4 ~D_M^2}),  \label{Mchi2}
        \end{eqnarray}
        where $T_M=M_S^2+\frac{1}{2}y_1^2 v^2+M_D^2+\frac{1}{2}y_2^2 v^2$, 
                   $D_M=\frac{1}{2}y_1 y_2 v^2-M_S M_D $.
     $\chi_1^0$ and $\chi_2^0$ are neutral fermion fermions in the diagonalized base with masses $M_{\chi_1^0}$ and $M_{\chi_2^0}$ respectively.
  
  The mixing angles $\theta_{L, R}$ can be solved as
   \begin{eqnarray}
 && \tan 2 \theta_{L}=\frac{\sqrt{2} v\left(M_{S} y_{1}+M_{D} y_{2}\right)}{M_{D}^{2}-M_{S}^{2}+\frac{v^{2}}{2}\left(y_{1}^{2}-y_{2}^{2}\right)} 
\\  
 && \tan 2 \theta_{R}=\frac{\sqrt{2} v\left(M_{S} y_{2}+M_{D} y_{1}\right)}{M_{D}^{2}-M_{S}^{2}+\frac{v^{2}}{2}\left(y_{2}^{2}-y_{1}^{2}\right)}
  \end{eqnarray}
  Writing $H=(0, (v+h)/\sqrt{2})^T$,
  the interaction Lagrangian of dark matter fields $\chi^0_{1,2}$ and the CP-even neutral Higgs field $h$ is obtained from
  Eq. (\ref{Lagrangian}) as 
   	\begin{equation}
  	\Delta {\cal L}
  	=-y_{A} \overline{\chi_{1}^{0}} \chi_{1}^{0} h -y_{B} \overline{\chi_{2}^{0}} \chi_{2}^{0} h  
  	  -[y_{C} \overline{\chi_{1}^{0}} P_R \chi_{2}^{0} h
  	  +y_{D} \overline{\chi_{1}^{0}} P_L \chi_{2}^{0} h +h.c.]  \label{Lagrangian-1}
  	\end{equation}	 
	where $P_{R,L}=\left(1\pm\gamma_{5}\right)/2$, 
	$y_{A} = (-y_{2} \cos \theta_{L} \sin \theta_{R}-y_{1} \sin \theta_{L} \cos \theta_{R})/\sqrt{2} $, 
	$y_{B} = (y_{2} \cos \theta_{R} \sin \theta_{L}+y_{1} \sin \theta_{R} \cos \theta_{L})/\sqrt{2}$ , 
	$y_{C} =(y_{2} \cos \theta_{L} \cos \theta_{R}-y_{1} \sin \theta_{L} \sin \theta_{R})/\sqrt{2} $ 
	and $y_{D} = (-y_{2}\sin \theta_{L} \sin \theta_{R} +y_{1} \cos \theta_{L} \cos \theta_{R})/\sqrt{2}$.

  \subsection{Effective potential and RG improved Effective potential}
  
  In the following, we will take $\phi$ to denote a neutral external field. $\phi/\sqrt{2}$ corresponds to
  the CP-even neutral component of the Higgs doublet in the SM.
  The tree-level potential of $\phi$ is
  \begin{eqnarray}
  V_0(\phi)=-\frac{m^2_\phi}{2} \phi^2+\frac{\lambda}{4} \phi^4, \label{treelevel-potential}
  \end{eqnarray}
  where $\lambda$ is the Higgs quartic self-coupling in the SM and $m_\phi$ the mass term.
  Coleman-Weinberg type quantum correction to the potential can be calculated using vacuum diagram by considering the
  quantum fluctuations around the external field $\phi$.  The one-loop contribution of extra fermion in the SDFDM
  model to the effective potential is calculated as

  \begin{eqnarray}
  V_{1}^{\mathrm{Ext}}= -\frac{1}{64 \pi^{2}} M_{\chi_{1}}^{4}(\phi)\left[\ln \frac{M_{\chi_{1}}^{2}(\phi)}{\mu^{2}}-3/2\right]-
                         \frac{1}{64 \pi^{2}} M_{\chi_{2}}^{4}(\phi)\left[\ln \frac{M_{\chi_{2}}^{2}(\phi)}{\mu^{2}}-3/2\right]
  \label{one-loop-SDFDM-potential}
  \end{eqnarray}
  where  $M^2_{\chi_1,\chi_2}(\phi)$ are obtained from Eqs. \eqref{Mchi1} and \eqref{Mchi2}
  by replacing $v$ with $\phi$.  $\mu$ is the renormalization scale chosen in this calculation.
  In the limit $\phi \gg v$,  we have approximately
  $M^2_{\chi_1,\chi_2}(\phi) \approx y^2_1 \phi^2/2, y^2_2\phi^2/2$.  
  
  We arrive at an one-loop effective potential as follows.
   \begin{equation}
  V_{eff}(\phi,\lambda_i,\mu)=V_{0}^{\mathrm{SM}}(\phi,\lambda)+V_{1}^{\mathrm{SM}}(\phi,\lambda_i,\mu)
  +V_{1}^{\mathrm{Ext}}(\phi, \lambda_i,\mu)
  \label{effpotenial}
  \end{equation}
  where $\lambda_i$ denotes various parameters in the model.
  $V_{0}^{\mathrm{SM}}(\phi,\lambda)$ is given in Eq. (\ref{treelevel-potential}). 
  $V_{1}^{\mathrm{SM}}(\phi,\lambda_i,\mu)$ is the one-loop contribution to the effective potential in the SM.
  The effective potential in the SM is known up to two-loop~\cite{Buttazzo:2013uya,Martin:2001vx}.
  $V_{1}^{\mathrm{Ext}}(\phi,\lambda_i,\mu)$ is given in Eq. (\ref{one-loop-SDFDM-potential}).

   In the vacuum stability analysis, we must consider the behavior of the effective potential for large external field.
   That is to say, we must deal with potentially large logarithms of the type $\log (\phi / \mu)$ for a neutral external field $\phi$. 
   The standard way to solve the problem is by means of the RG equation(RGE). $V_{eff}$ satisfies the RGE
  \begin{equation}
  \left(\mu \frac{\partial}{\partial \mu}+\beta_{i} \frac{\partial}{\partial \lambda_{i}}-\gamma \phi \frac{\partial}{\partial \phi}\right) V_{\mathrm{eff}}=0,
  \end{equation} 
  where $\beta_i$ is the $\beta$ function of parameter $\lambda_i$, and $\gamma$ the anomalous dimension of the scalar field.
  Straightforward application of this method leads to a solution~\cite{Ford:1992mv}
  \begin{equation}
  V_{\mathrm{eff}}\left(\mu, \lambda_{i}, \phi\right)=V_{\mathrm{eff}}\left(\mu(t), \lambda_{i}(t), \phi(t)\right)
  \end{equation}
  where
  \begin{equation}
  \label{mutphit}
  \begin{aligned} \mu(t) &=\mu e^{t} \\ \phi(t) &=e^{\Gamma(t)} \phi \end{aligned}
  \end{equation}
  with
  \begin{equation}
  \Gamma(t)=-\int_{0}^{t} \gamma\left(\lambda\left(t^{\prime}\right)\right) d t^{\prime}
  \end{equation}
  and $\lambda_{i}(t)$ the running coupling determined by the equation
  \begin{equation}
  \frac{d \lambda_{i}(t)}{d t}=\beta_{i}\left(\lambda_{i}(t)\right), \label{runningeq}
  \end{equation}
  with the boundary condition $\lambda_{i}(0)=\lambda_{i}$.
  So the RG improved effective potential can be written by simply  substituting $\mu$, $\lambda_{i}$, $\phi$  
  in the original effective potential with  $\mu(t)$, $\lambda_{i}(t)$, $\phi(t)$.

 The RG improved effective potential in the SDFDM model
 is obtained by implementing the substitution mentioned above into Eq.  (\ref{effpotenial}).  We have
  \begin{equation}
   V_{eff}(\phi , t) = V_{0}^{\mathrm{SM}}(\phi , t) + V_{1}^{\mathrm{SM}}(\phi , t)
  	+ V_{1}^{\mathrm{Ext}}(\phi , t) , \label{RGeffectivepotential}
  \end{equation}
  with
  \begin{eqnarray}
  V_{0}^{\mathrm{SM}}(\phi , t) &&= -\frac{m^2_\phi(t)}{2} \phi^2(t)+\frac{1}{4}  \lambda(t) \phi^{4}(t),  \label{SMpotential}\\
  V_{1}^{\mathrm{SM}}(\phi , t) &&=\sum_i \frac{ (-1)^i n_{i}}{64 \pi^{2}} M_{i}^{4}(\phi,t)\left[\ln \frac{M_{i}^{2}(\phi,t)}{\mu^{2}(t)}-c_{i}\right]
  \label{SMeff} \\
  V_{1}^{\mathrm{Ext}}(\phi , t) &&=\sum_i \frac{(-1)^i n_{i}}{64 \pi^{2}} M_{\chi_{i}}^{4}(\phi,t)\left[\ln \frac{M_{\chi_{i}}^{2}(\phi,t)}{\mu^{2}(t)}-3/2\right],  \label{SDFDMeff}
  \end{eqnarray}
  In Eq.~\eqref{SMeff} the index $i=H,G,f,W,Z$  runs over SM fields in the loop,  
  and $c_{H, G, f}=3 / 2, c_{W, Z}=5 / 6$.
  In Eq. \eqref{SDFDMeff} the index $i=\chi_1,\chi_2$  runs over extra neutral fermions in the SDFDM model.
  $n_{i}$ is the number of degrees of freedom of the fields. 
  In Eqs.  \eqref{SMeff},  and $M_{i}(\phi)$ is given by 
  \begin{equation}
  M_{i}^{2}(\phi,t)=\kappa_{i}(t) \phi^{2}(t)-\kappa_{i}^{\prime}(t)
  \end{equation}
  The values of $n_{i}$ , $\kappa_{i}$ and $\kappa_{i}^{\prime}$  in the SM
  can be found in Eq. (4) in Ref. \cite{Casas:1994qy} in the Landau gauge and in 
  Ref. \cite{DiLuzio:2014bua} both in the Fermi gauge and in the $R_{\xi}$ gauge.
  For new contributions in the SDFDM model, we have $n_i=1$,  
 and  $M^2_{\chi_1,\chi_2}(\phi,t)\approx y^2_1(t) \phi^2(t)/2, y^2_2(t) \phi^2(t)/2$.
  In Eqs. \eqref{SMeff} and \eqref{SDFDMeff}, $(-1)^i$ equals to $\pm 1$.
  For gauge and scalar bosons $(-1)^i$ take a positive sign, while for fermion fields it takes a negative sign.
  
  In the limit $\phi \gg v$,  Eq.~\eqref{RGeffectivepotential}  can be written approximately  as follows.
  \begin{equation}
  V_{\mathrm{eff}}(\phi, t) \approx \frac{\lambda_{\mathrm{eff}}(\phi, t)}{4} \phi^{4},
  \end{equation}
  where $\lambda_{\mathrm{eff}}$  is an effective coupling.
  In vacuum stability analysis, we generally take $\mu(t)=\phi$, so $\lambda_{\mathrm{eff}}(\phi , t)$ can be written as \cite{Degrassi:2012ry}
  \begin{equation}
  \begin{aligned}
  \lambda_{\mathrm{eff}}(\phi, t) \approx e^{4 \Gamma(t)}&\left\{\lambda(t)+\frac{1}{(4 \pi)^{2}} \sum_{i} N_{i} \kappa_{i}^{2}(t)\left(\log \kappa_{i}(t) e^{2 \Gamma(t)}-c_{i}\right)\right\} 
  \end{aligned}
  \label{lambdaeff}
  \end{equation}
  The values of coefficients $N_{i}$ , $\kappa_{i}$ , and $c_{i}$ appearing in Eq.~\eqref{lambdaeff} are listed 
 in Table.~\ref{effparamaters}.
 
 We note that the two-loop contributions of  strong coupling and the top Yukawa to the effective potential can be written in the $\lambda_{eff}$ as
  \begin{equation}
  \begin{aligned}
  \lambda_{\mathrm{eff}}^{2-loop}(\phi, t) \approx e^{4 \Gamma(t)} \frac{ y_{t}^{4}}{(4 \pi)^{4}}\left[8 g_{s}^{2}\left(3 r_{t}^{2}-8 r_{t}+9\right)-\frac{3}{2} y_{t}^{2}\left(3 r_{t}^{2}-16 r_{t}+23+\frac{\pi^{2}}{3}\right)\right],
  \end{aligned}
  \label{lambdaeff-twoloop}
  \end{equation}
where  $r_t = ln \frac{y^2_t}{2}+2\Gamma$. 
We can see in Eq.~\eqref{lambdaeff-twoloop} that the two-loop contributions from top loops are of the order of $y^6_t/(4\pi)^4$,
 while the one-loop terms are of the order of $y^4_t/(4\pi)^2$ as can be seen in (\ref{lambdaeff}). 
 The two-loop contributions from the new fermions in SDFDM model are similar.
 We expect that these new two-loop contributions would be much smaller than the one-loop contribution if the new Yukawa couplings $y_1$ and $y_2$
 are not much larger than the top Yukawa.
 So in this work, we do not take into account these two loop contributions from the new particles in the SDFDM model.  
 For similar reasons, we do not consider two-loop contributions of new fermions to the $\beta$ function. 
 More detailed analysis on this aspect, in particular for the case with very large Yukawa coupling, is outside the scope of the present article.

  \begin{table*}[h]
  		\begin{tabular}{|c|cccccccc  c  c|}
  			\hline
  			$p$ & $t$ & $W$ & $Z$ & $h$ & $G^{+}$ & $G_{0}$   & $C^{\pm}$  & $C_{Z}$ & $\chi_{1}$  &  $\chi_{2}$  \\ 
  			\hline
  			$N_i$ & $-12$ & $6$ & $3$ & $1$ & $2$ & $1$ & $-2$ & $-1$ & $-1$ & $-1$ \\ 
  			$c_i$ & $\frac{3}{2}$ & $\frac{5}{6}$ & $\frac{5}{6}$ &$\frac{3}{2}$& $\frac{3}{2}$& $\frac{3}{2}$&$\frac{3}{2}$ & $\frac{3}{2}$ & $\frac{3}{2}$ & $\frac{3}{2}$ \\ 
  			$\kappa_i$ & $\frac{y_t^2}{2}$ & $\frac{g^2}{4}$ & $\frac{g^2+g'^2}{4}$ & $3 \lambda$ 
  			& $\lambda+\frac{\overline{\xi}_{W}g^2}{4}$ &  $\lambda+\frac{\overline{\xi}_{Z}(g^2+g^{\prime})}{4}$  &  $\frac{\overline{\xi}_{W}g^2}{4}$& 
  			  $\frac{\overline{\xi}_{W}(g^2+g^{\prime})}{4}$  &  $\frac{y_{1}^2}{2}$ &  $\frac{y_{2}^2}{2}$ \\ 
  			\hline
  		\end{tabular}
  		\caption{The coefficients in Eq.~\eqref{lambdaeff} for the background $R_{\xi}$ gauge \cite{DiLuzio:2014bua}.
  			$\overline{\xi}_{W}$ and $\overline{\xi}_{Z}$ are the gauge-fixing parameters in the background $R_{\xi}$ gauge,
  			$G^+$  and $G^0$ the goldstone bosons, $C^{\pm}$ and $C_{Z}$ the ghost fields,
  			$\chi_{1}$ and $\chi_{2}$ are the dark matter particles in SDFDM model. 
  			For ${\bar \xi}_{W} = {\bar \xi}_{Z} = 0$, Eq.~\eqref{lambdaeff} reproduces 
  			the one-loop result in the Landau gauge, and for ${\bar \xi}_{W} = {\bar \xi}_{Z} = 1$, we get the result in the 't Hooft-Feynman gauge.}	
  	\label{effparamaters}
  \end{table*}

  \subsection{Running parameters in the $\MS$ scheme}
  To study the vacuum stability of a model at high energy scale, we need to know the value of coupling constants 
  at low energy scale and then run them to the Plank scale according to RGEs. 
  To determine these parameters at low energy scale, the threshold corrections must be taken into account. 
  In this article we work with  the modified minimal subtraction($\MS$) scheme
  and use the strategy in~\cite{Sirlin:1985ux, Buttazzo:2013uya} to evaluate one-loop threshold corrections
  and determine the initial values for RGE.
  The details of the corrections are summarized in Appendix ~\ref{sec:thersholdeffect}.
  Using these results, we find coupling constants in the $\MS$ scheme at $\mu=M_{t}$ scale which are different
  for the SM and for the SDFDM model.
  We list some of the results in Table.~\ref{initialvaluesforRGE}.
  Both the change of the Yukawa couplings $y_{1,2}$ and the change of mass term have an effect on the corrections.
  We can see in  Table.~\ref{effectofdifferentmasses} and  Table.~\ref{initialvaluesforRGE}  that changing the mass scale of dark matter particles
  does not give rise to change of the initial parameters as significant as that of changing Yukawa couplings. 
  Therefore, we will always choose mass parameters as given in Table.~\ref{initialvaluesforRGE}
  and concentrate on the impact of different Yukawa couplings $y_{1,2}$ in the remaining part of the article.
   With these initial values in Table.~\ref{initialvaluesforRGE}, we then run the parameters all the way up to $M_{Pl}$ scale.
  For RGE running, we use three-loop SM $\beta$ functions~\cite{Buttazzo:2013uya}.
  We also include one-loop contributions of new particles in the SDFDM model to the $\beta$ functions of these SM parameters.
  For new parameters in the SDFDM model, we use one-loop $\beta$ functions which can be extracted using $\texttt{PyR@TE 2}$\cite{Lyonnet:2016xiz}.
  The results are shown in Appendix~\ref{betafunction}.

\begin{figure}[!t]
 	\centering
 	\subfigure[\label{SDFDMlambdavsSM:a}]
 	{\includegraphics[width=.486\textwidth]{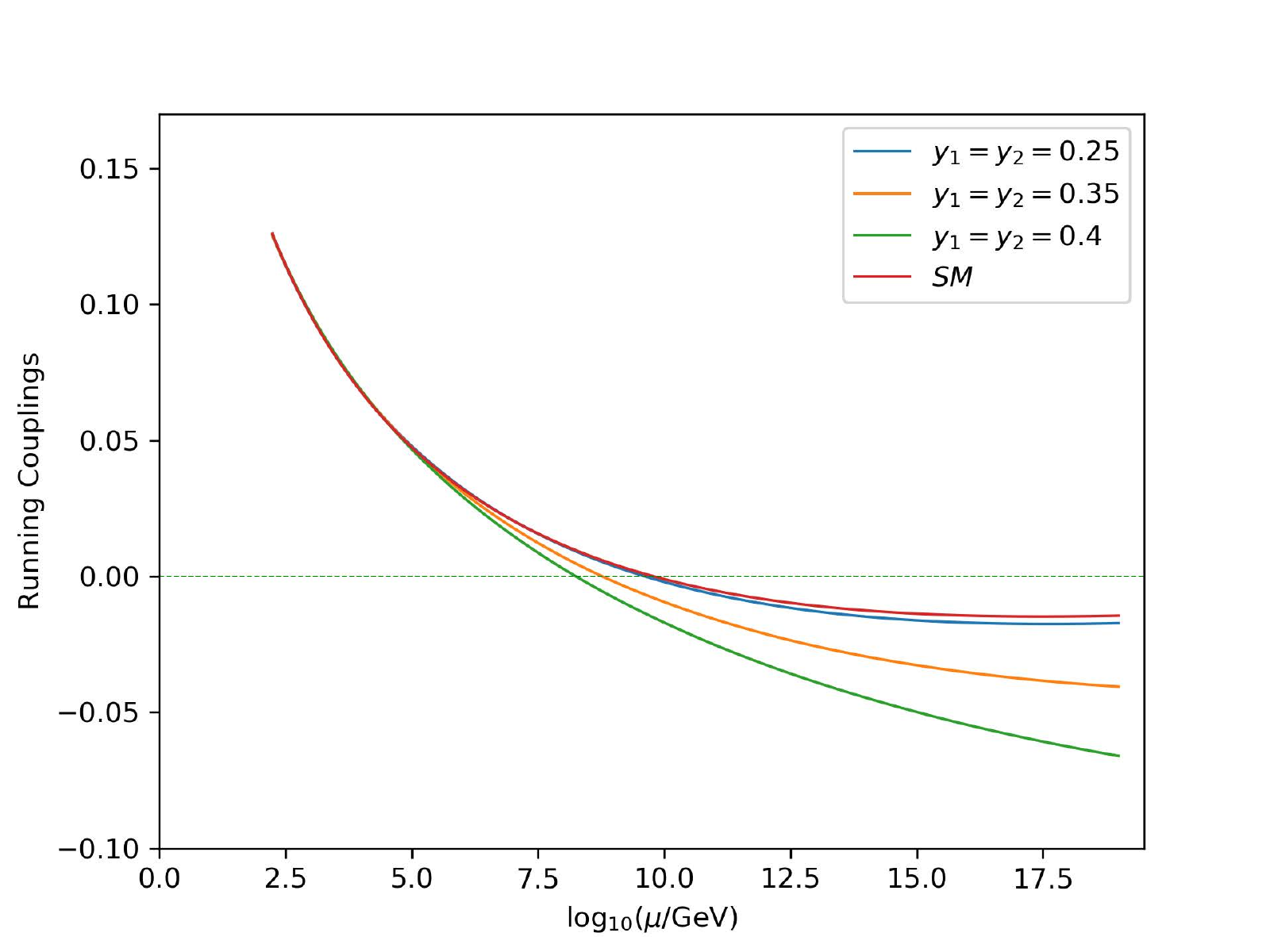}}
 	\subfigure[\label{lambdaeff:b}]
 	{\includegraphics[width=.486\textwidth]{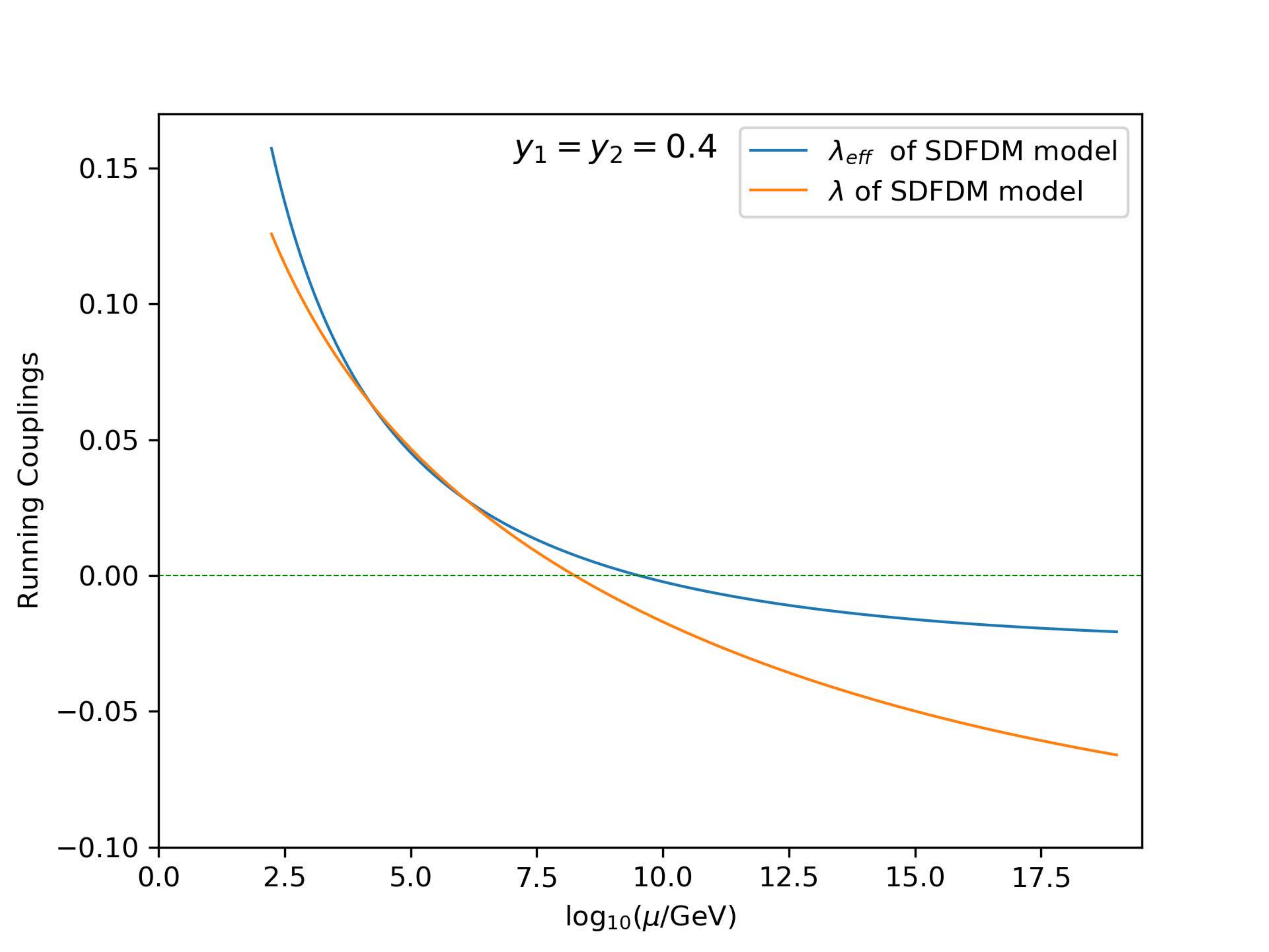}}
 	\caption{ (a) $\lambda(t)$ up to $M_{Pl}$ for the SM and for various Yukawa couplings in the SDFDM model;
	(b) Running $\lambda(t)$ and $\lambda_{eff}(t)$ up to $M_{Pl}$ scale for the SDFDM model.  
 	}
 	\label{lambdatitlediagrams}
 \end{figure}

  We can see that the evolution of $\lambda(t)$ both in the SM and in the SDFDM model in Fig.~\ref{SDFDMlambdavsSM:a}.
  We see that the $\lambda_{min}$ in the SDFDM model, the minimum of $\lambda(t)$ in the RGE running, is negative
  and is more negative than in the SM. This indicates that in the SDFDM model the EW vacuum is unstable and 
   lifetime of the EW vacuum could be much shorter owing to new physics effects.
   The greater the Yukawa couplings $y_{1}$ and $y_{2}$, the greater the destabilization effects of the SDFDM model.
  
  As shown in Eq.~\eqref{lambdaeff}, $\lambda_{eff}$ differs from $\lambda$.
  In the SM, the difference $\lambda_{eff} - \lambda$ is always positive and negligible near the Planck scale as shown in Ref. \cite{Degrassi:2012ry}.
   The situation is different in the SDFDM model.  As we can see in Fig.~\ref{lambdaeff:b},
    $\lambda_{eff} - \lambda$ is not negligible in the SDFDM model. 
    In fact,  $\lambda_{eff}$  is suppressed by the $e^{4 \Gamma(t)}$ factor in Eq.~\eqref{lambdaeff}
    which comes from the contribution of the anomalous dimension.
    As we can see, the instability scale $\Lambda_{I}$, the energy scale at which $\lambda_{eff}(t)$ or $\lambda(t)$ becomes zero,
    is larger when determined by $\lambda_{eff}(t)$. This is the case both in the SM and in the SDFDM model.
   
  \begin{table}[h]
 	\setlength{\tabcolsep}{.5em}
 	\renewcommand{\arraystretch}{1.2}
 	\begin{tabular}{c|cccc}
 		\multicolumn{5}{c}{Initial values in $\MS$ scheme for RGE running} \\
 		\hline
 		$\quad\quad\mu=M_t\quad\quad$ & $\quad\quad\lambda\quad\quad$ &\quad\quad$y_t\quad\quad$ &\quad\quad$g_2\quad\quad$ &$\quad\quad g_Y\quad\quad$ \\
 		\hline
 		$\mathrm{SM}_{\mathrm{LO}}$ & 0.12917 & 0.99561 &0.65294 & 0.34972 \\
 		$\mathrm{SM}_{\mathrm{NNLO}}$ & 0.12604 & 0.93690$^*$ & 0.64779 & 0.35830 \\
 		$\quad\quad\mathrm{SDFDM_{\mathrm{NLO}}^{\mathrm{BMP1}}}$ & 0.12549 & 0.93526$^*$ & 0.64573 & 0.35752 \\
 		$\quad\quad\mathrm{SDFDM_{\mathrm{NLO}}^{\mathrm{BMP2}}}$$\quad$ & 0.12554 & 0.93368$^*$ & 0.64574 & 0.35630 \\
 		$\quad\quad\mathrm{SDFDM_{\mathrm{NLO}}^{\mathrm{BMP3}}}$$\quad$ & 0.12586 & 0.93269$^*$ & 0.64573 & 0.35553 \\
 		$\quad\quad\mathrm{SDFDM_{\mathrm{NLO}}^{\mathrm{BMP4}}}$$\quad$ & 0.13126 & 0.92744$^*$ & 0.64573 & 0.35144 \\
 		\hline
 	\end{tabular}
 	\caption{All the parameters are renormalized at the top pole mass($M_t$) scale in the $\MS$ scheme.    
 		BMP1: $y_1=y_2=0.25$, $M_S=1000$~GeV, $M_D=1000$~GeV; BMP2: $y_1=y_2=0.35$, $M_S=1000$~GeV, $M_D=1000$~GeV;
 		BMP3: $y_1=y_2=0.4$, $M_S=1000$~GeV, $M_D=1000$~GeV; BMP4: $y_1=y_2=0.6$, $M_S=1000$~GeV, $M_D=1000$~GeV;
 		The superscript $*$ indicates that the NNNLO pure QCD effects are also included. BMPs means benchmark points.
 	}
 	\label{initialvaluesforRGE}
 \end{table}

\begin{table}[h]
\setlength{\tabcolsep}{.5em}
\renewcommand{\arraystretch}{1.2}
\begin{tabular}{c|cccc}
	\multicolumn{5}{c}{Effects of different masses on initial values} \\
	\hline
	$\quad\quad\mu=M_t\quad\quad$ & $\quad\quad\lambda\quad\quad$ &\quad\quad$y_t\quad\quad$ &\quad\quad$g_2\quad\quad$ &$\quad\quad g_Y\quad\quad$ \\
	\hline
	$M_S= M_D = 800$~GeV &0.12564 & 0.93402$^*$ & 0.64599 & 0.35650 \\
	$M_S= M_D = 1000$~GeV & 0.12554 & 0.93368$^*$ & 0.64574 & 0.35630 \\
	$M_S= M_D = 1200$~GeV & 0.12546 & 0.93340$^*$ & 0.64552 & 0.35613 \\
	\hline
\end{tabular}
\caption{$y_{1} = y_{2} = 0.35$ for all three cases with different masses of the new particles. 
	The superscript $*$ indicates that the NNNLO pure QCD effects are also included. }

\label{effectofdifferentmasses}
\end{table}

\section{Vacuum stability and lifetime of the vacuum}
\label{sec:vacuumstability}
As we have seen in the last section, RG improvement to the effective potential can be quite significant in SDFDM model. 
We need to consider the effects of RG improved effective potential in the calculation of the vacuum decay rate.
The decay rate of the false vacuum can be computed by finding a bounce solution to the field equations in Euclidean space
\cite{Langer, Coleman:1977py,Callan:1977pt}.
For a potential $U(\phi)$, the decay rate per unit time per unit volume, $\Gamma_t$, can be expressed as
\begin{eqnarray}
\Gamma_t= A_ t~e^{- S_{cl} }. \label{decayrate}
\end{eqnarray}
where $S_{cl}$ is the Euclidean action of bounce solution and $A_t$ is the quantum correction.
For fluctuation of $\phi$ field,  $A_t$ is given as
\begin{eqnarray}
A_t=\frac{S^2_{cl}}{4\pi^2} \bigg| \frac{\det'[-\partial^2+U^{''}(\phi_B)]}{\det[-\partial^2+U^{''}(\phi_0)]}\bigg|^{-1/2}, \label{decayrate-A}
\end{eqnarray}
where $\partial^2$ is operated on Euclidean space and $\det'$  the determinant omitting the zero mode contribution.
$\phi_0$ is the field value in the false vacuum which can be taken as zero as an approximation.
$\phi_B$ refers to the spherical symmetric bounce solution to the Euclidean field equation. $\phi_B$ satisfies
\begin{equation}
-\partial^2 \phi_B+U^{\prime}(\phi_B)=-\frac{d^{2} \phi_B}{d r^{2}}-\frac{3}{r} \frac{d \phi_B}{d r}+U^{\prime}(\phi_B)=0,
\end{equation}
where $U'$ means derivative of $U$ with respect to the field.
In the case under consideration,  $\phi/\sqrt{2}$ is the CP-even neutral component of the Higgs doublet in the SM.
If there are other particles coupled to bounce field, their contributions to the determinants should also be taken into account,
as happens in the SM and in the singlet-doublet fermion extension models considered in this article.

For a potential $U(\phi)=\frac{\lambda}{4}\phi^4$ with a negative $\lambda$, the calculation leads to~\cite{Isidori:2001bm}
\begin{eqnarray}
S_{cl}=\frac{8 \pi^{2}}{3\left|\lambda \right|} . \label{bounce-action-1}
\end{eqnarray}
In the SM, there is a mass of the Higgs field.
The Higgs mass can be safely neglected in this calculation because the bounce solution is dominated by the behavior at large field values,
so that the potential can be written approximately as a $\phi^4$ form.
In quantum theory, $\lambda$ is a quantity running with energy scale.  
To simplify calculation, $\lambda$ can be taken at a sufficiently large energy  scale $M$ so that $\lambda(M)$ is negative 
and varies slowly with energy scale. So $S_{cl}= 8 \pi^{2}/|\lambda(M)|$ in this case.
It has been shown that this scale dependence of $S_{cl}$ in the false vacuum decay rate  is cancelled 
when taking into account one-loop correction from the determinant~\cite{Isidori:2001bm, Chigusa:2018uuj}.

To fully take into account quantum corrections, we need to consider the effective action.
As long as the field varies slowly with space and time, we can compute the effective action using derivative expansion~\cite{Coleman:1973jx}.
Neglecting terms with higher derivative, we can write the effective action in Euclidean space for external field $\phi$ as
\begin{equation}
\label{effaction1}
S_{\mathrm{eff}}[\phi]=\int d^{4} x\left[\frac{1}{2}\left(\partial_{\mu} \phi\right)^{2} Z_{2}(\phi)+V_{\mathrm{eff}}(\phi)\right].
\end{equation}
$Z_2$ can be obtained from the $p^2$ terms in the Feynman diagrams as shown in Appendix.~\ref{derivativecorr}.
It is renormalized to make $Z_2(\phi=0)=1$ which makes the
kinetic term going back to the standard form when there is no external field.
The results in the 't Hooft-Feynman gauge are summarized in Table. \ref{selfenergy}  for the SM, 
and in Table. \ref{newselfenergy} for new contributions in SDFDM model. In the large $\phi$ limit, we can simplify the
result. We obtain $Z_2$ for the SM in Eq.~\eqref{Z2highphiSM},  and $Z_2$ for the SDFDM model in Eq.~\eqref{Z2highphiSDFDM}.
As we can see, the explicit dependences on $\phi$ are cancelled in these results.

\begin{figure}[!t]
	\centering
	\subfigure[\label{Z2factorSM}]
	{\includegraphics[width=.486\textwidth]{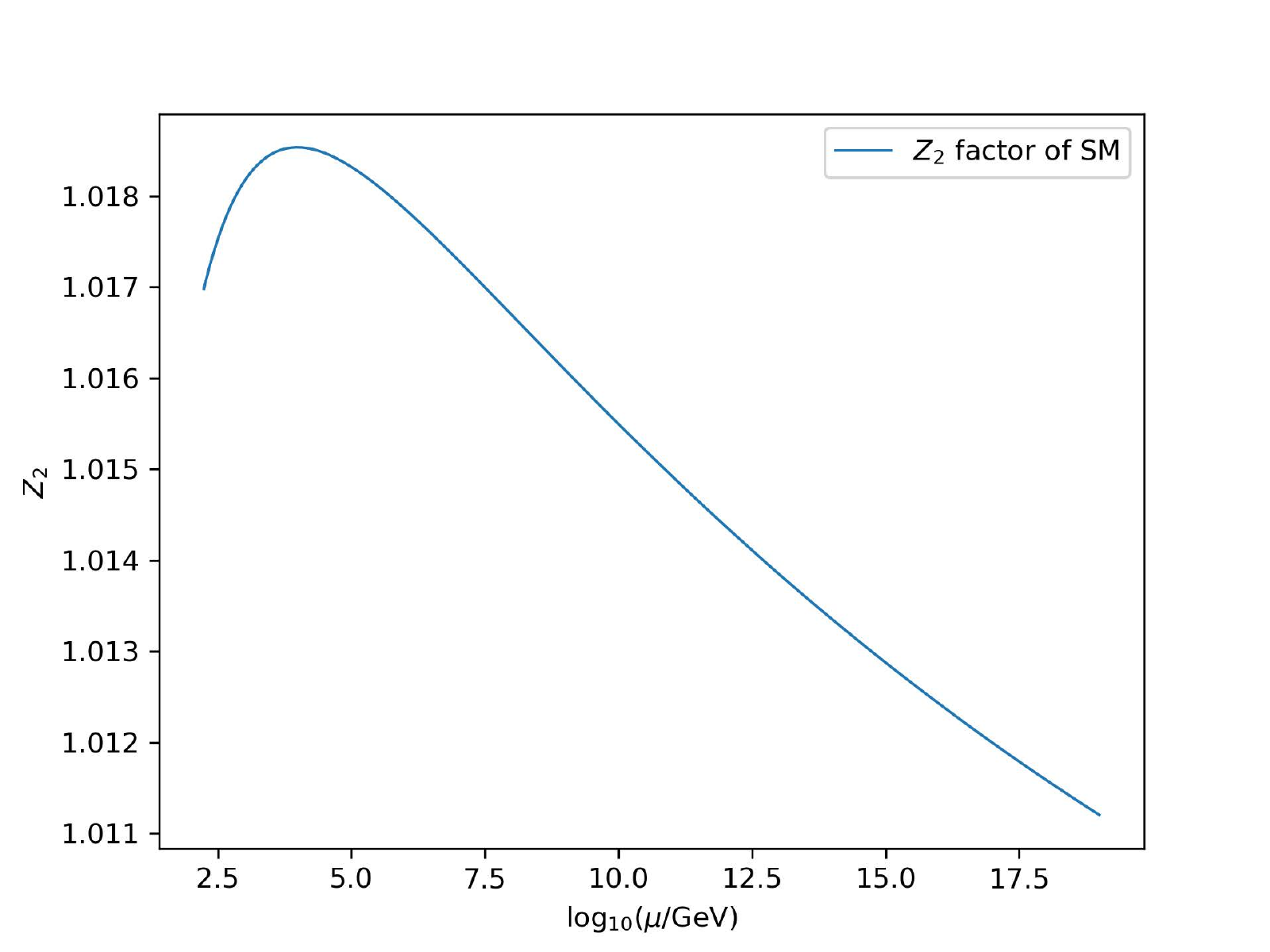}}
	\subfigure[\label{Z2factorSDFDM}]
	{\includegraphics[width=.486\textwidth]{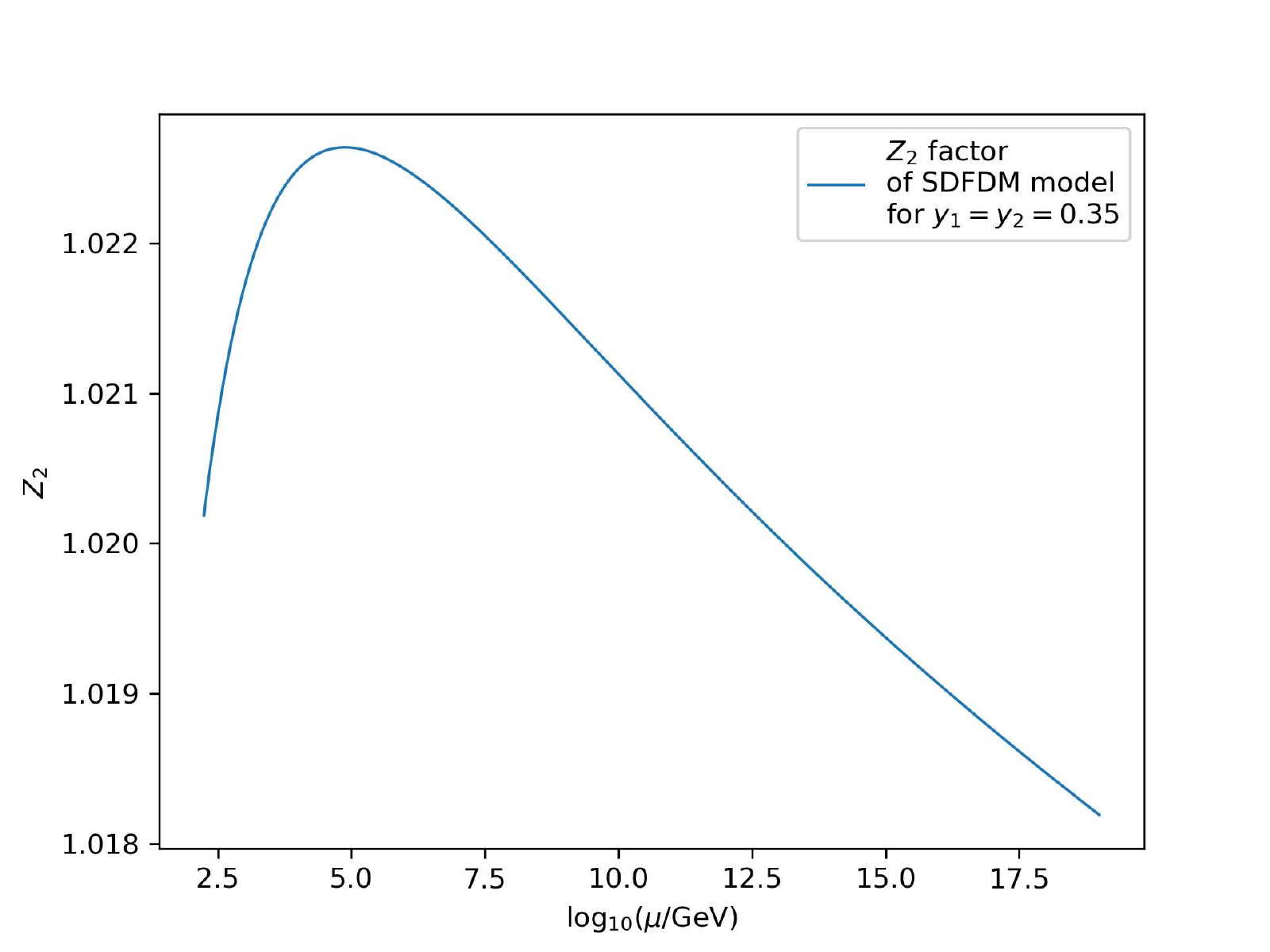}}
	\caption{ (a)The behavior of $Z_2$ at large energy scale in the SM; 
		(b) The behavior of $Z_2$ at large energy scale with $y_{1} = y_{2} = 0.35$ and $M_{S} = M_{D} = 1000$ GeV in the SDFDM model.}
	\label{Z2largephi:fig}
\end{figure}

RG improvement of the kinetic term can be studied similar to the effective potential. 
The kinetic term in the effective action is the one-particle irreducible self-energy $\Gamma_2$.
It satisfies the RG equation
\begin{eqnarray}
  \left(\mu \frac{\partial}{\partial \mu}+\beta_{i} \frac{\partial}{\partial \lambda_{i}}-2 \gamma \right) \Gamma_2(\phi)=0.
\end{eqnarray}
The equation can be solved in a way similar to solving $V_{eff}(\phi)$. 
Solving this equation gives rise to $Z_2(\phi,t)$ with
 all parameters $\lambda_i$ in $Z_2(\phi)$ substituted by $\lambda_i(t)$ and with an $e^{2\Gamma(t)}$ factor in the kinetic term. So we arrive at an Euclidean action
\begin{eqnarray}
S=\int d^4 x~\big[ e^{2 \Gamma(t) } Z_2(\phi,t) \frac{1}{2} (\partial_\mu \phi)^2+ e^{4 \Gamma(t)}  \frac{{\tilde \lambda}(t)}{4} \phi^4 \big],
\label{action-2}
\end{eqnarray}
where $\widetilde{\lambda}$ is only different from  Eq.~\eqref{lambdaeff} by a factor $e^{4 \Gamma(t)}$, that is
\begin{equation}
\label{lambdatilde}
\widetilde{\lambda}(t) = \lambda(t)+\frac{1}{(4 \pi)^{2}} \sum_{i} N_{i} \kappa_{i}^{2}(t)\left(\log \kappa_{i}(t) e^{2 \Gamma(t)}-c_{i}\right).
\end{equation}

The Euclidean equation of bounce solution becomes
\begin{eqnarray}
-Z_{2} ~\partial^2 {\tilde \phi}_B   + \widetilde{\lambda} ~{\tilde \phi}^{3}_B ~e^{2 \Gamma(t)}= 0. \label{bounce-eq-1}
\end{eqnarray}
From the bounce action in Eq. (\ref{bounce-action-1}), one can immediately deduce that the bounce action becomes
\begin{eqnarray}
\label{bounceaction}
S_{cl}=e^{2\Gamma} Z_2 \times \frac{8 \pi^2}{3 |{\tilde \lambda}| e^{2\Gamma} /Z_2 }=
(Z_2)^2 \frac{8 \pi^2}{ 3|{\tilde \lambda}|}. \label{bounce-action-2}
\end{eqnarray}
$S_{cl}$ depends on $Z_2$ but is independent of the $e^{\Gamma(t)}$ factor.
Similar to the case of obtaining Eq. (\ref{bounce-action-1}),  
running parameters in Eqs. (\ref{bounce-eq-1}) and (\ref{bounce-action-2}) 
are understood to be at an arbitrary large energy scale $M$. The leading dependence on $M$ in 
the decay rate would be cancelled by including quantum correction from the determinant, similar to analysis
in Ref. ~\cite{Isidori:2001bm, Chigusa:2018uuj}

 Similarly, one can find that $(S_{cl})^2$  factor in Eq. (\ref{decayrate-A}), which comes from the zero mode contribution, becomes
$[8 \pi^2/(3 |{\tilde \lambda}| e^{2\Gamma} /Z_2 ) ]^2$.  The ratio of determinants in Eq. (\ref{decayrate-A}) becomes
$\big| \det'[- e^{2\Gamma} Z_2 \partial^2+ 3{\tilde \lambda} e^{4\Gamma}{\tilde \phi}_B^2]/\det[-e^{2\Gamma} Z_2\partial^2 ]\big|^{-1/2}$
which equals to  $\big| \det'[-  \partial^2+3({\tilde \lambda}/Z_2) e^{2\Gamma} {\tilde \phi}_B^2]/\det[-\partial^2 ]\big|^{-1/2} \times (e^{2\Gamma} Z_2)^2$ when including effects omitting four zero modes.  It is easy to see that if taking $\phi_B=e^{\Gamma} {\tilde \phi}_B$
the non-zero eigenvalues of operator $- \partial^2+3({\tilde \lambda}/Z_2) e^{2\Gamma} {\tilde \phi}_B^2$ for
${\tilde \phi}_B$  satisfying Eq.  (\ref{bounce-eq-1}) would be the same of the operator
$- \partial^2+3({\tilde \lambda}/Z_2)  \phi_B^2$  for $\phi_B$ satisfying 
\begin{eqnarray}
-Z_{2} ~\partial^2  \phi_B   + \widetilde{\lambda} ~\phi^{3}_B = 0. \label{bounce-eq-2}
\end{eqnarray}
So eventually we find that the decay rate is again expressed by Eq. (\ref{decayrate}) but with $S_{cl}$ expressed by Eq. (\ref{bounce-action-2})
and with
\begin{eqnarray}
A_t=\frac{S^2_{cl}}{4\pi^2} \bigg| \frac{\det'[-\partial^2+3({\tilde \lambda}/Z_2)  \phi_B^2 ]}{\det[-\partial^2]} \bigg|^{-1/2}, \label{decayrate-B}
\end{eqnarray}
in which $\phi_B$ satisfies Eq. (\ref{bounce-eq-2}).
We see that the final result depends on $Z_2$ but does not depend on $e^{\Gamma(t)}$.
The  factor $e^{\Gamma(t)}$ comes from the wave function renormalization but can be associated with an arbitrariness in relating $\phi$ with  a
renormalization scale. So it is not surprising to see that the physical result does not depend on it.
One can actually re-define, from the very beginning, the external field $\phi$ of the Euclidean action in Eq. (\ref{action-2}) 
in the path integral and arrive at this conclusion.

Note that the idea that physical quantities should not depend on $e^{\Gamma}$  has been expressed in \cite{Espinosa:2015qea}. 
In this article, the authors re-define the field and introduce canonically normalized field in effective action. 
In our convention, the canonically normalized field
$\phi_{can}$ is related to $\phi$ with the equation $d \phi_{can}/d\phi= e^{\Gamma} \sqrt{Z_2}$. 
The solution of $\phi_{can}$ can be found approximately as 
$\phi_{can}=e^{\Gamma}\sqrt{Z_2} \phi+\gamma e^{\Gamma} \sqrt{Z_2} \phi $ in which the derivative with respect to 
$\sqrt{Z_2}$, that gives contribution of higher order, has been neglected.  As can be found numerically, the anomalous dimension $\gamma$ is at most
a few of a thousand in high energy scale if the new Yukawa couplings are not too large. 
Consequently, we can take $\phi_{can}=e^{\Gamma}\sqrt{Z_2} \phi$  approximately in high energy scale, which agrees with the result
in \cite{Espinosa:2015qea} up to the factor $\sqrt{Z_2}$.
The coupling $\lambda_{can}$,  introduced  in \cite{Espinosa:2015qea} using the canonically normalized field, can be found 
approximately as ${\widetilde \lambda}/(Z_2)^2$ in our case. This agrees with the result in (\ref{bounce-action-2}) which also 
depends on ${\widetilde \lambda}/(Z_2)^2$.  So the real quartic coupling which controls physical quantities is $\lambda_{can}$, 
not $\lambda_{eff}$, or just ${\widetilde \lambda}$ if $Z_2$ is close to one.  As will be shown in detail later, $Z_2$ is indeed close to one
and the difference between ${\widetilde \lambda}$ and $\lambda$ in the SM is also very small at high energy scale. 
In the SDFDM model,  ${\widetilde \lambda}-\lambda$ could be smaller than $\lambda_{eff}-\lambda$ shown in Fig.~\ref{lambdaeff:b}.
However,  ${\widetilde \lambda}-\lambda$ can still be significant in some cases, as will be shown later.
So a careful analysis of vacuum stability and vacuum decay in extensions of the SM should use ${\widetilde \lambda}$  and take the 
relevant quantum corrections into account.

 $Z_2$ is a running parameter. As we can see in Fig.~\ref{Z2largephi:fig}, $Z_2$ has a small deviation from unity
at high energy scale, both in the SM and in the SDFDM model. 
So the decay rate of false vacuum is mainly controlled by the behavior of ${\tilde \lambda}(t)$. 
In the SDFDM model, the scale dependence appearing in $S_{cl}$ is also cancelled by one-loop contribution from the determinant.
This energy scale can be taken conveniently at $\Lambda_B$, the scale of bounce, so that $S_{cl}(\Lambda_B)$ takes care of the major 
contribution in the exponential~\cite{Isidori:2001bm, Chigusa:2018uuj, Chigusa:2017dux, Degrassi:2012ry}. 
$\Lambda_{\mathrm{B}}$ is determined as the scale at which the vacuum decay rate is maximized. 
In practice, this roughly corresponds to the scale at which
 the negative ${\tilde \lambda}(\Lambda_{\mathrm{B}})$  is at the minimum.
If $\Lambda_{\mathrm{B}}>M_{\mathrm{Pl}}$ , we can only obtain a lower bound on the tunneling probability by setting 
$\lambda\left(\Lambda_{\mathrm{B}}\right)=\lambda\left(M_{\mathrm{Pl}}\right)$.

In this way, the vacuum decay probability $P_0$ in our universe up to the present time
can be expressed as~\cite{Degrassi:2012ry,Khan:2014kba} 
 \begin{equation}
\mathcal{P}_{0}=0.15 \frac{\Lambda_{B}^{4}}{H_{0}^{4}} e^{-S\left(\Lambda_{\mathrm{B}}\right)}
\end{equation}
where $H_{0}=67.4 ~\mathrm{km} ~\mathrm{sec}^{-1} ~\mathrm{Mpc}^{-1}$ is the Hubble constant at the present time.
$S\left(\Lambda_{\mathrm{B}}\right)$ is the action of the bounce of size $R=\Lambda_{\mathrm{B}}^{-1}$.

\begin{table}[!t]
	\setlength{\tabcolsep}{.5em}
	\renewcommand{\arraystretch}{1.2}
	\begin{tabular}{|c|ccc|cccc|}
		\hline
		\multirow{2}{*}{} &\multicolumn{3}{c}{Result with $\lambda(t)$}& \multicolumn{4}{c}{Result with ${\tilde \lambda}(t)$ in effective potential}\\
		\hline
		&$\lambda_{min}$ & $\log_{10}(\mu_{min}/\mathrm{GeV})$ & $\log_{10}(\mathcal{P}_0)$
		& $\widetilde{\lambda}_{min}$ &$Z_{2}(\Lambda_{B})$& $\log_{10}(\mu_{min}/\mathrm{GeV})$ & $\log_{10}(\mathcal{P}_0)$\\
		\cline{1-3}
		\hline
		$SM$ & -0.0148 & 17.46 & -535.34 & -0.0150 & 1.0116 &18.07 & -543.35 \\
		$\mathrm{BMP1}$ & -0.0176 & $17.60$ & -413.72 & -0.0165 & 1.0152 & 18.23& -474.68 \\
		$\mathrm{BMP2}$ & -0.0406 & $M_{Pl}$ & -38.98 & -0.0346 & 1.0182 & $M_{Pl}$& -99.73 \\
		$\mathrm{BMP3}$ & -0.0661 & $M_{Pl}$ & unstable &-0.0539 &1.0231 & $M_{Pl}$& unstable \\
		\hline
	\end{tabular}
	\caption{
		The results computed by using  $\lambda(t)$ and $\widetilde{\lambda}(t)$ are presented.
		Three benchmark models are
		BPM1($y_{1} = y_{2} = 0.25, M_{S}  =M_{D} = 1000$ GeV ),
		BPM2($y_{1} = y_{2} = 0.35, M_{S}  =M_{D} = 1000$ GeV),
		BPM3($y_{1} = y_{2} = 0.4, M_{S}  =M_{D} = 1000$ GeV ).
		$\lambda_{\mathrm{min}}$ is the minimal value of the running $\lambda$.
		$\widetilde{\lambda}_{min}$ is the minimal value of the running $\widetilde{\lambda}$.
		$\mu_{min}$ is the energy scale when minimal value of $\lambda$ or ${\tilde \lambda}$ is achieved.
		$\mathcal{P}_0$ represents the EW vacuum decay probability.
	}
	\label{SDFDMbmp}
\end{table}

In vacuum stability analysis, we call the vacuum stable if the potential at large $\phi$ keeps positive. 
This requires $\widetilde{\lambda} > 0$ for energy scale up to the Planck scale. 
If $\widetilde{\lambda}<0$ at an energy scale but with $P_{0} < 1$, it means 
that the lifetime of the false vacuum is greater than the age of the Universe. In this case 
we call the vacuum metastable. Other scenarios can be similarly defined.  In summary, we list them as follows.
\begin{itemize}
	\item Stable: $\widetilde{\lambda} > 0$ for $\mu<M_\mathrm{Pl}$;
	\item Metastable: $ \widetilde{\lambda}(\Lambda_\mathrm{B}) < 0$ and $\mathcal{P}_0<1$;
	\item Unstable: $ \widetilde{\lambda}(\Lambda_\mathrm{B}) < 0$ and $\mathcal{P}_0>1$;
	\item Non-perturbative: $|\lambda|>4 \pi$ before the Planck scale
\end{itemize}

Note that we classify states of EW vacuum in a way different
from Ref.~\cite{Degrassi:2012ry,Wang:2018lhk},  since $\widetilde{\lambda}(t)$  differs from $\lambda(t)$ by 
one-loop Coleman-Weinberg type corrections. 
 As will be shown, $\widetilde{\lambda}(t)$  can be different from $\lambda(t)$ significantly in the SDFDM model.
  We further note that the effective action we have used has an imaginary part. The present work actually works on real part of
 the effective action and discusses the effect of the distortion of the bounce solution in the presence of quantum correction to the effective action.
  A discussion on the effect of the imaginary part of the effective action would be interesting, e.g. as in Ref. \cite{Andreassen:2016cvx}.
  In the present article, we will not elaborate on this topic.

\begin{figure}[!t]
	\centering
	\subfigure[\label{SMlambdavstildelambda:b}]
	{\includegraphics[width=.486\textwidth]{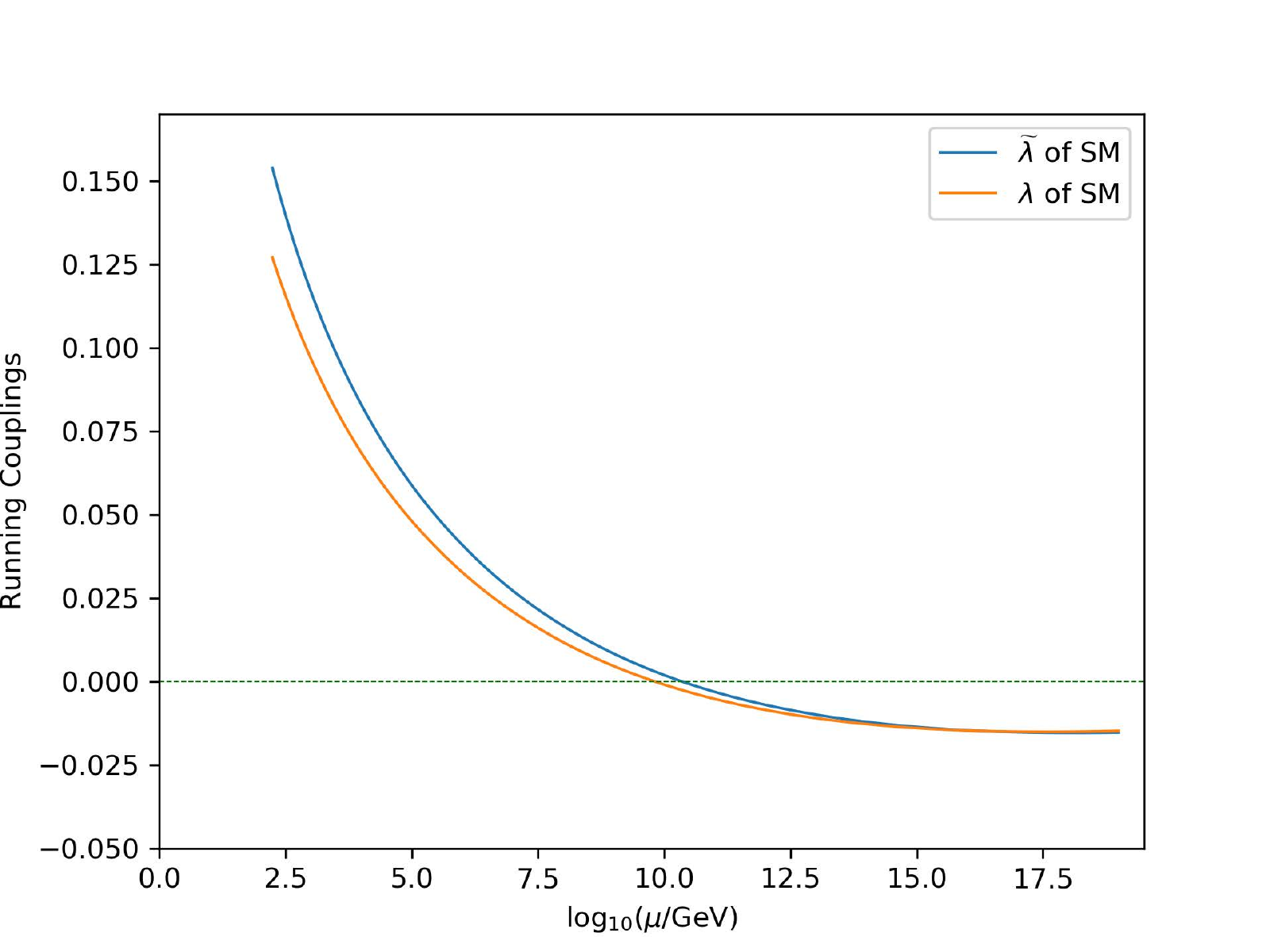}}
	\subfigure[\label{lamevolutionbdaeffSDFDM:a}]
	{\includegraphics[width=.486\textwidth]{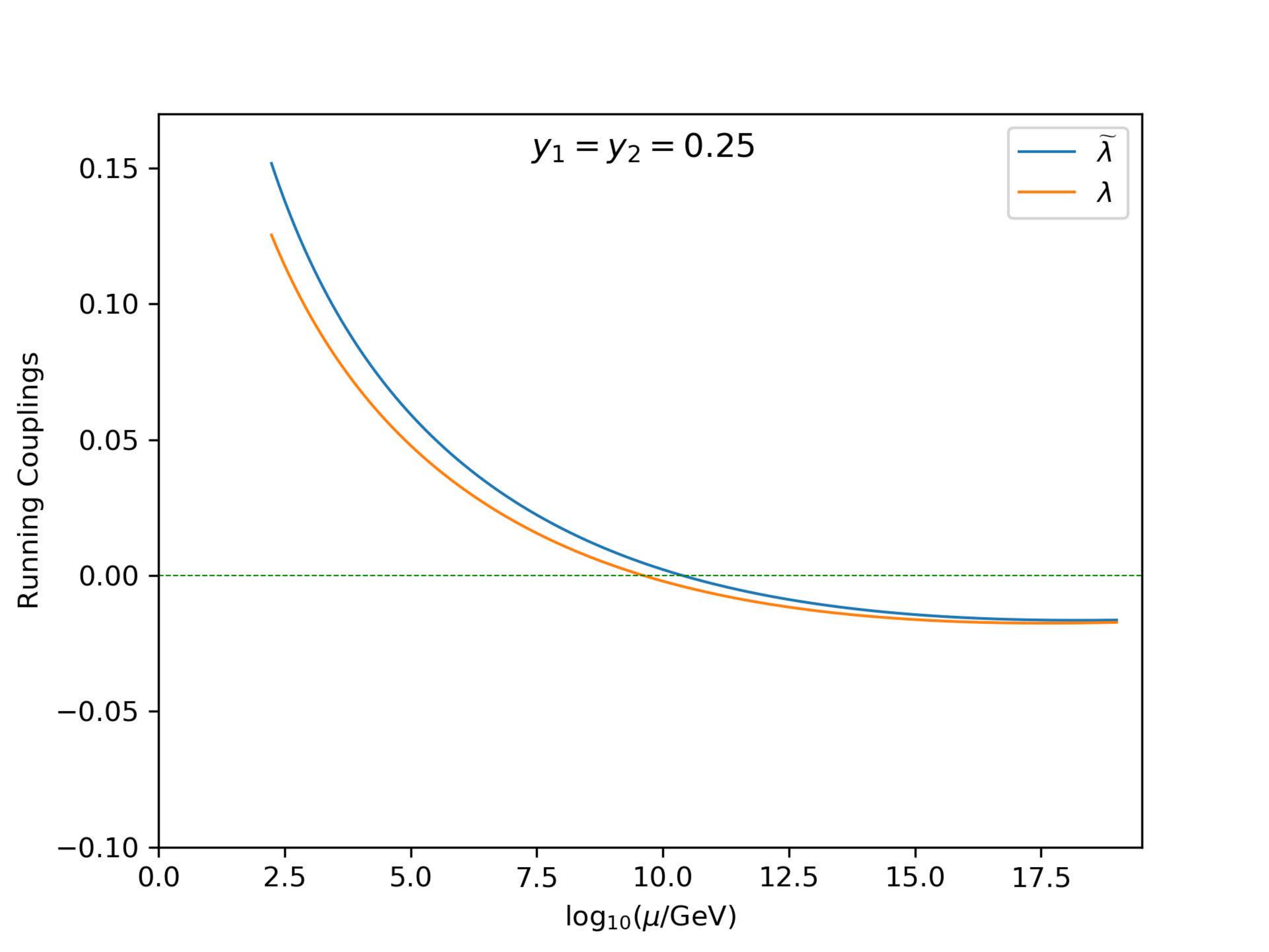}}
	\caption{(a)Comparison between $\lambda$ and $\widetilde{\lambda}$  in the SM; 
	(b) Comparison between  $\lambda$ and ${\tilde \lambda}$ with $y_{1} = y_{2} = 0.25$ and $M_{S}  =M_{D} = 1000$ GeV in the SDFDM model. 
	}
	\label{SDFDMlambdaeffdiagrams}
\end{figure}

\begin{figure}[!t]
	\centering
	\subfigure[\label{lambdatitlde1:a}]
	{\includegraphics[width=.486\textwidth]{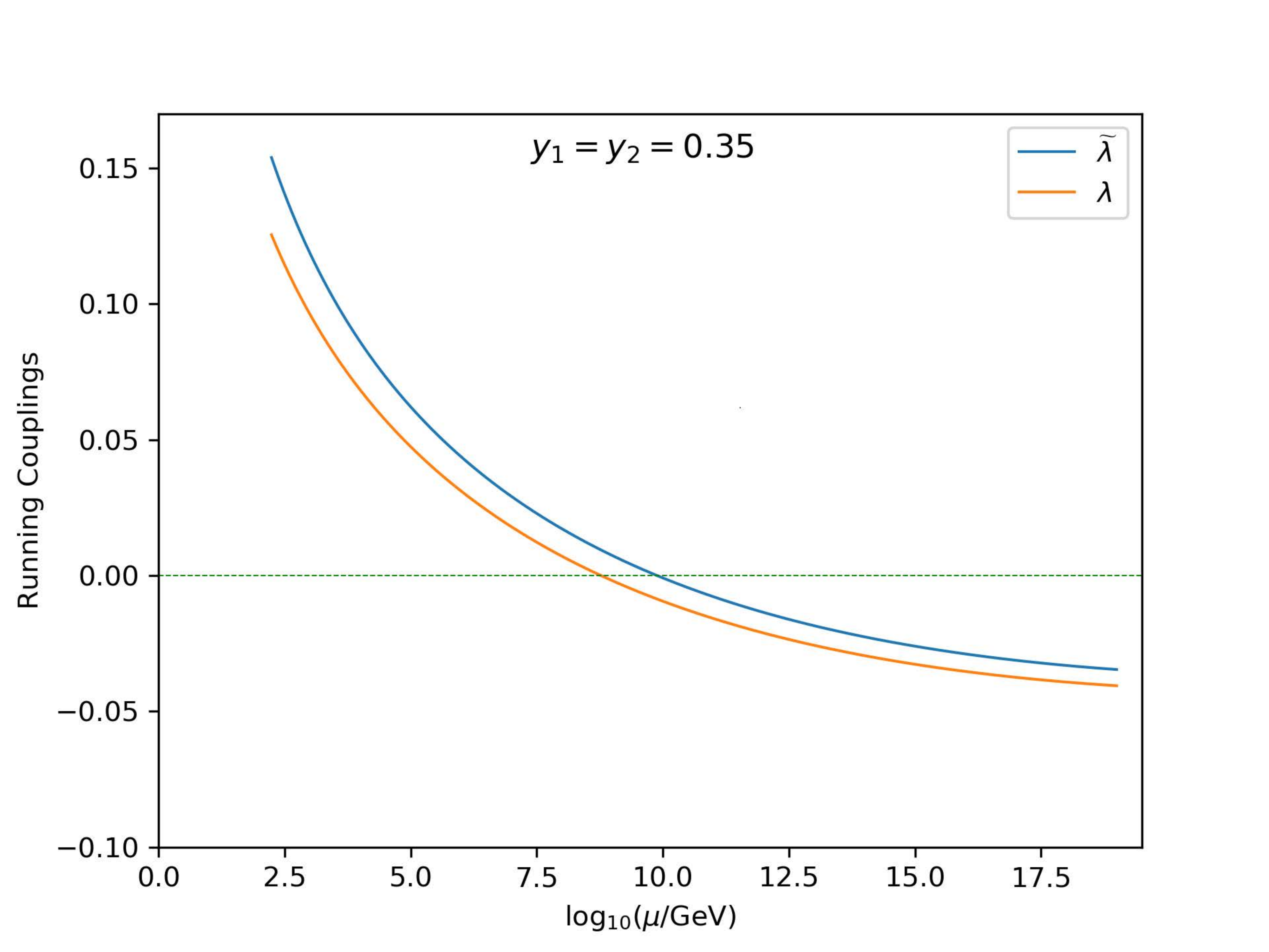}}
	\subfigure[\label{lambdatitlde2:b}]
	{\includegraphics[width=.486\textwidth]{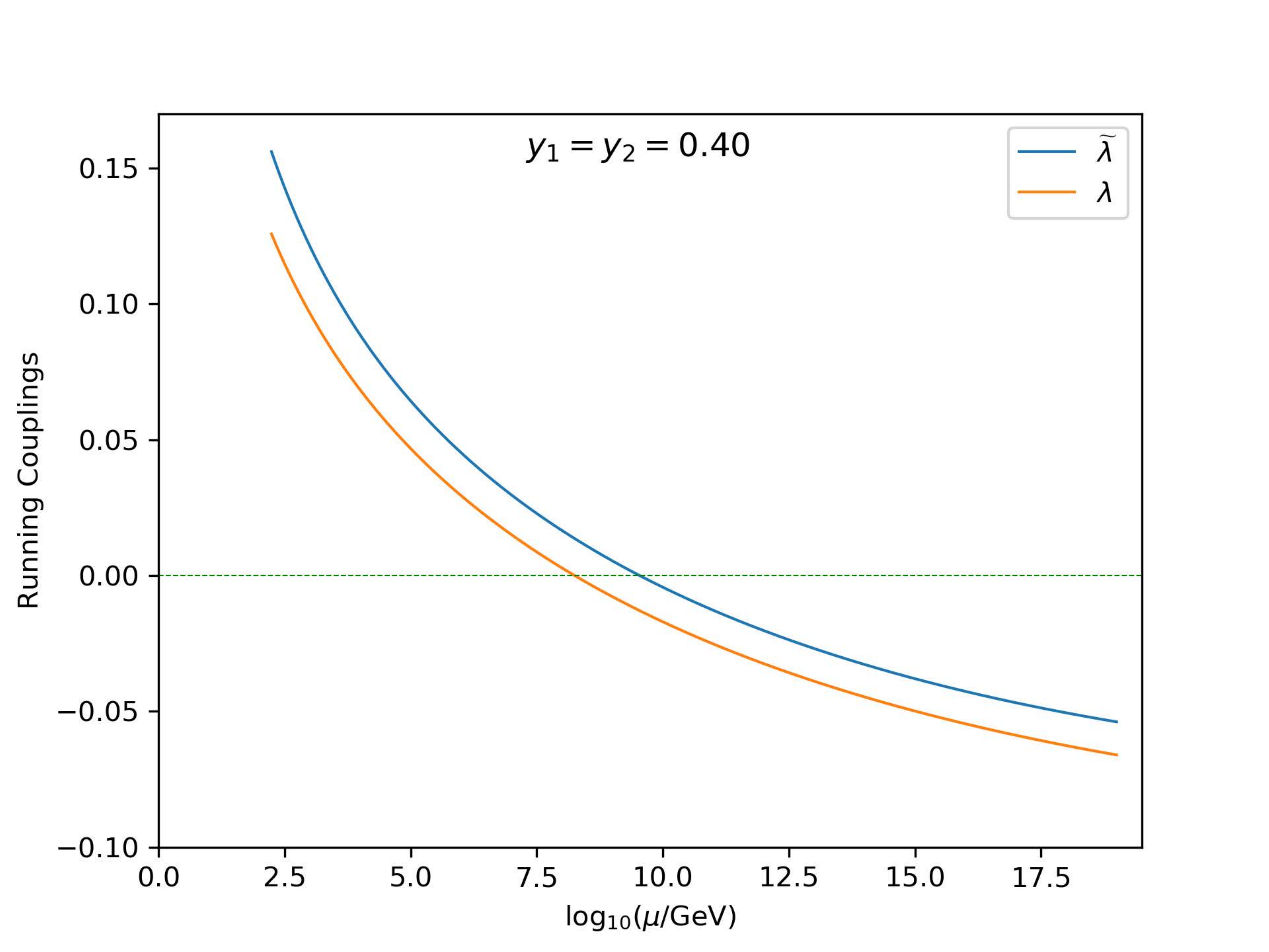}}
	\caption{
	Comparison between  $\lambda(t)$ and ${\tilde \lambda}(t)$ 
	in the SDFDM model.  (a) $y_{1} = y_{2} = 0.35$ and $M_{S}  =M_{D} = 1000$ GeV;
	(b) $y_{1} = y_{2} = 0.4$ and $M_{S}  =M_{D} = 1000$ GeV.
	}
	\label{lambdatitlediagrams}
\end{figure}

Now we come to discuss the tunneling probability.
As shown in Eqs. (\ref{decayrate}),  (\ref{bounce-action-2}) and (\ref{decayrate-B}), the decay rate of false vacuum
depends on $Z_2$ and ${\tilde \lambda}(t)$ when including one-loop correction to the effective action.
As mentioned before, the decay rate is mainly controlled by the behavior of ${\tilde \lambda}(t)$.
We first compare $\widetilde{\lambda}(t)$ and  $\lambda(t)$ in the SM. 
In the SM,  $\widetilde{\lambda}(t)$ and  $\lambda(t)$  are very close at high energy scale, as shown in Fig.~\ref{SMlambdavstildelambda:b}.
 They  both approach the minimum before the Planck scale. 
Both the values of their minima and the energy scales of the minima are very close, as can be seen in Table \ref{SDFDMbmp}.
This means that the one loop corrections to effective potential have little effects on the tunneling probability in the SM.
In the SDFDM model, the situation can be different. As can be seen in Fig. \ref{lamevolutionbdaeffSDFDM:a},
$\widetilde{\lambda}(t)$ and  $\lambda(t)$ at high energy scale are not as close as in the SM.
In this plot, $\widetilde{\lambda}(t)$ and  $\lambda(t)$ all approach their minima before the Planck scale. 
But their values at the minima and the energy scales of the minima are not as close as in the SM, as can be seen in Table \ref{SDFDMbmp}.
 In Fig~\ref{lambdatitlediagrams}, we give more plots with larger $y_1$ and $y_2$.
 In these cases,  the difference between $\widetilde{\lambda}(t)$ and  $\lambda(t)$ is more significant.
 The larger the Yukawa coupling $y_{1}$ and $y_{2}$, the larger the difference.
 We can see that the difference between ${\widetilde \lambda}$ and $\lambda$ in Fig.  \ref{lambdatitlde2:b} 
 is not as significant as the difference between $\lambda_{eff}$ and $\lambda$ in Fig.~\ref{lambdaeff:b}.
However,  ${\widetilde \lambda}-\lambda$ is still significant in this case.
 In these cases in Fig. \ref{lambdatitlediagrams}, both $\widetilde{\lambda}(t)$ and  $\lambda(t)$ have no minimum for energy scale below the Planck scale.
 The energy scale of bounce, $\Lambda_B$, is chosen as the Planck scale for these two cases.
 We note that the positive sign of  $\widetilde{\lambda} - \lambda$ shown in Fig~\ref{lambdatitlediagrams} 
 means that the lifetime calculated using $\widetilde{\lambda}$ in these plots is longer than that computed by using $\lambda$.

In Table. \ref{SDFDMbmp}, we list more numerical results for the SM and for some benchmark points in the SDFDM model. 
As a comparison, we also list the results just using $\lambda(t)$. We can see that using ${\tilde \lambda}(t)$ and $Z_2$ in the
effective action leads to some differences in the probability of false vacuum decay. 
 For the case of the SM, we can see that the lifetime of EW vacuum computed using effective action is slightly longer than 
 that computed just using $\lambda(t)$ although ${\tilde \lambda}(t)$ and $\lambda(t)$ are very close at high energy scale.
 This is caused mainly by the presence of $Z_2$ in the effective action. 
  In the SM,  $(Z_{2})^2$ term in Eq.~\eqref{bounce-action-2} is about 1.02 which makes $S_{cl}$  slightly larger 
  and leads to a smaller decay rate. 
  In SDFDM model, the difference between  $\widetilde{\lambda}$ and $\lambda$ is significant, and the $Z_2$ factor increases with the increase of the Yukawa couplings $y_1$ and $y_2$.
 Therefore, both the $Z_{2}$ factor and the increasing value of $\widetilde{\lambda} - \lambda$ makes the lifetime  
 calculated using effective action longer than that computed just using $\lambda$.

\begin{figure}[!t]
	\centering
	\subfigure[\label{y1y2stable:a}]
	{\includegraphics[width=.486\textwidth]{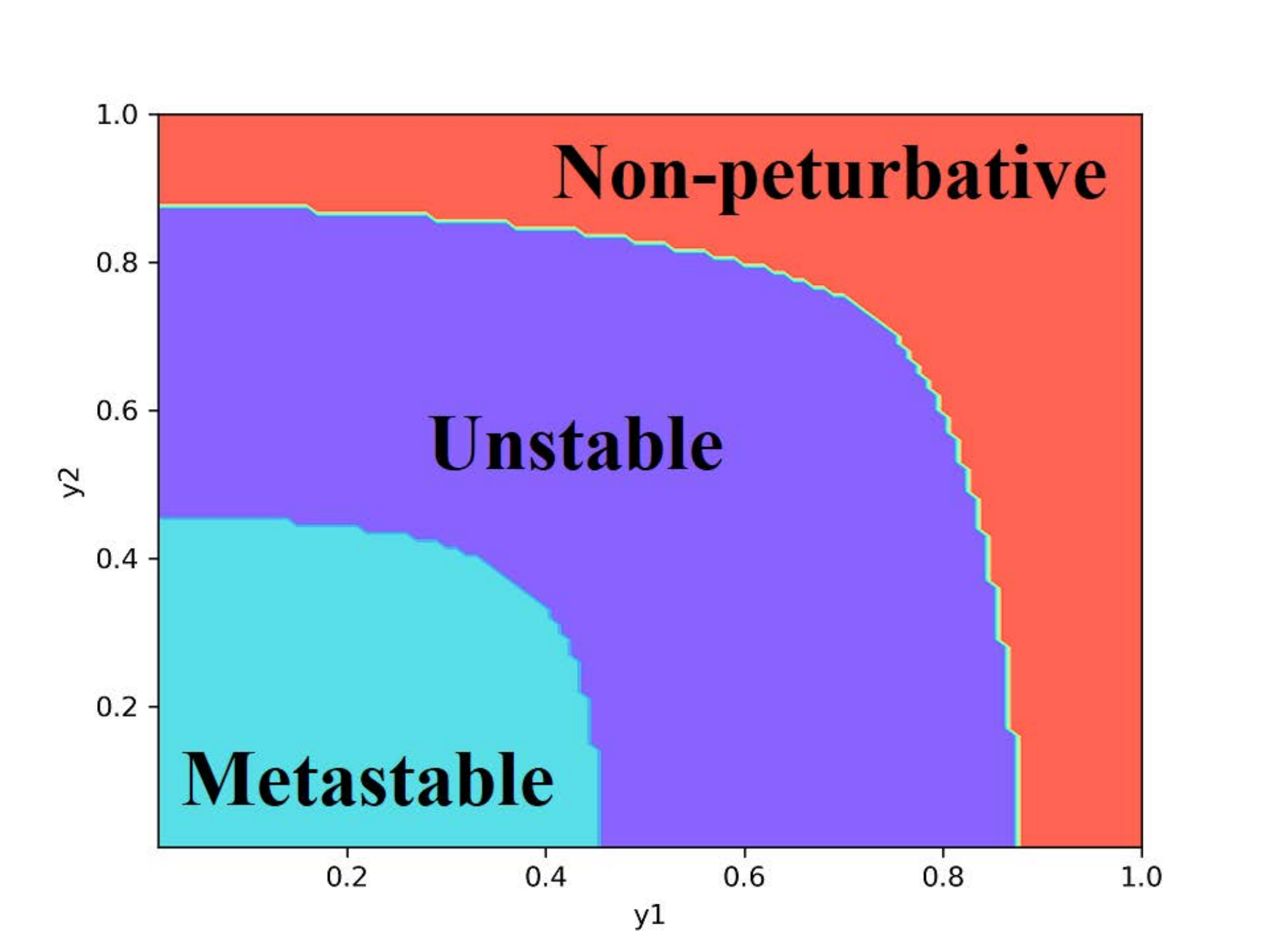}}
	\subfigure[\label{y1y2stableeff:b}]
	{\includegraphics[width=.486\textwidth]{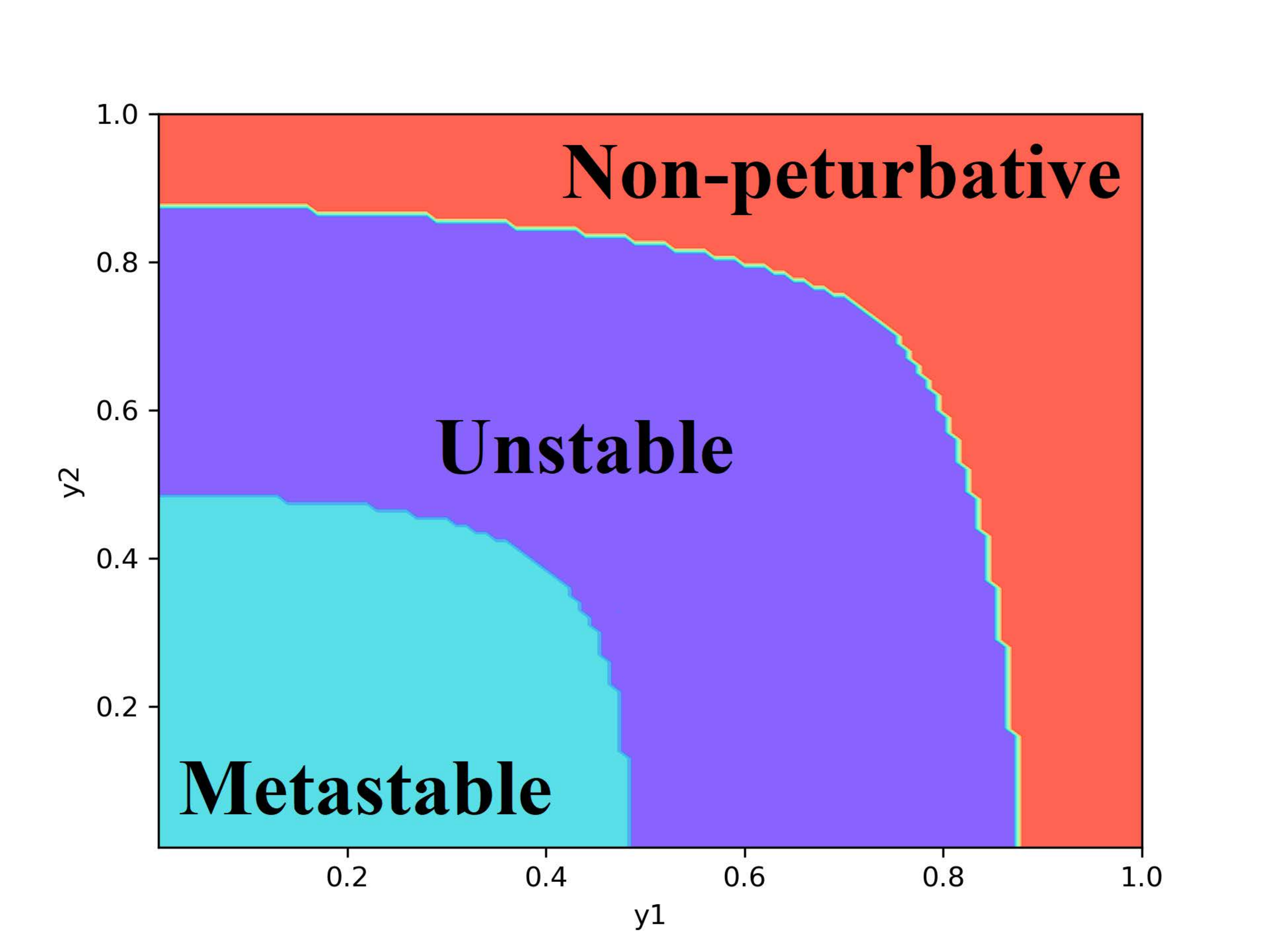}}
	\caption{Status of the EW vacuum in the $y_{1}-y_{2}$ plane with $M_{S}  = M_{D} = 1000$ GeV .
	The left panel is given by using the $\phi^4$ potential and the running $\lambda(t)$,  and the right panel
		is computed by using effective action and ${\tilde \lambda}(t)$.
		The green, blue, red regions indicate that the EW vacuum is metastable, unstable and non-perturbative.
	}
	\label{comparey1y2plane}
\end{figure}

In Fig.~\ref{comparey1y2plane}, we compare the two ways of obtaining the tunneling probability.
The green(blue) region indicates that the EW vacuum is metastable(unstable), and the red 
region means that the EW vacuum is non-perturbative.
We find that the one-loop effect on effective action slightly enlarges the parameter space for the vacuum to be metastable.

The parameter space of the singlet-doublet fermion dark matter model 
is constrained by phenomenological considerations  of dark matter, such as the direct detection and the constraint from the dark matter relic density.
It is found in~\cite{Yaguna:2015mva} that the dark matter Yukawa couplings must be very small, i.e. $|y_{1}|, |y_{2}| \lesssim 4 \times 10^{-3}$, and 
that the masses of dark matter particle are constrained to be
$M_{\chi_{1}} \lesssim 733$ GeV and $|M_{\chi_{2}}- M_{\chi_{1}}|/ M_{\chi_{1}} \lesssim 0.1$.
The vacuum stability analysis of this model does not give a constraint on the parameter space stronger than these phenomenological constraints.

\section{ General singlet-doublet fermion extension model }
\label{GSDM}
More general singlet-doublet fermion extension of the SM can be considered.
In general, we can add N copies of $\mathrm{SU}(2)$ doublet fermions $\psi_{Ln,Rn}$($n=1,\cdots, N$) 
and Q copies of singlet fermions $S_{Lq,Rq}$($q=1,\cdots,Q$).  The relevant Lagrangian can be written as
\begin{equation}
\begin{aligned} 
\mathcal{L}_{\mathrm{general}}&=\sum_{n=1}^{N} \overline{\psi}_n i  \slash \hspace{-0.25cm} D \psi_n + \sum_{q=1}^{Q} \overline{S}_q i \slash \hspace{-0.2cm} \partial S_q
-\sum_{n=1}^{N}\overline{\psi}_{Ln} ({\widetilde M}_{D})_{nn} \psi_{Rn}-\sum_{q=1}^{Q} \overline{S}_{Lq} ({\widetilde M}_{S})_{qq} S_{Rq}\\
&-\sum_{n=1}^{N}\sum_{q=1}^{Q}\left( Y_{Rnq} \overline{\psi}_{Ln} \tilde{H} S_{Rq}+Y_{Lnq} \overline{\psi}_{Rn} \tilde{H} S_{Lq} \right)+h.c,
\label{Lag-GSDM}
\end{aligned}
\end{equation}
where we have chosen to work in the base that the mass matrices $ {\widetilde M}_{D}$ and $ {\widetilde M}_{S}$ are diagonal and real. 

After the EW symmetry breaking,  the mass matrix of the charged components of $\psi_{Ln,Rn}$($\psi_{Ln,Rn}^{-}$)  are not changed.
We simply denote $\psi_{Ln,Rn}^{-}$ as $\chi^-_{Ln,Rn}$.
The N copies of neutral component of $\psi_{Ln,Rn}$($\psi_{Ln,Rn}^{0}$) and Q copies of $S_{Lq,Rq}$ get a mass term.
Introducing  $S_{L,R}=(S_{L1,R1},\cdots, S_{LQ,RQ})^T$ and $\psi_{L,R}=(\psi_{L1,R1},\cdots, \psi_{LN,RN})^T$, we can write the mass term as
\begin{equation}
\begin{pmatrix} {\bar S}_L & {\bar \psi^0}_L  \end{pmatrix} M  \begin{pmatrix} S_{R} \\ \psi^0_{R} \end{pmatrix} + h.c.
\end{equation}
with the mass matrix $M$ given as
\begin{equation}
M=
\begin{pmatrix}
{\widetilde M}_S &\frac{v}{\sqrt{2}}Y^{+}_{L}\\
\frac{v}{\sqrt{2}}Y_{R}& {\widetilde M}_D. \\
\end{pmatrix}
\end{equation}
Here $Y_{L}$ and $Y_R$ are the  $N\times Q$  matrices of Yukawa coupling given in (\ref{Lag-GSDM}).

Performing a field transformation 
\begin{equation}
 \begin{pmatrix} S_{L,R} \\ \psi^0_{L,R} \end{pmatrix} = U_{L,R} \chi^0_{L,R}
\end{equation}
using two unitary matrices $U_{L}$ and $U_R$ with $\chi^{0} = (\chi^{0}_1,\chi^{0}_2,\cdots,\chi^0_{N+Q})^T$ ,
the mass matrix $M$ can be diagonalized and becomes
\begin{equation}
M^d=
diag\left\{{M_{\chi^0_{1}},M_{\chi^0_{2}},\cdots,M_{\chi^0_{N+Q}}}\right\}
= U_{L}^{\dagger} M U_{R}
\end{equation}
where $M_{\chi^0_i}$ is the mass of $\chi^0_{i}$ field.
The interaction Lagrangian of $\chi^0_{i}$ and the CP-even neutral Higgs field $h$ is obtained as
\begin{equation}
\Delta {\cal L}_{h} = -
 \overline{\chi^{0}} Y P_R \chi^{0} h
-\overline{\chi^{0}} Y^{\dagger} P_L \chi^{0} h  \label{Lagrangian-general}
\end{equation}	 
where  $Y$ is the matrix of Yukawa coupling
\begin{equation}
Y =\frac{1}{\sqrt{2}} U^{+}_L
\begin{pmatrix}
0&Y^{+}_{L}\\
Y_{R}&0\\
\end{pmatrix}U_{R}.
\label{generalYmatrix}
\end{equation}

The interactions Lagrangian of $\chi^0_{i}$, $\chi^{-}_{n}$ and the gauge bosons become
\begin{equation}
\begin{aligned} 
\Delta {\cal L}_{W,Z} =& 
\sum_{n=1}^{N} \left[\frac{g_{2}}{{\sqrt{2}}}\left(U_{L}^{+}\right)_{i, n+Q}\overline{\chi_{i}^{0}} \gamma^{\mu} P_L \chi_{n}^{-} W^{+}_{\mu} +\frac{g_{2}}{{\sqrt{2}}}\left(U_{L}\right)_{n+Q, i} \overline{\chi_{n}^{-}} \gamma^{\mu} P_L\chi_{i}^{0} W^{-}_{\mu}  \right.\\
&\left. +\frac{g_{2}}{2cos\theta_{w}}\left(U_{L}^{+}\right)_{i, n+Q}\left(U_{L}\right)_{n+Q,j}\overline{\chi_{i}^{0}} \gamma^{\mu} P_L \chi_{j}^{0} Z_{\mu} + (L\longrightarrow R)  \right] \\
&+\sum_{n=1}^{N}\overline{\chi_{n}^{-}}\left(-g_2 sin\theta_w ~\slash \hspace{-0.25cm} A_{\mu}-\frac{g_2 cos2\theta_w}{2 cos\theta_w}  \  \slash
\hspace{-0.25cm} Z_{\mu} \right)\chi_{n}^{-}.
\label{gaugeinteraction}	
\end{aligned}
\end{equation}	
Here $\theta_{w}$ is the Weinberg angle.

In numerical analysis we consider three typical models.
\subsection{ Model \uppercase\expandafter{\romannumeral1}}
This model includes N copies of  $\mathrm{SU}(2)$ doublet fermions  and  N copies of singlet fermions.
We assume the mass matrix matrices ${\widetilde M}_S$ and  ${\widetilde M}_D$ are proportional to unit matrix.
We also assume the Yukawa coupling $Y_L$ and $Y_R$ are diagonal and are proportional to the unit matrix.
The relevant Lagrangian is
\begin{equation}
\begin{aligned} 
 \mathcal{L}_{\mathrm{Model    \ \uppercase\expandafter{\romannumeral1}}}&=
  \sum_{n=1}^{N} \left( \overline{\psi}_n i  \slash \hspace{-0.25cm} D \psi_n +  \overline{S}_n i \slash \hspace{-0.2cm} \partial S_n
 -M_{D}\overline{\psi}_{n} \psi_{n}-M_{S} \overline{S}_{n} S_{n}\right)\\
 &-\sum_{n=1}^{N}\left( y_{1} \overline{\psi}_{Ln} \tilde{H} S_{Rn}+y_{2} \overline{\psi}_{Rn} \tilde{H} S_{Ln} \right)+h.c
\end{aligned} 
\label{CASE1Lagrangian}
\end{equation} 
 This model basically introduces N generations of  singlet-doublet fermions and there are
  no couplings between generations. So the mass matrix can be diagonalized in the same way as in Eq. (\ref{Mmatrix}) for each generation
  and there are N copies of neutral fermions $\chi_{1}$ and $\chi_{2}$ in the diagonalized base with masses given in (\ref{Mchi1}) and (\ref{Mchi2}).

\subsection{Model \uppercase\expandafter{\romannumeral2}}
In this model we add N copies of  $\mathrm{SU}(2)$ doublet fermions $\psi_{L, R}=\left(\psi_{L,R}^{0}, \psi_{L,R}^{-}\right) ^T$ and only one copy of singlet fermion $S_{L, R}$.  The doublet fermions are all coupled with the only singlet.  We assume that  ${\widetilde M}_D$ are proportional to the unit matrix
and all generations of doublet fermions couple with the singlet fermion with the same strength.
\begin{equation}
\begin{aligned} 
 \mathcal{L}_{\mathrm{Model    \ \uppercase\expandafter{\romannumeral2}}}&=
  \sum_{n=1}^{N} \left( \overline{\psi}_n i  \slash \hspace{-0.25cm} D \psi_n +  \overline{S} i \slash \hspace{-0.2cm} \partial S
 -M_{D}\overline{\psi}_{n} \psi_{n}-M_{S} \overline{S} S\right)\\
 &-\sum_{n=1}^{N}\left( y_{1} \overline{\psi}_{Ln} \tilde{H} S_{R}+y_{2} \overline{\psi}_{Rn} \tilde{H} S_{L} \right)+h.c .
\end{aligned} 
\label{CASE2Lagrangian}
\end{equation} 

After EW symmetry breaking, the mass matrix $M$ is obtained as 
\begin{equation}
M=
\begin{pmatrix}
M_{S} & \frac{y_{2}v}{\sqrt{2}}  & \cdots   & \frac{y_{2}v}{\sqrt{2}}  \\
\frac{y_{1}v}{\sqrt{2}}& M_{D}  & \cdots   & 0  \\
\vdots & \vdots  & \ddots   & \vdots  \\
\frac{y_{1}v}{\sqrt{2}} & 0  & \cdots\  & M_D  \\
\end{pmatrix}
\end{equation}
In a suitable base, the singlet $S$ can be considered coupled only to one of the linear combinations of $\psi_n$, 
i.e. $\Psi=\frac{1}{\sqrt{N} }\sum_{n=1}^N \psi_n$, with effective couplings $\sqrt{N} y_1$ and  $\sqrt{N} y_2$  in the new base.
Other orthogonal linear combinations of $\psi_n$ do not couple to the singlet fermion.
So the mass matrix can be diagonalized to a form 
\begin{equation}
M^d=diag\left\{M_{\chi_1^0},M_{\chi_2^0}, M_D,\cdots,M_D\right\},
\end{equation}
where the $M_{\chi_1^0}$ and $M_{\chi_2^0}$ are the masses of the two neutral fields, $\chi^0_{1}$ and $\chi^0_{2}$, which couple to the neutral Higgs field.
We can obtain
\begin{eqnarray}
&& M_{\chi_1^0}^2=\frac{1}{2}(T_N-\sqrt{T_N^2-4 ~D_N^2}),   \label{MNchi1} \\
&& M_{\chi_2^0}^2=\frac{1}{2}(T_N+\sqrt{T_N^2-4 ~D_N^2}),  \label{MNchi2}
\end{eqnarray}
where $T_N=M_S^2+\frac{N}{2}y_1^2 v^2+M_D^2+\frac{N}{2}y_2^2 v^2$, 
$D_N=\frac{N}{2}y_1 y_2 v^2-M_S M_D $.

\subsection{Model \uppercase\expandafter{\romannumeral3}}
In the third model,  we consider extending the SM by adding N copies of singlet fermions $S_{L,R}$ with only one copy of  $\mathrm{SU}(2)$ doublet fermion.
The singlet fermions are all coupled with the only doublet.  We assume that  ${\widetilde M}_S$ are proportional to the unit matrix
and all generations of singlet fermions couple with the doublet fermion with the same strength.

The mass matrix is given as:
\begin{equation}
M=
\begin{pmatrix}
M_{S} & \cdots  & 0   & \frac{y_{2}v}{\sqrt{2}}  \\
\vdots& \ddots   & \vdots   & \frac{y_{2}v}{\sqrt{2}}  \\
0 & \cdots & M_{S}   & \vdots  \\
\frac{y_{1}v}{\sqrt{2}} & \frac{y_{1}v}{\sqrt{2}}  & \cdots\  & M_D  \\
\end{pmatrix}.
\end{equation}
Similar to the case in Model \uppercase\expandafter{\romannumeral2}, 
the doublet $\psi$ can be considered coupled only to one of the linear combinations of $S_q$, 
i.e. $\frac{1}{\sqrt{N} }\sum_{q=1}^N S_q$, with effective couplings $\sqrt{N} y_1$ and  $\sqrt{N} y_2$  in a suitable base.
Other orthogonal linear combinations of $S_n$ do not couple to the doublet fermion.
So the mass matrix can be diagonalized to a form 
\begin{equation}
M^d=diag\left\{M_{\chi_1^0},M_{\chi_2^0}, M_S,\cdots,M_S\right\}
\end{equation}
The $M_{\chi_1^0}$ and $M_{\chi_2^0}$ are the masses of the two neutral fields, $\chi^0_{1}$ and $\chi^0_{2}$, 
which couple to the neutral Higgs field. The expressions of $M_{\chi^0_{1}}$ and $M_{\chi^0_{2}}$  are the same as 
in~\eqref{MNchi1} and \eqref{MNchi2}.

\subsection{Vacuum Stability in general singlet-doublet fermion extension models}
In this section, we study the vacuum stability in the three models just introduced using RG improved effective action.
For Model \uppercase\expandafter{\romannumeral1}, we can write down immediately the contribution to the RG improved effective 
potential following Eq.~\eqref{SDFDMeff}. We get
\begin{equation}
V_{1}^{\mathrm{Ext}}(\phi , t) =N\sum_i \frac{(-1)^i n_{i}}{64 \pi^{2}} M_{\chi_{i}}^{4}(\phi,t)\left[\ln \frac{M_{\chi_{i}}^{2}(\phi,t)}{\mu^{2}(t)}-3/2\right]. \label{case1eff}
\end{equation}
For Model \uppercase\expandafter{\romannumeral2} and Model \uppercase\expandafter{\romannumeral3}, 
the RG improved effective potential can be obtained by simply substituting the neutral fermions masses \eqref{MNchi1} and \eqref{MNchi2} into Eq.~\eqref{SDFDMeff}.

Threshold corrections to couplings in general singlet-doublet fermion extension model are given in Appendix.~\ref{sec:thersholdeffectgeneral}.
In Table.~\ref{thersholdcase1}, Table. ~\ref{thersholdcase2} and Table.~\ref{thersholdcase3}, we show some
numerical  results of couplings in the $\MS$ scheme for the three models shown above.
New contributions to $Z_2$ factor in three models are shown Appendix~\ref{sec:thersholdeffectgeneral}.
The one-loop $\beta$ functions in the three models are given in Appendix~\ref{betamodel1}, \ref{betamodel2}, \ref{betamodel3}.

\begin{table}[!t]
	\setlength{\tabcolsep}{.5em}
	\renewcommand{\arraystretch}{1.2}
	\begin{tabular}{c|cccc}
		\multicolumn{5}{c}{Threshold Effects for different N in  $\mathrm{Model \ \uppercase\expandafter{\romannumeral1}}$ } \\
		\hline
		$\quad\quad\mu=M_t\quad\quad$ & $\quad\quad\lambda\quad\quad$ &\quad\quad$y_t\quad\quad$ &\quad\quad$g_2\quad\quad$ &$\quad\quad g_Y\quad\quad$ \\
		\hline
		$N=2$ & 0.12495 & 0.93361$^*$ & 0.64367 & 0.35859 \\
		$N=5$ & 0.12330 & 0.92868$^*$ & 0.63750 & 0.35902 \\
		$N=7$ & 0.12221 & 0.92539$^*$ & 0.63338 & 0.35932 \\
		\hline
	\end{tabular}
	\caption{Threshold corrections to couplings when $y_{1} = y_{2} = 0.25$ for $\mathrm{Model \ \uppercase\expandafter{\romannumeral1}}$, $M_D=M_S=1000$~GeV.
		The superscript $*$ indicates that the NNNLO QCD effects are included. }
	
	\label{thersholdcase1}
\end{table}

\begin{table}[!t]
	\setlength{\tabcolsep}{.5em}
	\renewcommand{\arraystretch}{1.2}
	\begin{tabular}{c|cccc}
		\multicolumn{5}{c}{Threshold Effects for different N in  $\mathrm{Model \ \uppercase\expandafter{\romannumeral2}}$ } \\
		\hline
		$\quad\quad\mu=M_t\quad\quad$ & $\quad\quad\lambda\quad\quad$ &\quad\quad$y_t\quad\quad$ &\quad\quad$g_2\quad\quad$ &$\quad\quad g_Y\quad\quad$ \\
		\hline
		$N=2$ & 0.12545 & 0.93480$^*$ & 0.64367 & 0.35686 \\
		$N=5$ & 0.12644 & 0.93164$^*$ & 0.63750 & 0.35340 \\
		$N=7$ & 0.12835 & 0.92954$^*$ & 0.63714 & 0.34576 \\
		\hline
	\end{tabular}
	\caption{Threshold corrections to couplings when $y_{1} = y_{2} = 0.2$ for $\mathrm{Model \ \uppercase\expandafter{\romannumeral2}}$,  
	$M_D=M_S=1000$~GeV.
		The superscript $*$ indicates that the NNNLO QCD effects are included. }
	
	\label{thersholdcase2}
\end{table}

\begin{table}[!t]
	\setlength{\tabcolsep}{.5em}
	\renewcommand{\arraystretch}{1.2}
	\begin{tabular}{c|cccc}
		\multicolumn{5}{c}{Threshold Effects for different N in  $\mathrm{Model \ \uppercase\expandafter{\romannumeral3}}$ } \\
		\hline
		$\quad\quad\mu=M_t\quad\quad$ & $\quad\quad\lambda\quad\quad$ &\quad\quad$y_t\quad\quad$ &\quad\quad$g_2\quad\quad$ &$\quad\quad g_Y\quad\quad$ \\
		\hline
		$N=2$ & 0.12545 & 0.93479$^*$ & 0.64573 & 0.35809 \\
		$N=5$ & 0.12644 & 0.93164$^*$ & 0.64573 & 0.35563 \\
		$N=7$ & 0.12835 & 0.92954$^*$ & 0.64573 & 0.35400 \\
		\hline
	\end{tabular}
	\caption{Threshold corrections to couplings when $y_{1} = y_{2} = 0.2$ for $\mathrm{Model \ \uppercase\expandafter{\romannumeral3}}$,
	$M_D=M_S=1000$~GeV.
		The superscript $*$ indicates that the NNNLO QCD effects are  included. }
	
	\label{thersholdcase3}
\end{table}

\begin{figure}[!t]
	\centering
	\subfigure[\label{case1lambda025}]
	{\includegraphics[width=.486\textwidth]{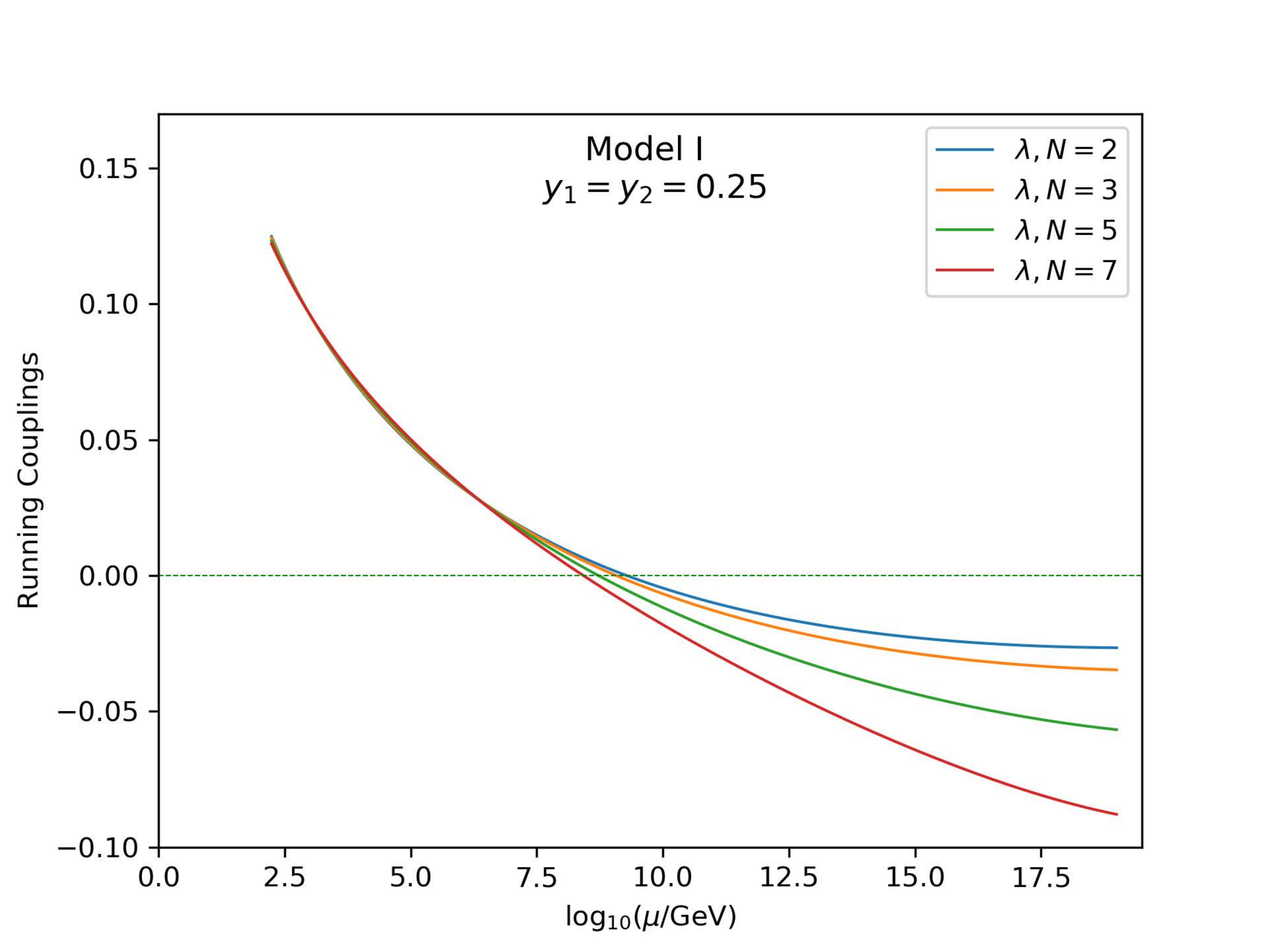}}
	\subfigure[\label{case1lambdatilde025}]
	{\includegraphics[width=.486\textwidth]{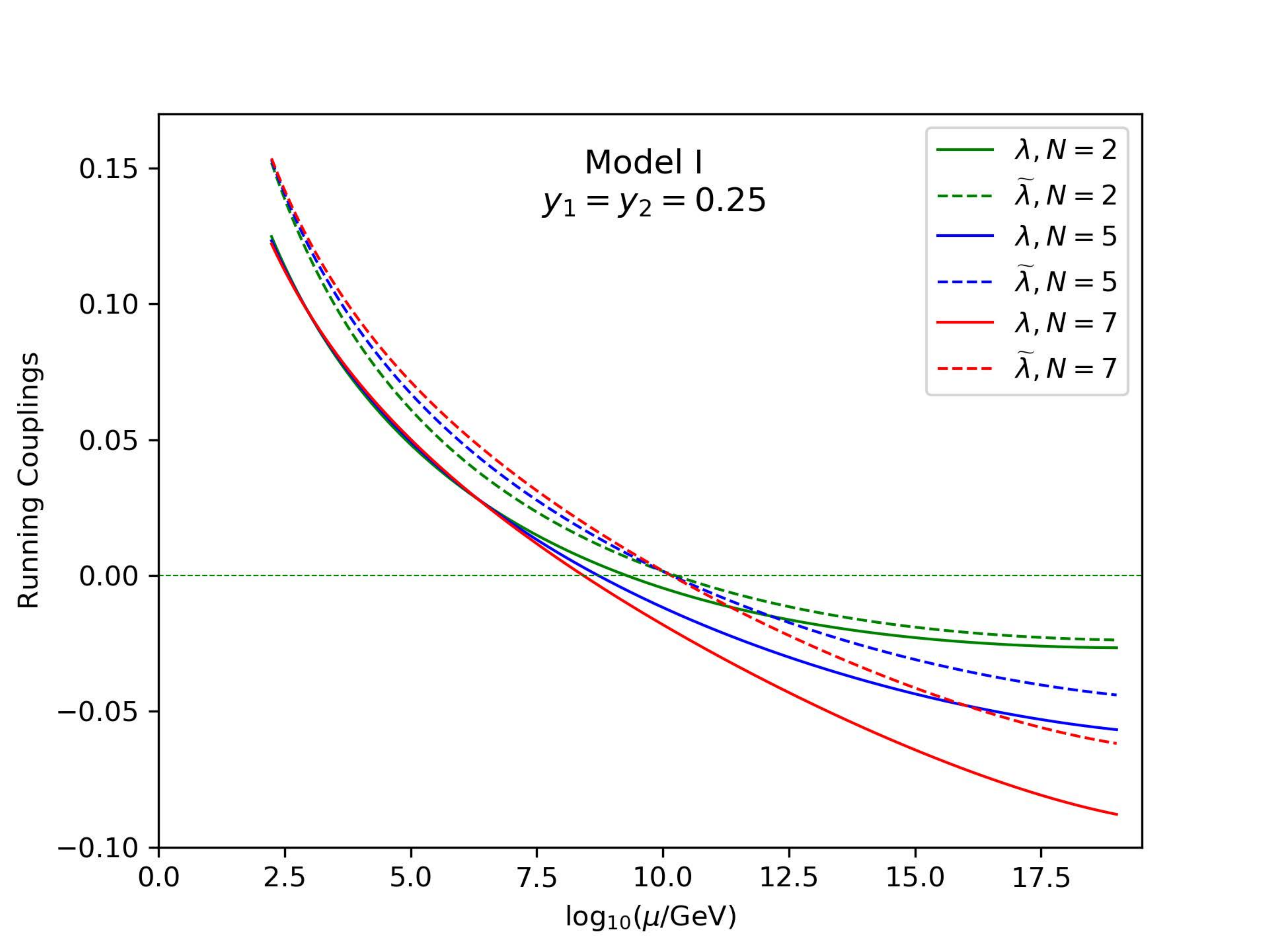}}
	\caption{ (a) $\lambda(t)$ up to $M_{Pl}$ for different values of N in Model  \uppercase\expandafter{\romannumeral1}.
	The value of Yukawa couplings $y_1$ and $y_2$ are $0.25$;
	(b) Running $\lambda(t)$ and ${\tilde \lambda}(t)$  up to $M_{Pl}$ scale for $N=2,5,7$ in Model  \uppercase\expandafter{\romannumeral1}.  
	}
	\label{case1lambda}
\end{figure}

\begin{figure}[!t]
	\centering
	\subfigure[\label{case2lambda020}]
	{\includegraphics[width=.486\textwidth]{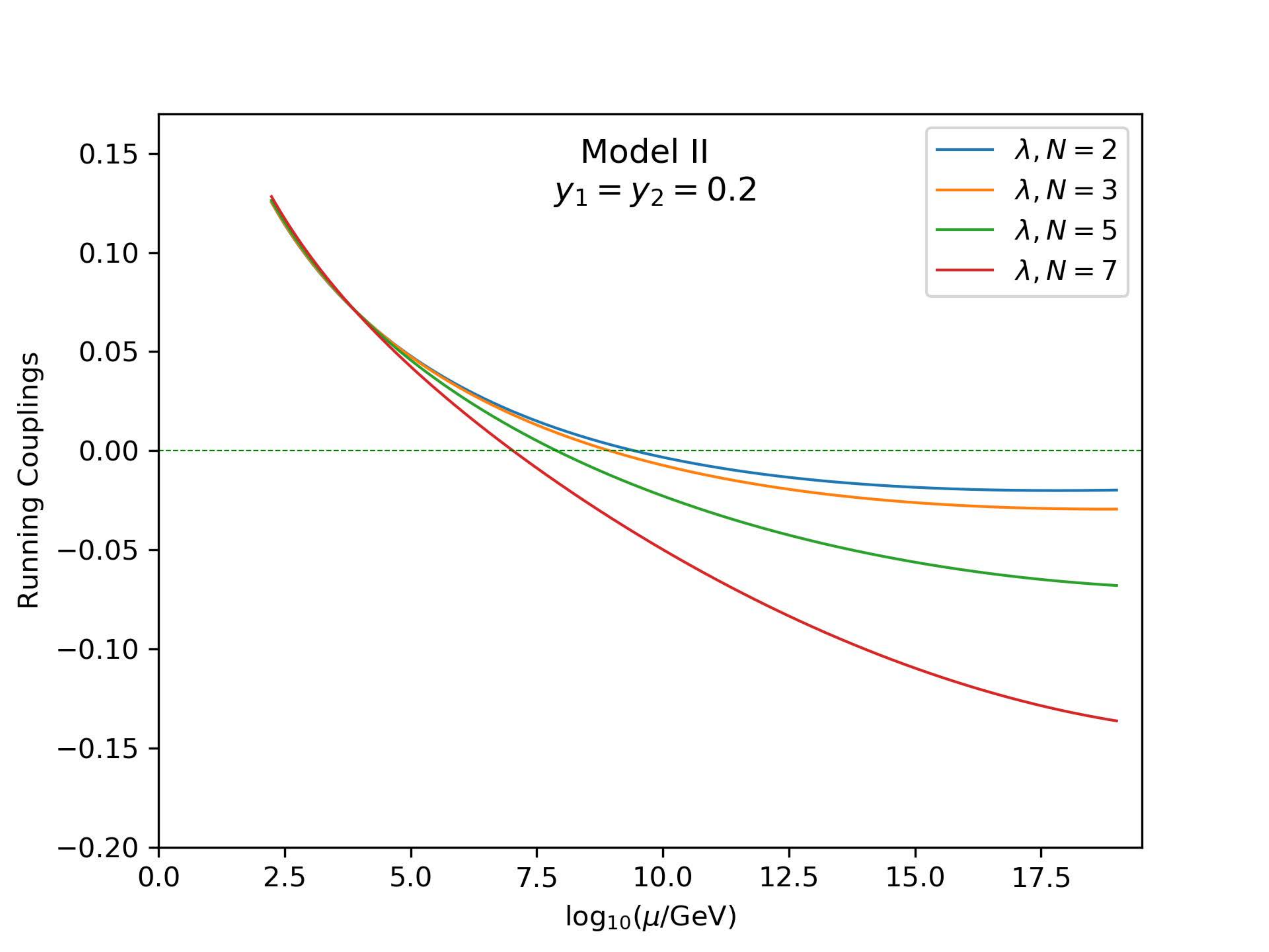}}
	\subfigure[\label{case2lambdatilde020}]
	{\includegraphics[width=.486\textwidth]{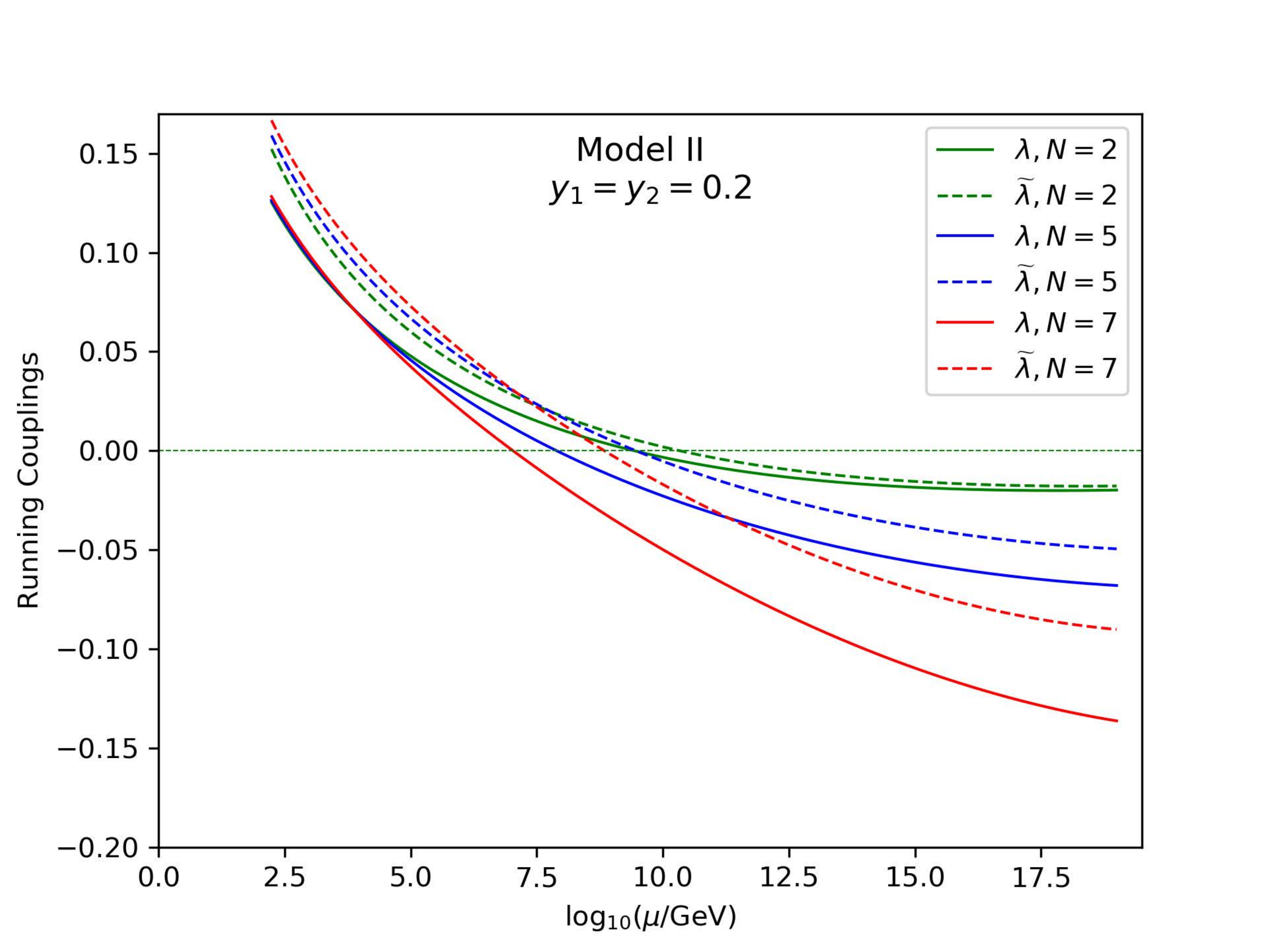}}
	\caption{ (a) $\lambda(t)$ up to $M_{Pl}$ for different values of N in Model \uppercase\expandafter{\romannumeral2}.
	The value of Yukawa couplings $y_1$ and $y_2$ are $0.2$;
	(b) Running $\lambda(t)$ and ${\tilde \lambda}(t)$  up to $M_{Pl}$ scale for $N=2, 5, 7$ in Model  \uppercase\expandafter{\romannumeral2}.  
	}
	\label{case2lambda}
\end{figure}

\begin{figure}[!t]
	\centering
	\subfigure[\label{case3lambda020}]
	{\includegraphics[width=.486\textwidth]{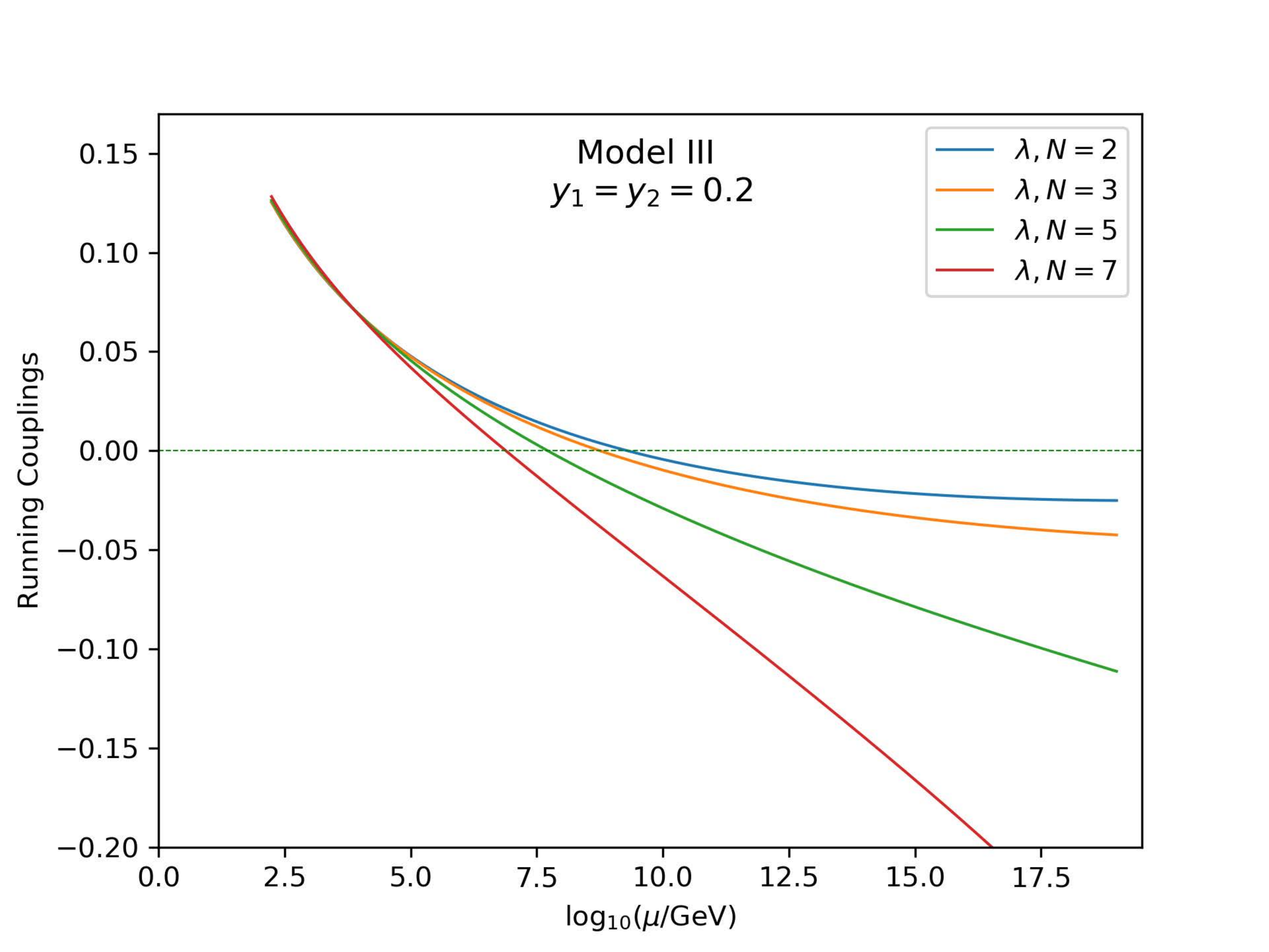}}
	\subfigure[\label{case3lambdatilde020}]
	{\includegraphics[width=.486\textwidth]{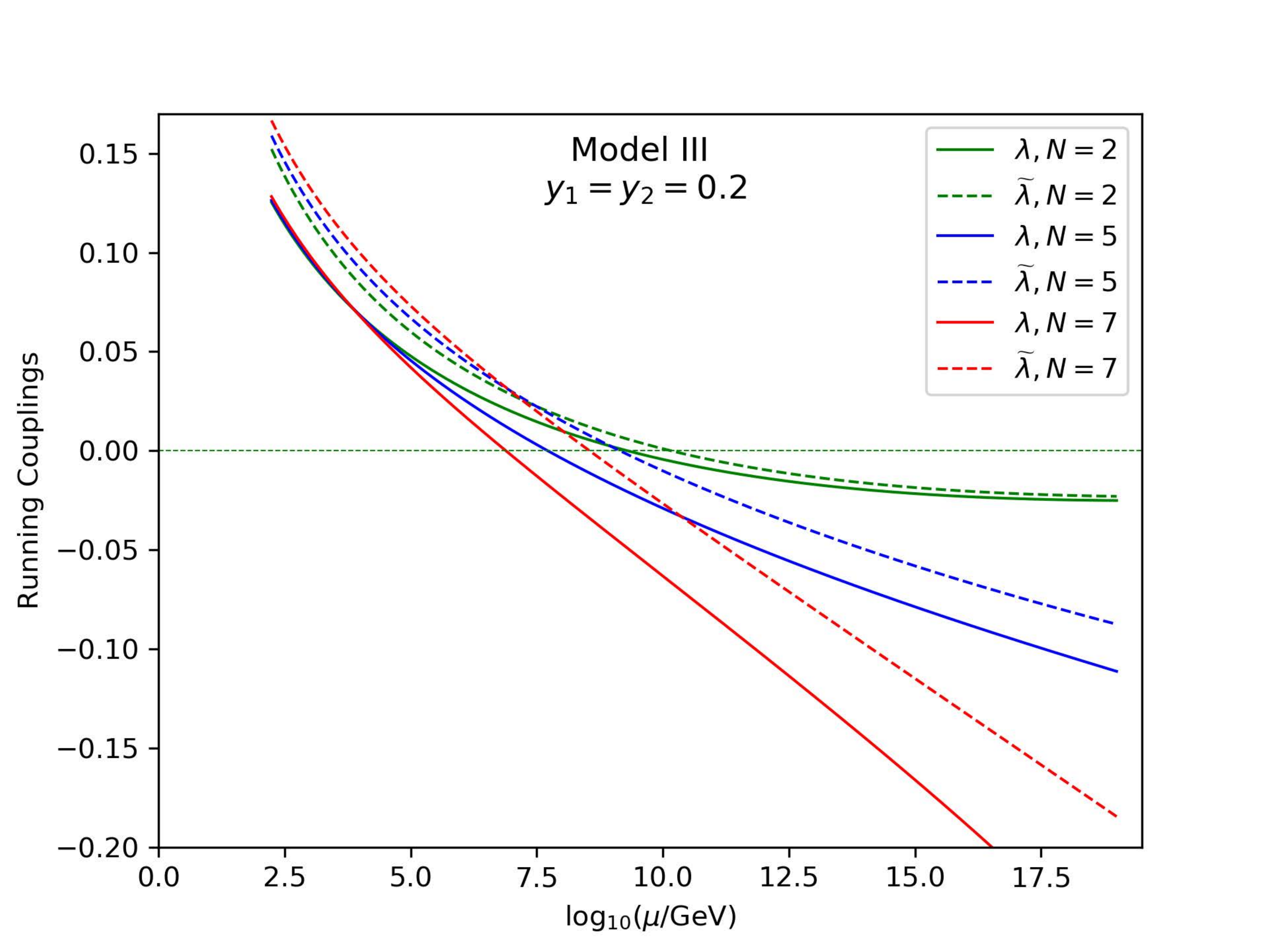}}
	\caption{ (a) $\lambda(t)$ up to $M_{Pl}$ for different values of N in Model  \uppercase\expandafter{\romannumeral3}.
	The value of Yukawa couplings $y_1$ and $y_2$ are $0.2$;
		(b) Running $\lambda(t)$ and ${\tilde \lambda}(t)$ up to $M_{Pl}$ scale for $N=2, 5, 7$ in Model \uppercase\expandafter{\romannumeral3}.  
	}
	\label{case3lambda}
\end{figure}

\begin{figure}[!t]
	\centering
	\subfigure[\label{case1stability}]
	{\includegraphics[width=.486\textwidth]{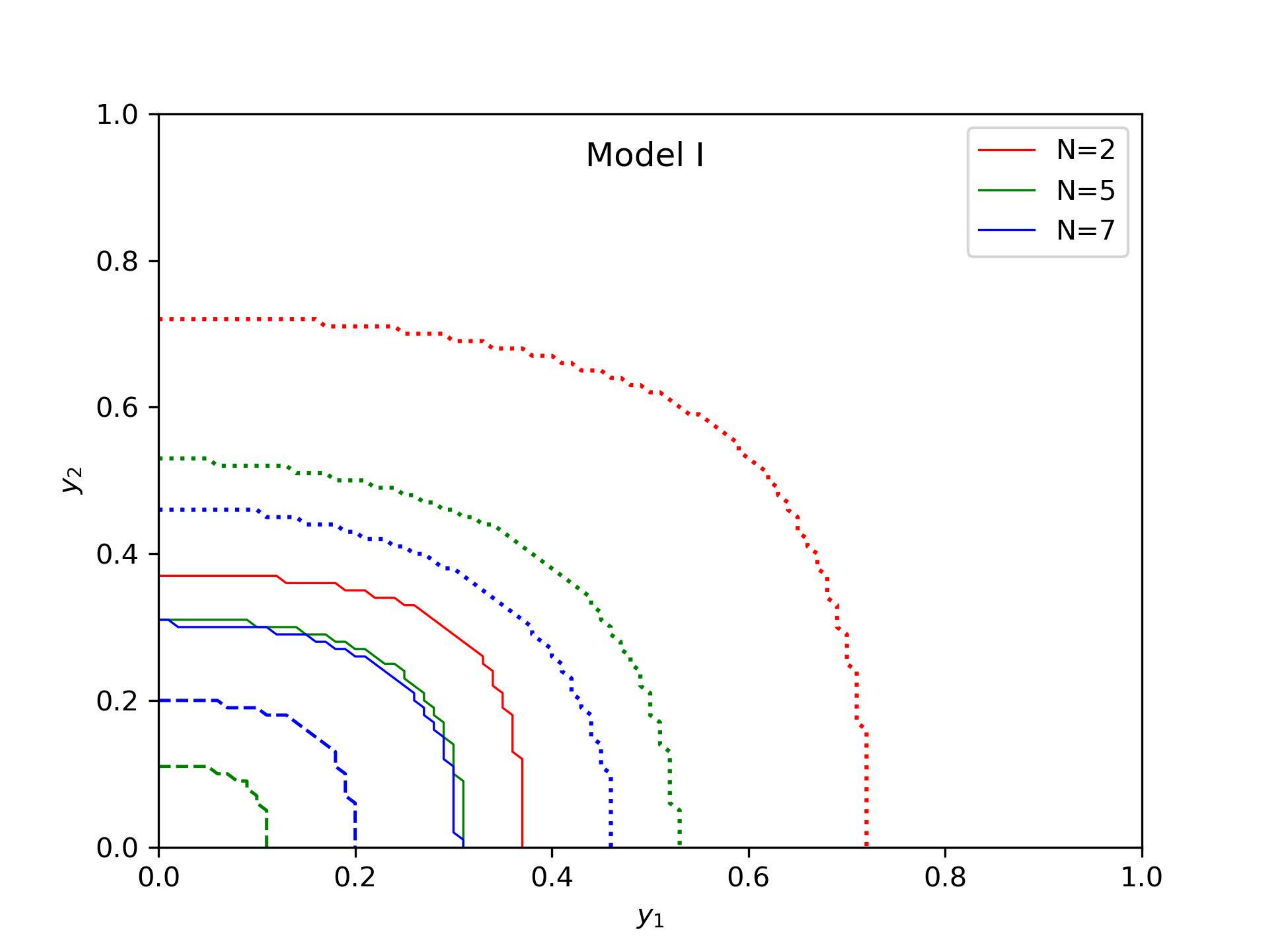}}
	\subfigure[\label{case2stability}]
	{\includegraphics[width=.486\textwidth]{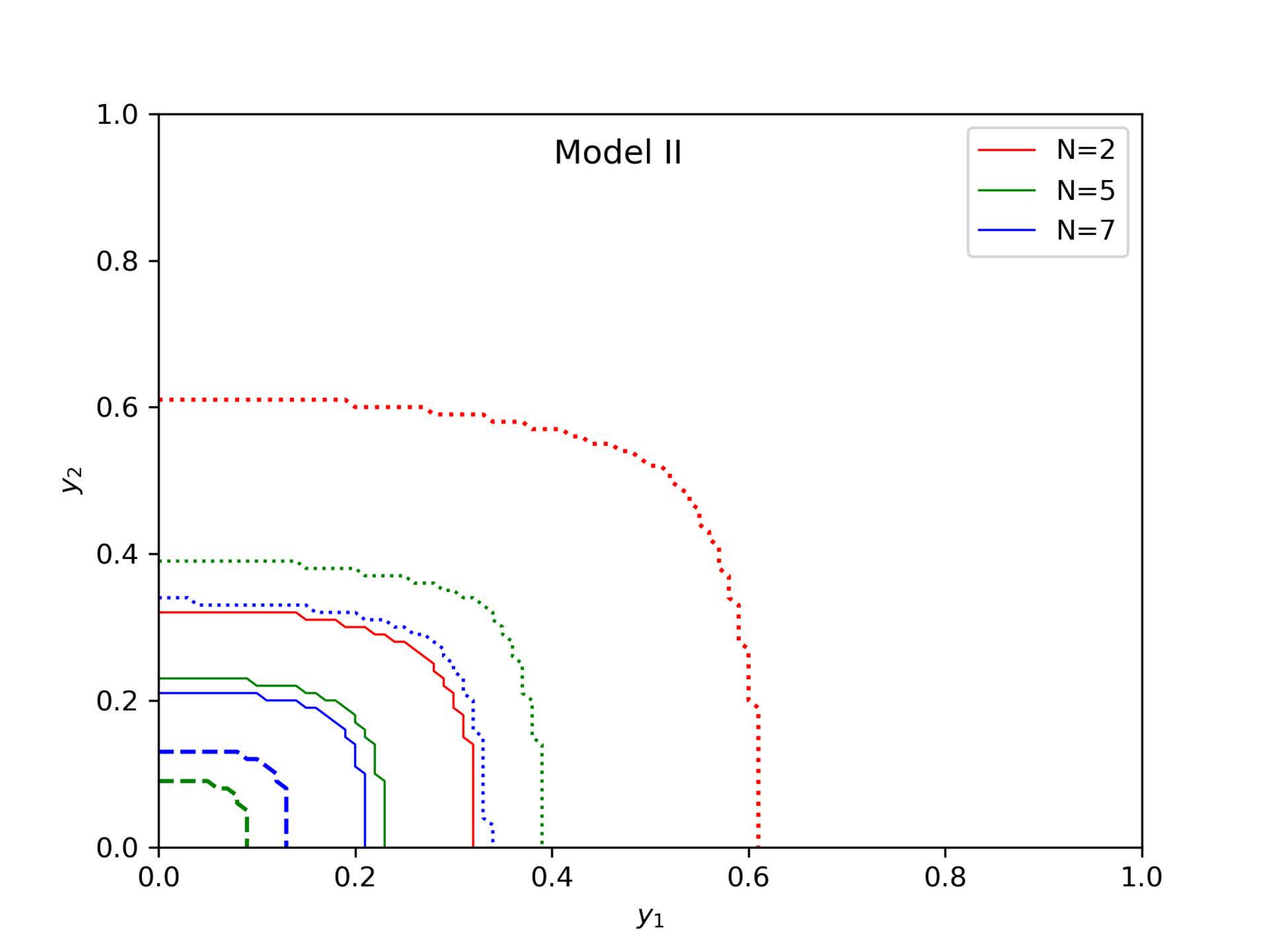}}
	\caption{ (a) Status of the EW vacuum in the $y_{1}-y_{2}$ plane for $N = 2, 5, 7$ in Model  \uppercase\expandafter{\romannumeral1};
	(b) Status of the EW vacuum in the $y_{1}-y_{2}$ plane for $N = 2, 5, 7$ in Model  \uppercase\expandafter{\romannumeral2}.  
	In both cases, the dashed line is the stable bound, the solid line  the metastable bound, the dotted line the unstable bound.
	}
	\label{case12stability}
\end{figure}

\begin{figure}[!t]
	\centering
	\subfigure[\label{case3stability}]
	{\includegraphics[width=.486\textwidth]{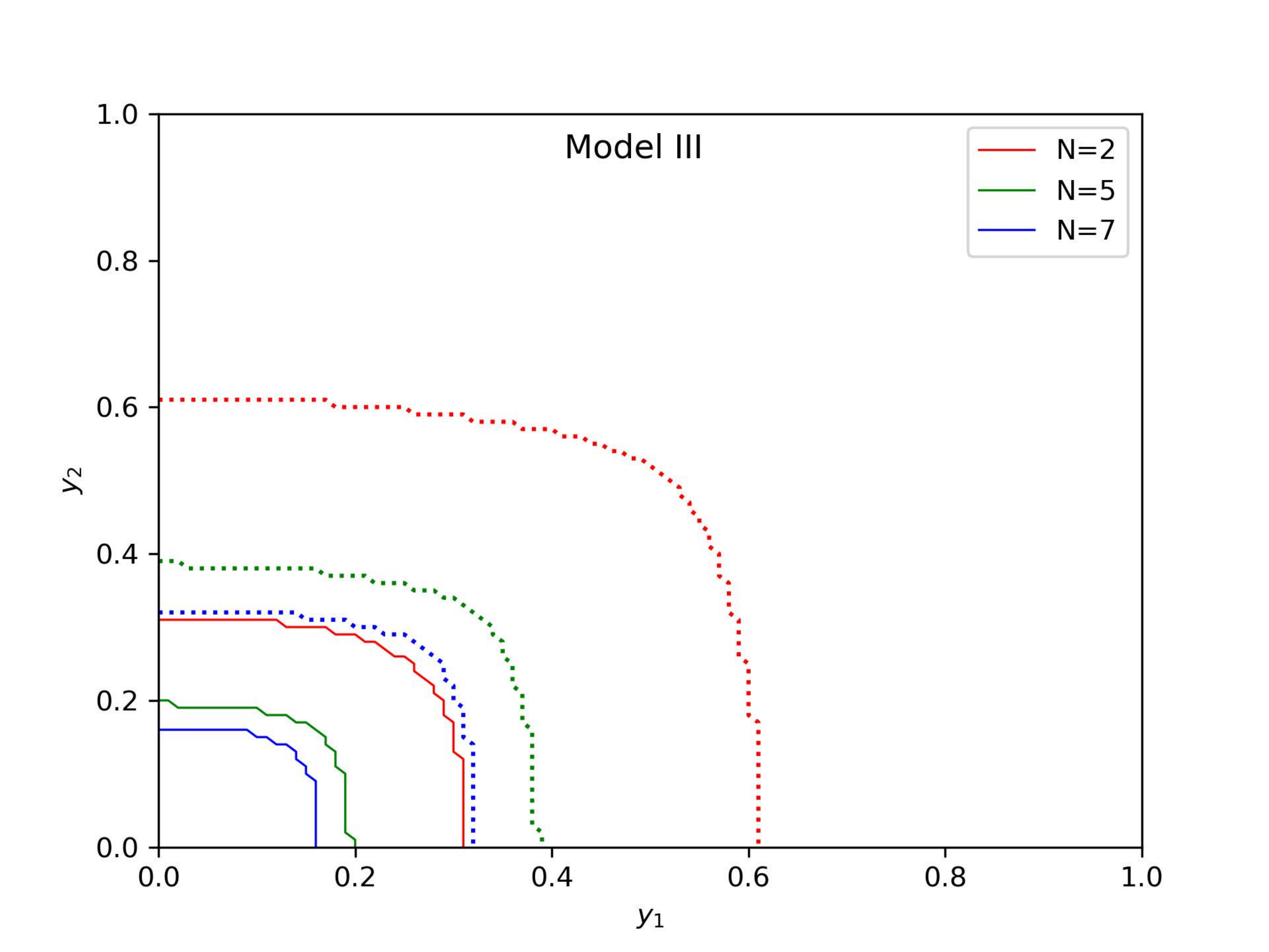}}
	\subfigure[\label{betafunctioncase1}]
	{\includegraphics[width=.486\textwidth]{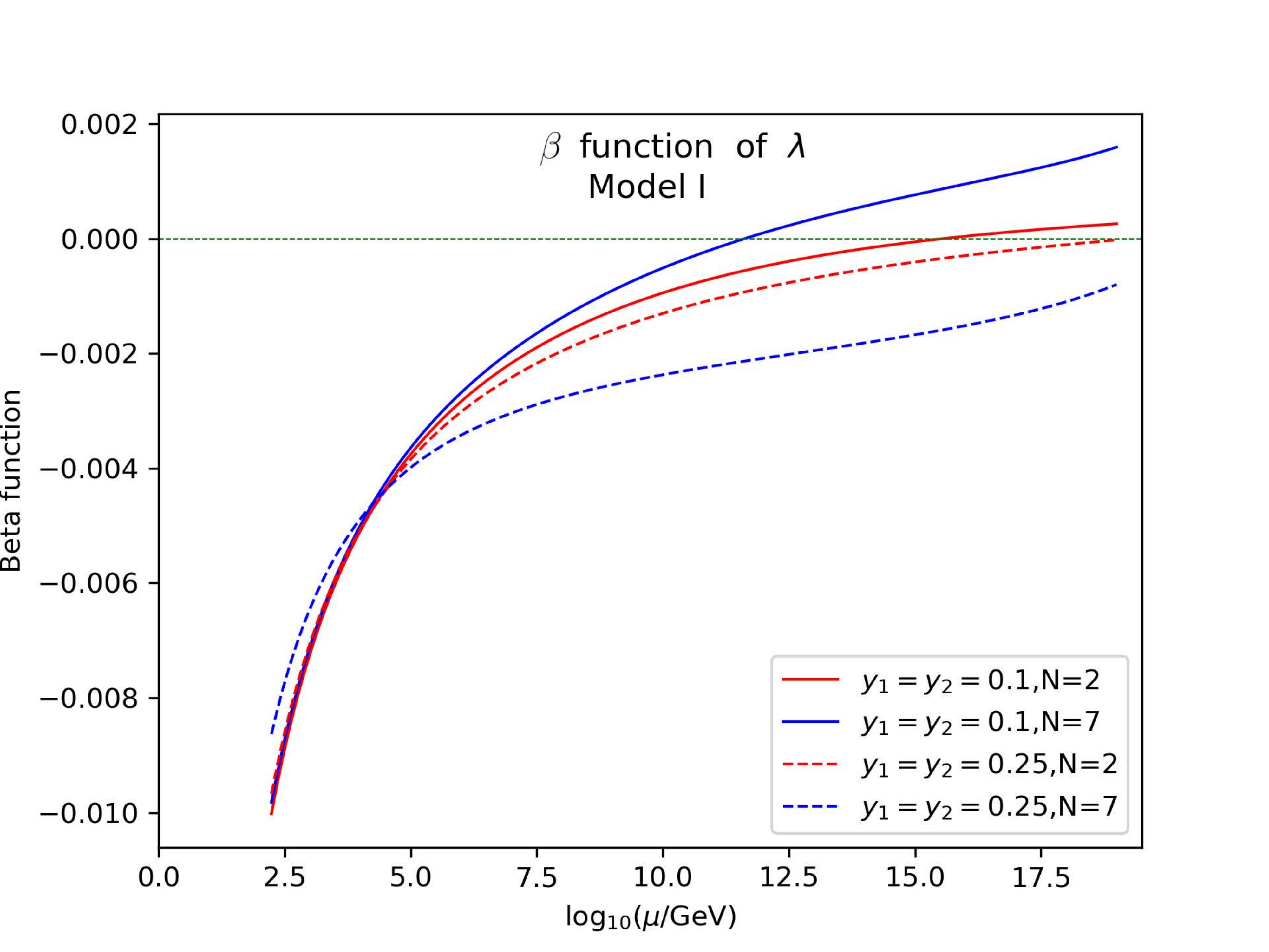}}
	\caption{(a)Status of the EW vacuum in the $y_{1}-y_{2}$ plane for $N = 2, 5, 7$ in Model  \uppercase\expandafter{\romannumeral3}.
	The solid line is the metastable bound, the dotted line the unstable bound.
	(b) $\beta$ function of $\lambda$ in Model  \uppercase\expandafter{\romannumeral1}.
	The solid line is for $y_1 = y_2 = 0.1$, the dashed line for $y_1 = y_2 = 0.25$.
	}
\end{figure}

\begin{figure}[H]
	\centering
	\subfigure[\label{case1lambdaydiff}]
	{\includegraphics[width=.486\textwidth]{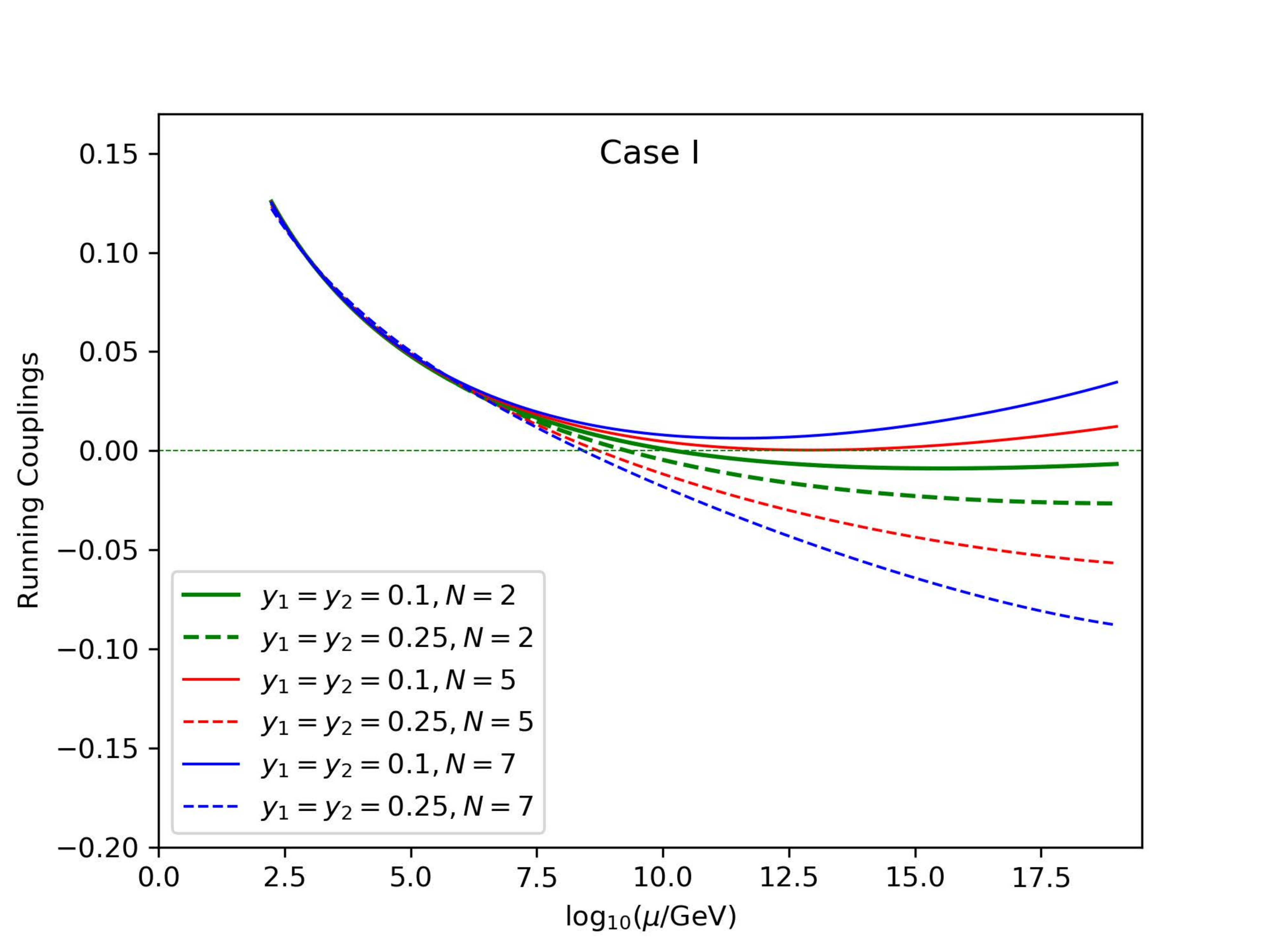}}
	\subfigure[\label{case2lambdaydiff}]
	{\includegraphics[width=.486\textwidth]{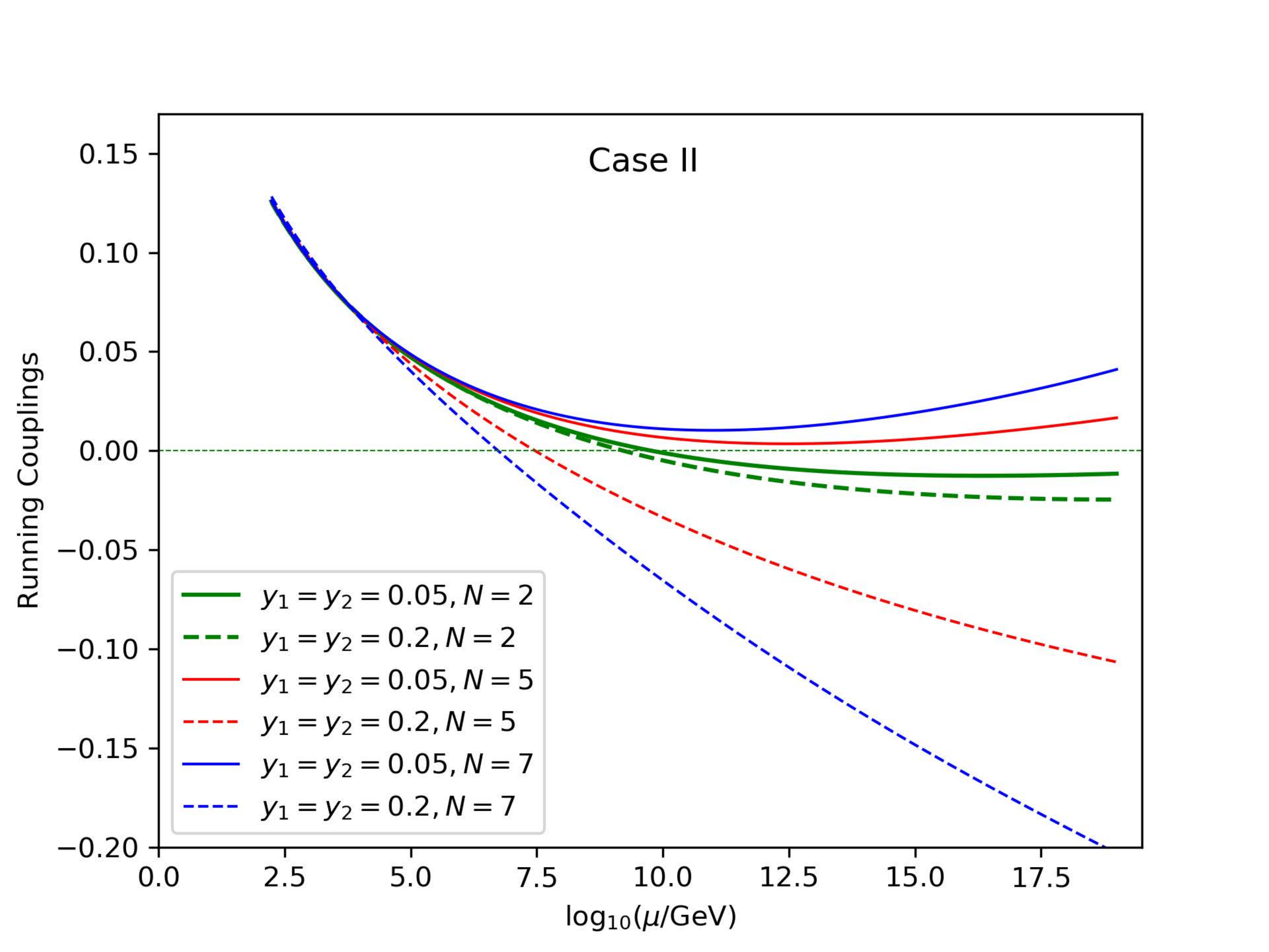}}
	\caption{
		(a)$\lambda(t)$ up to $M_{Pl}$ for different values of Yukawa couplings in Model  \uppercase\expandafter{\romannumeral1},
		The solid lines are for $y_1=y_2=0.1$ and the dashed lines for $y_1=y_2=0.25$.
		(b) $\lambda(t)$ up to $M_{Pl}$ for different values of Yukawa couplings in Model  \uppercase\expandafter{\romannumeral2}.
		The solid lines are for $y_1=y_2=0.05$ and the dashed lines are for $y_1=y_2=0.2$.
	}
	\label{lambdaydiff}
\end{figure}

We can see the evolution of $\lambda(t)$ in three models in Fig.~\ref{case1lambda025},  Fig.~\ref{case2lambda020} and Fig.~\ref{case3lambda020}.
Here we choose $y_1=y_2=0.25$ in Model\ \uppercase\expandafter{\romannumeral1} and $y_1=y_2=0.2$ in 
Model \uppercase\expandafter{\romannumeral2}  and Model \uppercase\expandafter{\romannumeral3}.
We can see that the minimum of $\lambda(t)$ decreases as N is increased for these parameters in all these cases.
In Fig.~\ref{case1lambdatilde025},~\ref{case2lambdatilde020} and Fig.~\ref{case3lambdatilde020},we compare the evolution of $\lambda(t)$ and $\tilde{\lambda}$ in all models.The $\tilde{\lambda}$ is bigger than $\lambda$ due to the one-loop Coleman-Weinberg type corrections.
The difference between $\tilde{\lambda}$ and $\lambda$ increases with the increase of N.

We study the status of EW vacuum in the $y_1-y_2$ plane for different N in the three models. 
Note that in  Fig.~\ref{case1stability} and Fig.~\ref{case2stability}, there are two types of lines for for $N=2$, 
i.e. the solid line for the metastable bound, the dotted line for the unstable bound.
For $N=5$ and $7$, there are three types of lines, i.e.
and for $N=5,7$, the dashed line for the stable bound, the solid line in the middle for the metastable bound and
the dotted line for the unstable bound.
In Fig.~\ref{case3stability},  there are two types of lines for all cases of $N=2, 5$ and $7$, 
i.e.  the solid line for the metastable bound and the dotted line for the unstable bound.

We can see that in both Model  \uppercase\expandafter{\romannumeral1} and Model  \uppercase\expandafter{\romannumeral2}
 the vacuum becomes stable when $y_1$ and $y_2$ are small and N is large. This is quite different from the result when N is small.
 In comparison, we can also see that in Model \uppercase\expandafter{\romannumeral3},  there are no such regions in the parameter space 
 for which the EW vacuum becomes stable when N is large. 
 This happens because the extra copies of fermion doublets in Model  \uppercase\expandafter{\romannumeral1} and Model  \uppercase\expandafter{\romannumeral2} give positive contributions to the $\beta$ functions of $g_1$ and $g_2$,
 as can be seen in \eqref{betafunctiong1model1}, \eqref{betafunctiong2model1}, \eqref{betafunctiong1model2} and \eqref{betafunctiong2model2}.
 For larger $N$, this effect drives $g_1$ and $g_2$ running to larger values with faster rate.
 When $y_1$ and $y_2$ are small, the contribution of larger $g_1$ and $g_2$ can even make the $\beta$ function of $\lambda$ turning into positive value, 
 as can be seen in Fig.~\ref{betafunctioncase1}.
 If increasing the value of Yukawa coupling, the $\beta$ function can be turned into negative again,
 as can be seen in Fig.~\ref{betafunctioncase1}.
 
 The impact of these running effects can be seen clearly in Fig.~\ref{case1lambdaydiff} and ~\ref{case2lambdaydiff}.
 In Fig.~\ref{case1lambdaydiff}, we can see that $\lambda_{min} > 0$ for $y_1=y_2=0.1$ and $N=5,7$. 
 This means that the EW vacuum becomes stable for these parameters in Model  \uppercase\expandafter{\romannumeral1}.
 On the contrary, $\lambda_{min} < 0$ when $y_1=y_2=0.25$ even for $N=5$ and $7$.
 We can also see in Fig.~\ref{case2lambdaydiff} that $\lambda_{min} > 0$ when $y_1=y_2=0.05$ and for both $N=5$ and $7$.
 This means that the EW vacuum becomes stable for these parameters in Model  \uppercase\expandafter{\romannumeral2}.
 On the other hand, $\lambda_{min} < 0$ when $y_1=y_2=0.2$ in Model  \uppercase\expandafter{\romannumeral2}.
 The results here are consistent with the results indicated in Fig.~\ref{case12stability}.
 When Yukawa couplings become bigger, $\lambda_{min} < 0$ and the vacuum become unstable in both models.
 
 We note that 
In Fig.~\ref{case1stability}, Fig.~\ref{case2stability} and Fig.~\ref{case3stability},
we can also find that the metastable region and unstable region become smaller with the increase of N.
This occurs because when $y_1$ and $y_2$ are bigger, the new physics effects of extra fermions would dominate the running of $\lambda$. 
 More copies of extra fermions would make $\lambda$ running faster to a negative value.
 We can similarly compare two ways of obtaining the tunneling probability, as done for SDFDM model in Fig. \ref{comparey1y2plane}.
 The one-loop effective action again slightly modifies the parameter space presented in Fig.~\ref{case1stability}, Fig.~\ref{case2stability} and Fig.~\ref{case3stability}.

\section{Conclusion}
In summary, we have studied the one-loop Coleman-Weinberg type effective potential of the Higgs boson
in a single-doublet fermion dark matter extension of the SM. We have calculated the threshold effect of these
fermions in this model beyond the SM and have studied the RG running of parameters in the $\MS$ scheme.
We have studied the RG improvement to the effective potential. 
We have studied the vacuum stability using the RG improved effective potential.

Using the method of derivative expansion, we have studied the quantum correction to the effective action.
We have calculated the renormalization on the kinetic term in the effective action in the case with external field.
We have studied the RG improvement of the kinetic term. Using the RG improved kinetic term and the RG improved
effective potential, we calculate the decay rate of the false vacuum. We find that the factor arising from the anomalous
dimension which appears in the kinetic term and the effective potential cancels in the decay rate.
Taking all these considerations into account, we find that the decay rate of the false vacuum is slightly changed by the effective action.

We have also done all these studies in general singlet-doublet fermion extension models.
We perform a numerical analysis in three typical extension models with N copies of single-doublet fermions, N copies of doublet fermions 
and N copies of singlet fermions separately. We find that several copies of fermion doublet can make the $\beta$ function of $\lambda$ becoming
positive in some regions of parameter space when Yukawa couplings of these extra fermions are small.
Consequently, in models with small value of Yukawa couplings and large number of copies of fermion doublet, the EW vacuum can become stable.
For large value of the Yukawa coupling, the EW vacuum can again be turned into metastable or unstable.
We also find that the difference between Higgs self-coupling $\lambda$ and  ${\tilde \lambda}$,  the effective self-coupling after including
Coleman-Weinberg type correction, becomes larger when the number of copies of singlet fermions or doublet fermions is increased. 
In the general singlet-doublet fermion extension models, the decay rate of the false vacuum is also slightly changed by the effective action.

\begin{acknowledgments}
This work is supported by National Science Foundation of China(NSFC), grant No. 11875130.
\end{acknowledgments}

\appendix

\section{ Threshold effect and parameters in the $\MS$ scheme}
\label{sec:thersholdeffect}
\subsection{General strategy for one-loop matching}
\label{generalstrategy}
To study the vacuum stability of a model at high energy scale, we need to know the value of coupling constants at low energy scale and then run them to the Plank scale according to RGEs.To determine these parameters at low energy scale, the threshold corrections must be taken into account.
In this article we use the strategy in~\cite{Sirlin:1985ux,Buttazzo:2013uya} to evaluate one-loop corrections.
All the loop calculations are performed in the $\MS$ scheme in which all the parameters are gauge invariant and have gauge-invariant renormalisation group
equations~\cite{Caswell:1974cj}.

A parameter in the $\MS$ scheme, e.g. $\theta(\overline{\mathrm{\mu}})$,  can be obtained from renormalized parameter $\theta$ in physical scheme
which is directly related to physical observables. 
The connection  between $\theta$ and  $\theta(\overline{\mathrm{\mu}})$ to one-loop order, can be found by noting that the unrenormalized $\theta_{0}$ is related to the renormalized couplings by
\begin{equation}
	\theta_{0}=\theta-\delta \theta=\theta(\overline{\mu})-\delta \theta_{\mathrm{MS}}
\end{equation}
where $\delta\theta$ and $\delta\theta_{\MS}$ are the corresponding counterterms.
By definition $\delta\theta_{\MS}$ subtracts only the divergent part proportional to $1 /{\epsilon}+\gamma-\mathrm{ln}(4\pi)$
in dimensional regularization with $d=4-2\epsilon$ being the space-time dimension. 
Since the divergent parts in the $\delta\theta$ and $\delta\theta_{\MS}$ counterterms 
are of the same form,  $\theta(\overline{\mu})$ can be rewritten as
\begin{equation}
\label{thetaMS}
	\theta(\overline{\mu})=\theta-\delta \theta|_{\mathrm{fin}}
\end{equation}
where the subscript fin denotes the finite part of the quantity $\delta \theta$, obtained after subtracting the terms proportional to 
$1 /{\epsilon}+\gamma-\mathrm{ln}(4\pi)$. Difference at two-loop order has been neglected in this expression.

The physical parameters which would be used in Eq. (\ref{thetaMS}), such as  $\mu^2$ and $\lambda$, the quadratic and quartic couplings
in the Higgs potential, the vacuum expectation value $v$, the top Yukawa coupling $y_t$, 
the gauge couplings $g_2$ and $g_Y$ of $\mathrm{SU}(2)_\mathrm{L} \times \mathrm{U}(1)_\mathrm{Y}$ group, 
can be determined from physical observables, such as the pole mass of the Higgs boson $\left(M_{h}\right)$,  
the pole mass of the top quark $\left(M_{t}\right)$, the pole mass of the Z boson $\left(M_{Z}\right)$,  the pole mass of the W boson $\left(M_{W}\right)$,
and the Fermi constant $\left(G_{\mu}\right)$. These physical observables are listed in Table.~\ref{observables}. 
If knowing the corresponding counterterms in the physical scheme,  the $\MS$ couplings are then obtained using \eqref{thetaMS}.
For example, if knowing $\delta \lambda$, we then obtain $\lambda(\overline{\mu})$  in the $\MS$ scheme.
More details are explained as follows.

 \begin{table}[!t]
	\setlength{\tabcolsep}{.5em}
	\renewcommand{\arraystretch}{1.2}
	\begin{tabular}{c|c}
		\multicolumn{2}{c}{Input values of SM observables} \\
		\hline
		$\quad$$\quad$Observables$\quad$$\quad$ & Values\\
		\hline
		$M_W$ & $80.384\pm0.014$ GeV \\
		$M_Z$ & $91.1876\pm0.0021$ GeV \\
		$M_h$ & $125.15\pm0.24$ GeV  \\
		$M_t$ & $173.34\pm0.76$ GeV  \\
		$v=(\sqrt{2}G_\mu)^{-1/2}$ & $\quad$$246.21971\pm0.00006$ GeV $\quad$ \\
		$\alpha_3({M}_Z)$ & $0.1184\pm0.0007$\\
		\hline
	\end{tabular}
	\caption{Input values of physical observables used to fix the SM fundamental parameters $\lambda$, $m$, $y_t$, $g_2$, and $g_Y$. $M_W$, $M_Z$, $M_h$, and $M_t$ are
		the pole masses of the $W$ boson, of the $Z$ boson, of the Higgs boson, and of the top quark, respectively. $G_\mu$ is the Fermi constant for $\mu$ decay, and $\alpha_3$ is the $SU(3)_c$ gauge coupling at the scale $\mu=M_Z$ in the $\MS$ scheme.}
	\label{observables}
\end{table}

We follow Ref.~\cite{Sirlin:1985ux} to fix the notation.
We write the classical Higgs potential in bare quantities as 
\begin{equation}
\label{Vunrenormalized}
V=-\mu_{0}^{2} \Phi^{\dagger} \Phi+\lambda_{0}\left(\Phi^{\dagger} \Phi\right)^{2}
\end{equation}
with
\begin{equation}
\Phi=\left(\begin{array}{c}{\phi^{+}} \\ {\sqrt{\frac{1}{2}}\left(\phi_{1}+i \phi_{2}+v_{0}\right)}\end{array}\right)
\end{equation}
Setting $\lambda_{0}=\lambda-\delta \lambda, v_{0}=v-\delta v, \mu_{0}^{2}=\mu^{2}-\delta \mu^{2}$,
where $\lambda$, $v$ and $\mu$ are regarded as renormalized quantities, we write
\begin{equation}
V=V_{(r)}-\delta V
\end{equation}
with
\begin{equation}
\begin{aligned} V_{(\mathrm{r})}=& \lambda\left[\phi^{+} \phi^{-}\left(\phi^{+} \phi^{-}+\phi_{1}^{2}+\phi_{2}^{2}\right)+\frac{1}{4}\left(\phi_{1}^{2}+\phi_{2}^{2}\right)^{2}\right] \\ &+\lambda v \phi_{1}\left[\phi_{1}^{2}+\phi_{2}^{2}+2 \phi^{+} \phi^{-}\right]+2 \lambda v^{2} \frac{1}{2} \phi_{1}^{2} \end{aligned}
\end{equation}
and
\begin{equation}
\label{deltaV}
\begin{aligned} \delta V=& \delta \lambda\left[\left(\phi^{+} \phi^{-}\right)\left(\phi^{+} \phi^{-}+\phi_{1}^{2}+\phi_{2}^{2}\right)+\frac{1}{4}\left(\phi_{1}^{2}+\phi_{2}^{2}\right)^{2}\right] \\ &+[\lambda \delta v+v \delta \lambda] \phi_{1}\left[\phi_{1}^{2}+\phi_{2}^{2}+2 \phi^{+} \phi^{-}\right]+\delta \tau\left[\phi^{+} \phi^{-}+\frac{1}{2} \phi_{2}^{2}\right] \\ &+\delta M_{\mathrm{h} }^{2} \frac{1}{2}\phi_{1}^{2}+v \delta \tau \phi_{1} \end{aligned}
\end{equation}
where
\begin{eqnarray}
\delta M_{\mathrm{h}}^{2} &&= 3 v^{2} \delta \lambda+6 \lambda v \delta v-\delta \mu^{2},\label{countertermMh} \\
\delta \tau &&= v^{2} \delta \lambda+2 \lambda v \delta v-\delta \mu^{2}.\label{countertermtu}
\end{eqnarray}
$v$ is determined at tree-level by $G_\mu$ as shown in Table. \ref{observables}.

	In order to determine $\delta \lambda$, $\delta v$ and $\delta \mu^{2} $ we need three constraints.
	The strategy is to adjust $\delta \tau$ so that the $v \delta \tau \phi_{1}$ term in Eq.~\ref{deltaV} cancels the tadpole diagrams.
	Calling $iT$ the sum of the tadpole diagrams with the external legs extracted, we have the condition
	\begin{equation}
	\delta \tau=-T / v.
	\label{constrainttu}
	\end{equation}
	A second constraint is conveniently obtained by demanding that the coefficient  of the term proportional to $\frac{1}{2} \phi_{1}^{2}$ in $V_{(\mathrm{r})}$ be the physical mass of the Higgs boson. So we have
	\begin{equation}
	M_{\mathrm{h}}^{2} = 2 \lambda v^{2} \label{MHdefine}
	\end{equation}
	and $\delta M_{\mathrm{h}}^{2}$ is fixed by condition of on-shell renormalization, i.e.
	\begin{eqnarray}
	\delta M_{\mathrm{h}}^{2}=\operatorname{Re} \Pi_{\mathrm{hh}}\left(M_{\mathrm{h}}^{2}\right),\label{MHdefine-a}
	\end{eqnarray}
	where $\Pi_{\mathrm{hh}}\left(M_{\mathrm{h}}^{2}\right)$ is the Higgs boson self-energy
	evaluated on shell.
	A third constraint is provided by Eq.~(9b) of Ref.~\cite{Sirlin:1980nh}
	\begin{equation}
	\delta M_{W}^{2}=\operatorname{Re} \Pi_{w w}\left(M_{W}^{2}\right),
	\label{constrainMw}
	\end{equation}
	where $\Pi_{w w}\left(M_{W}^{2}\right)$ is  the W boson self-energy evaluated on shell.
	Recalling that the W-mass counterterm is given by~\cite{Sirlin:1980nh}
	\begin{equation}
	\delta M_{\mathrm{W}}^{2}=\frac{1}{2}\left(v^{2} g_{2} \delta g_{2}+g_{2}^{2} v \delta v\right),
	\label{countertermMw}
	\end{equation}
	$\delta v$ is obtained using this expression with  $\delta g_2$ known from other condition which
	can be found in Eq. (28a) of \cite{Sirlin:1980nh}.
	Putting $\delta v$, Eqs. (\ref{MHdefine-a}) and (\ref{countertermMw}) into (\ref{countertermMh}) and (\ref{countertermtu}), one can then obtain $\delta \lambda$ and $\delta \mu^2$.
	They are as follows.
	\begin{eqnarray}
	\delta \mu^{2}=&&\frac{1}{2}\left[\operatorname{Re} \Pi_{\mathrm{hh}}\left(M_{\mathrm{h}}^{2}\right)+3 T / v\right] ,\\
	\delta \lambda / \lambda=&&\left[\operatorname{Re} \Pi_{\mathrm{hh}}\left(M_{\mathrm{h}}^{2}\right)+T / v\right] / M_{\mathrm{h}}^{2}-\operatorname{Re} \Pi_{\mathrm{ww}}\left(M_{\mathrm{W}}^{2}\right) / M_{\mathrm{W}}^{2}+2 \delta g_{2} / g_{2}, \\
	\delta v / v=&&\operatorname{Re} \Pi_{\mathrm{ww}}\left(M_{\mathrm{W}}^{2}\right) /\left(2 M_{\mathrm{W}}^{2}\right)-\delta g_{2} / g_{2}
	\end{eqnarray}
	We can get the expressions of the  counterterms of the other parameters in a similar way. 
	
	Ignoring the contribution of higher order, we list the one-loop results of counterterms as follows.
	\begin{eqnarray}
	\label{90}
	\delta^{(1)} \lambda &&=\frac{G_{\mu}}{\sqrt{2}} M_{h}^{2}\left\{\Delta r_{0}^{(1)}+\frac{1}{M_{h}^{2}}\left[\frac{T^{(1)}}{v}+\operatorname{Re} \Pi_{hh}\left(M_{h}^{2}\right)\right]\right\}, \\
	\label{92}
	\delta^{(1)} y_{t} &&=2\left(\frac{G_{\mu}}{\sqrt{2}} M_{t}^{2}\right)^{1 / 2}\left(\frac{\operatorname{Re} \Pi_{t t}\left(M_{t}^{2}\right)}{M_{t}}+\frac{\Delta r_{0}^{(1)}}{2}\right), \\
	\label{93}
	\delta^{(1)} g_{2} &&=\left(\sqrt{2} G_{\mu}\right)^{1 / 2} M_{W}\left(\frac{\operatorname{Re} \Pi_{w w}\left(M_{W}^{2}\right)}{M_{W}^{2}}+\Delta r_{0}^{(1)}\right) ,\\
	\label{94}
	\delta^{(1)} g_{Y}&&=\left(\sqrt{2} G_{\mu}\right)^{1 / 2} \sqrt{M_{Z}^{2}-M_{W}^{2}}\left(\frac{\operatorname{Re} \Pi_{zz}\left(M_{Z}^{2}\right)-\operatorname{Re} \Pi_{ww}\left(M_{W}^{2}\right)}{M_{Z}^{2}-M_{W}^{2}}+\Delta r_{0}^{(1)}\right),
	\end{eqnarray}
	where superscripts $1$ in these equations  indicate that they are results at one-loop order.
	$\Delta r_{0}^{(1)}$ in the above equations can be written as a sum of several terms~\cite{Buttazzo:2013uya}
	\begin{equation}
	\label{deltar-0}
	\Delta r_{0}^{(1)}=V_{W}^{(1)}-\frac{A_{W W}^{(1)}}{M_{W}^{2}}+\frac{\sqrt{2}}{G_{\mu}} \mathcal{B}_{W}^{(1)}+\mathcal{E}^{(1)}
	\end{equation}
	where  $A_{WW}$ is the $W$ boson self-energy at zero momentum,
	 $V_W$  the vertex contribution in the muon decay process, 
	${\cal B}_W$  the box contribution, ${\cal E}$ a term due to the renormalization of external legs.
	They are all computed at zero external momentum.
	Thus we eventually get the $\MS$ parameter to one-loop order as follows~\cite{Buttazzo:2013uya,Wang:2018lhk}.
	\begin{eqnarray}
	\lambda(\overline{\mathrm{\mu}}) &&=\frac{G_{\mu}}{\sqrt{2}} M_{h}^{2}-\delta^{(1)}\left.\lambda\right|_{\mathrm{fin}},  
	\label{one-loop-lambda-1}\\
	y_{t}(\overline{\mathrm{\mu}}) &&=2\left(\frac{G_{\mu}}{\sqrt{2}} M_{t}^{2}\right)^{1 / 2}-\delta^{(1)}\left.y_{t}\right|_{\mathrm{fin}}, 
	\label{one-loop-lambda-2}\\
	g_{2}(\overline{\mathrm{\mu}})&& =2\left(\sqrt{2} G_{\mu}\right)^{1 / 2} M_{W}-\delta^{(1)}\left.g_{2}\right|_{\mathrm{fin}}, 
	\label{one-loop-lambda-3}\\
	g_{Y}(\overline{\mathrm{\mu}})&& =2\left(\sqrt{2} G_{\mu}\right)^{1 / 2} \sqrt{M_{Z}^{2}-M_{W}^{2}}-\delta^{(1)}\left.g_{Y}\right|_{\mathrm{fin}}. \label{one-loop-lambda-4}
	\end{eqnarray}

\subsection{$\MS$ parameters in the SDFDM model}
\label{oneloopmatching}
\begin{figure}[!t]
	\centering
	\subfigure[]
	{\includegraphics[width=.24\textwidth]{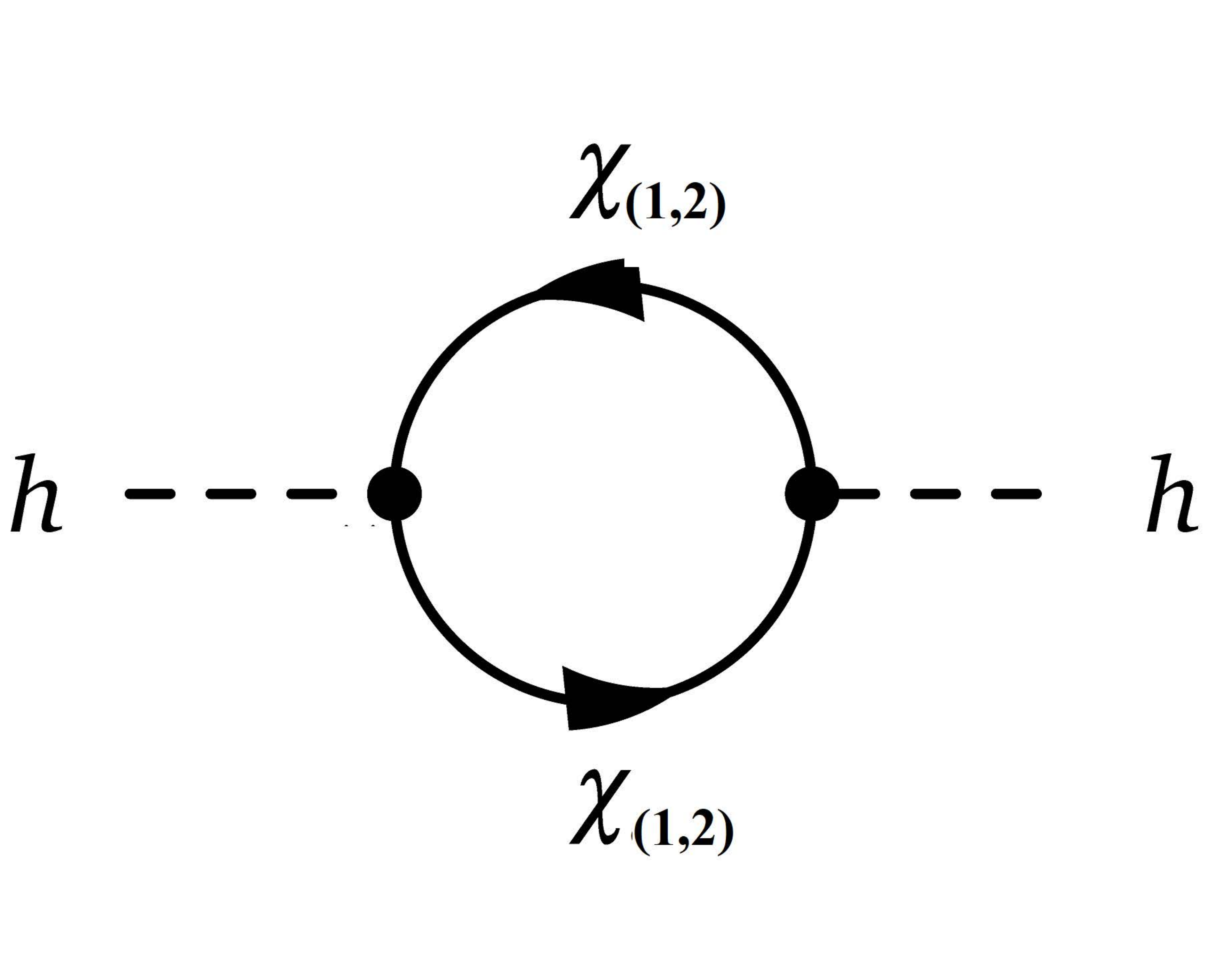}} ~~
	\quad
	\subfigure[]
	{\includegraphics[width=.24\textwidth]{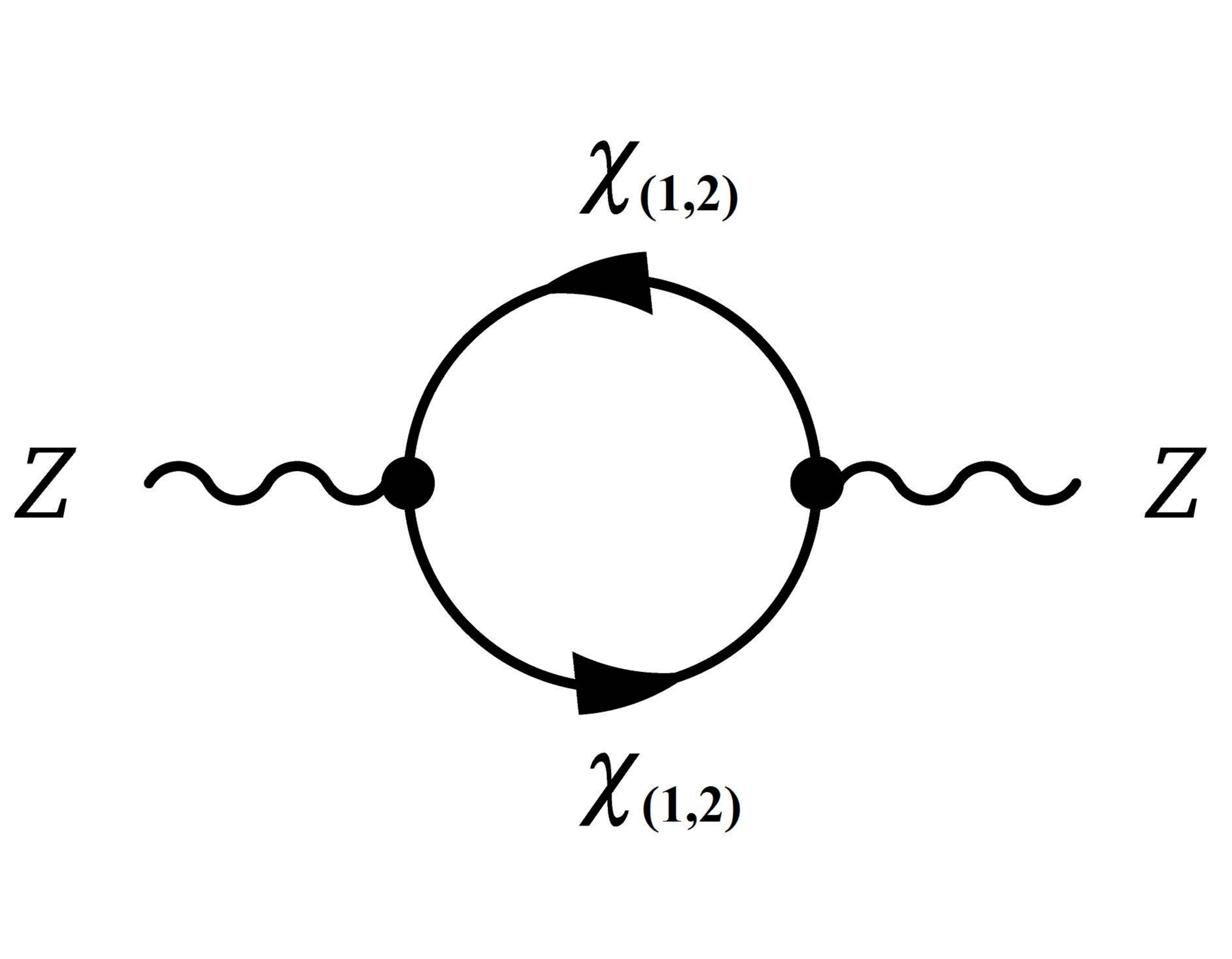}} ~~
	\quad
	\subfigure[]
	{\includegraphics[width=.24\textwidth]{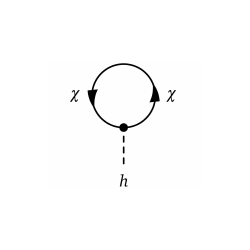}}~~~~~
	\quad
	\subfigure[]
	{\includegraphics[width=.24\textwidth]{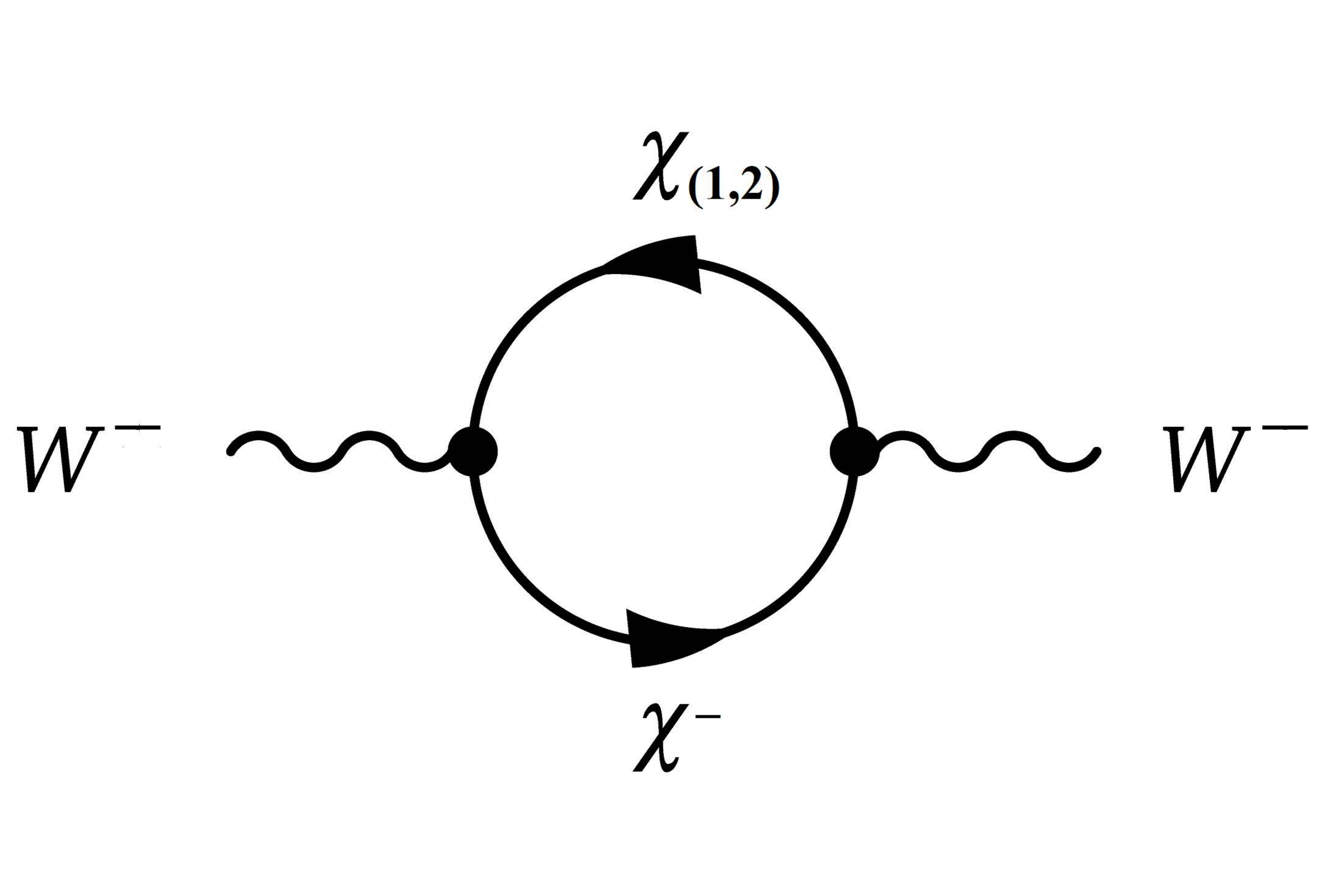}} ~~
	\quad
	\subfigure[]
	{\includegraphics[width=.24\textwidth]{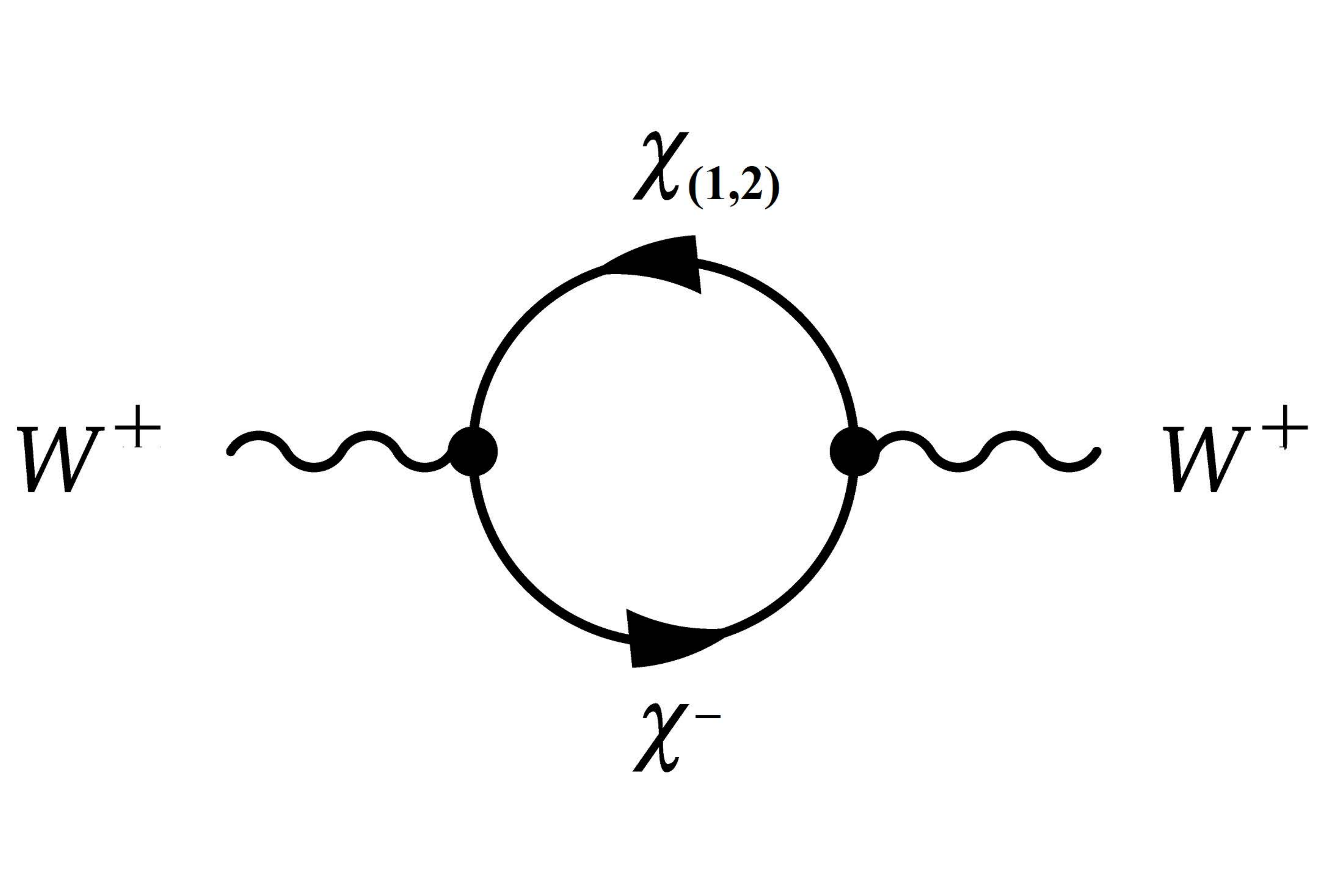}} 
	\caption{Contributions of extra fermions to the self-energy for (a) the Higgs boson, (b)$Z$  boson, and (d)(e) $W$ bosons, 
	as well as to (c) the tadpole of the Higgs boson. $\chi_{(1,2)}$  are dark sector fermions.}
	\label{feynmadiagrams}
\end{figure}

To determine the initial values of running couplings, we use the equations given in the  last section.
Since the threshold corrections have been done to NNLO in the SM, we 
only need to calculate the contribution of extra fermions in the SDFDM model.
All the relevant Feynman diagrams  for computing $\delta^{(1)}\left.\lambda\right|_{\mathrm{fin}}$ 
with extra fermions are listed in Fig. \ref{feynmadiagrams}.

As  singlet-doublet fermions in SDFDM model do not couple to SM leptons, Eq.~\eqref{deltar-0} can be further simplified as
\begin{equation}
\Delta r_{0}=-\frac{A_{W W}}{M_{W}^{2}}.
\end{equation}
Summing over all the loop contributions and using the matching conditions, we get coupling constants in the ${\MS}$ scheme 
at $M_{t} = 173~ \mathrm{GeV}$ energy scale and for the SDFDM model respectively. 

	We summarize here the one-loop corrections to $\lambda$ from new particles in SDFDM model by using Eq.~\eqref{90}.
	We write $\delta^{(1)} \lambda_{SDFDM}$ in terms of finite parts of the Passarino-Veltman functions
	\begin{equation}
	A_{0}(M)=M^{2}\left(1-\ln \frac{M^{2}}{\overline{\mu}^{2}}\right), 
	~ B_{0}\left(M_{1}, M_{2},p\right)=-\int_{0}^{1} \ln \frac{x M_{1}^{2}+(1-x) M_{2}^{2}-x(1-x) p^{2}}{\overline{\mu}^{2}} d x .
	\end{equation}
	The one-loop result is
	\begin{equation}
	\begin{aligned} 
	\delta^{(1)} \lambda|_{fin} =& \frac{G_{\mu}}{\sqrt{2} (4\pi)^2} \left\{ y_{A}^2[4A_{0}(M_{\chi_1^0})-2(M^2_h-4M^2_{\chi_{1}^0})B_{0}(M_{\chi_1^0},M_{\chi_1^0},M_{h})] \right.\\ 
	& +y_{B}^2[4A_{0}(M_{\chi_2^0})-2(M^2_h-4M^2_{\chi_{2}^0})B_{0}(M_{\chi_2^0},M_{\chi_2^0},M_{h})]\\
	& +2y_{C}^2[A_{0}(M_{\chi_1^0})+A_{0}(M_{\chi_2^0})-(M^2_h-M^2_{\chi_{1}^0}-M^2_{\chi_{2}^0})B_{0}(M_{\chi_1^0},M_{\chi_2^0},M_{h})]\\
	& +2y_{D}^2[A_{0}(M_{\chi_1^0})+A_{0}(M_{\chi_2^0})-(M^2_h-M^2_{\chi_{1}^0}-M^2_{\chi_{2}^0})B_{0}(M_{\chi_1^0},M_{\chi_2^0},M_{h})]\\
	& \left. +8 y_{C} y_{D} M_{\chi_1^0} M_{\chi_2^0} B_{0}(M_{\chi_1^0},M_{\chi_2^0},M_{h}) \right\}\\
	& + \frac{G_{\mu}}{\sqrt{2} (4\pi)^2v}\left[-4y_{A} M_{\chi_1^0} A_{0}(M_{\chi_1^0})-4y_{B} M_{\chi_2^0} A_{0}(M_{\chi_2^0})\right]\\
	& + \frac{G_{\mu}}{\sqrt{2}}M_{h}^2 \left.\Delta r_{0}^{(1)}\right|_{\text {fin }} 
	\end{aligned}
	\label{one-loop-lambda-1a}
	\end{equation}
	where $y_{A,B,C,D}$ has been given in the text following Eq. (\ref{Lagrangian-1}) and 
	\begin{equation}
	\begin{aligned}
	\left.\Delta r_{0}^{(1)}\right|_{\text {fin }}=& \frac{1}{(4 \pi v)^{2}} \left\{ (sin^2 \theta_{L} + sin^2\theta_{R}) \left[ \frac{2M^2_{\chi_{1}^0}}{M^2_{\chi^{-}}-M^2_{\chi_{1}^0}}A_{0}(M_{\chi_1^0})-\frac{2M^2_{\chi^{-}}}{M^2_{\chi^{-}}-M^2_{\chi_{1}^0}}A_{0}(M_{\chi^{-}})+ M^2_{\chi^{-}}+M^2_{\chi_{1}^0} \right] \right.\\
	&+ 8 sin\theta_{L} sin\theta_{R} \left[ \frac{M_{\chi_1^0} M_{\chi^{-}}}{M^2_{\chi^{-}}-M^2_{\chi_{1}^0}}\left( A_{0}(M_{\chi^{-}})-A_{0}(M_{\chi_1^0})\right) \right]\\
	&+ (cos^2 \theta_{L} + cos^2\theta_{R}) \left[ \frac{2M^2_{\chi_{2}^0}}{M^2_{\chi^{-}}-M^2_{\chi_{2}^0}}A_{0}(M_{\chi_2^0})-\frac{2M^2_{\chi^{-}}}{M^2_{\chi^{-}}-M^2_{\chi_{2}^0}}A_{0}(M_{\chi^{-}})+ M^2_{\chi^{-}}+M^2_{\chi_{2}^0} \right]\\
	&+\left. 8 cos\theta_{L} cos\theta_{R} \left[ \frac{M_{\chi_2^0} M_{\chi^{-}}}{M^2_{\chi^{-}}-M^2_{\chi_{2}^0}}\left( A_{0}(M_{\chi^{-}})-A_{0}(M_{\chi_2^0})\right) \right] \right\}	
	\end{aligned}
	\end{equation}
Plugging Eq.  (\ref{one-loop-lambda-1a}) into Eq. (\ref{one-loop-lambda-1}) we obtain $\lambda$ at one-loop order in the SDFDM model.
Contributions of extra fermions to 
$\delta^{(1)}\left.y_{t}\right|_{\mathrm{fin}}$, $\delta^{(1)}\left.g_{2}\right|_{\mathrm{fin}}$ and $\delta^{(1)}\left.g_{Y}\right|_{\mathrm{fin}}$ can
be similarly obtained.  Plugging them into Eqs. (\ref{one-loop-lambda-2}),  (\ref{one-loop-lambda-3}) and (\ref{one-loop-lambda-4})
we obtain relevant parameters at one-loop order in the SDFDM model.
Using these parameters in the $\MS$ scheme, we then carry out the calculation of the effective action in the $\MS$ scheme. 

\section{one-loop $\beta$ and $\gamma$ function in the SDFDM model}
\label{betafunction}

The $\beta$-function and the anomalous dimension can be decomposed into two parts:
\begin{align}
&\beta^{\text {total }}=\beta^{\mathrm{SM}}+\beta^{\mathrm{SDFDM}},
&\gamma^{\text {total }}=\gamma^{\mathrm{SM}}+\gamma^{\mathrm{SDFDM}}
\end{align}
where $\beta^{\mathrm{SM}}$ and $\gamma^{\mathrm{SM}}$ are the $\beta$ function and the anomalous dimension in the SM, 
while $\beta^{\mathrm{SDFDM}}$ and $\gamma^{\mathrm{SDFDM}}$ are the contributions from new particles in the SDFDM model.

The $\beta$ functions in the SM are known to three-loop~\cite{Buttazzo:2013uya}.
We list the one-loop results  as follows.
\begin{eqnarray}
\beta^{\mathrm{SM}}\left(g_{1}\right)=&&\frac{1}{(4 \pi)^{2}}\left(\frac{41}{10}\right) g_{1}^{3},\label{betafunctiong1SM}\\
\beta^{\mathrm{SM}}\left(g_{2}\right)=&&\frac{1}{(4 \pi)^{2}}\left(-\frac{19}{6}\right) g_{2}^{3},\label{betafunctiong2SM}\\
\beta^{\mathrm{SM}}\left(g_{3}\right)=&&\frac{1}{(4 \pi)^{2}}(-7) g_{3}^{3},\\
\beta^{\mathrm{SM}}\left(y_{t}\right)=&&\frac{y_{t}}{(4 \pi)^{2}}\left[\frac{9 y_{t}^{2}}{2}+\frac{3 y_{b}^{2}}{2}+y_{\tau}^{2}-8 g_{3}^{2}-\frac{9 g_{2}^{2}}{4}-\frac{17 g_{1}^{2}}{20}\right],\\
\beta^{\mathrm{SM}}\left(y_{b}\right)=&&\frac{y_{b}}{(4 \pi)^{2}}\left[\frac{3 y_{t}^{2}}{2}+\frac{9 y_{b}^{2}}{2}+y_{\tau}^{2}-8 g_{3}^{2}-\frac{9 g_{2}^{2}}{4}-\frac{g_{1}^{2}}{4}\right],\\
\beta^{\mathrm{SM}}\left(y_{\tau}\right)=&&\frac{y_{\tau}}{(4 \pi)^{2}}\left[3 y_{t}^{2}+3 y_{b}^{2}+\frac{5 y_{\tau}^{2}}{2}-\frac{9 g_{2}^{2}}{4}-\frac{9 g_{1}^{2}}{4}\right], 
\end{eqnarray}
\begin{eqnarray}
\label{betafunctionlambdaSM}
\begin{aligned}
\beta^{\mathrm{SM}}(\lambda)&&=\frac{1}{(4 \pi)^{2}}\left[2\lambda\left(12 \lambda+6 y_{t}^{2}+6 y_{b}^{2}+2 y_{\tau}^{2}-\frac{9 g_{2}^{2}}{2}-\frac{9 g_{1}^{2}}{10}\right)\right.\\
&&\left.-6 y_{t}^{4}-6 y_{b}^{4}-2 y_{\tau}^{4}+\frac{9 g_{2}^{4}}{8}+\frac{27 g_{1}^{4}}{200}+\frac{9 g_{2}^{2} g_{1}^{2}}{20}\right].
\end{aligned}
\end{eqnarray}

In this article we focus on the SDFDM model with Dirac type mass.
Here we show one-loop contributions of new particles in the SDFDM model 
to the $\beta$ functions of the SM parameters and the one-loop $\beta$ functions of new parameters in the SDFDM model.
They can be extracted using the python tool $\texttt{PyR@TE 2}$\cite{Lyonnet:2016xiz}.
They are as follows.  

The $\beta$ functions of the SM parameters receive one-loop contributions of new particles in the SDFDM model
as follows.
\begin{eqnarray}
\beta^{\mathrm{SDFDM}}\left(g_{1}\right)=&&\frac{1}{(4 \pi)^{2}}\left(\frac{2}{5}\right) g_{1}^{3},\\
\beta^{\mathrm{SDFDM}}\left(g_{2}\right)=&&\frac{1}{(4 \pi)^{2}}\frac{2}{3} g_{2}^{3},\\
\beta^{\mathrm{SDFDM}}\left(y_{\tau}\right)=&&\frac{1}{(4 \pi)^{2}}\left(y_{1}^{2}+y_{2}^{2}\right) y_{\tau},\\
\beta^{\mathrm{SDFDM}}\left(y_{b}\right)=&&\frac{1}{(4 \pi)^{2}}\left(y_{1}^{2}+y_{2}^{2}\right) y_{b},\\
\beta^{\mathrm{SDFDM}}\left(y_{t}\right)=&&\frac{1}{(4 \pi)^{2}}\left(y_{1}^{2}+y_{2}^{2}\right) y_{t},\\
\beta^{\mathrm{SDFDM}}(\lambda)=&&\frac{1}{(4 \pi)^{2}}\left[-2 y_{1}^{4} -2 y_{2}^{4}+4 \lambda\left(y_{1}^{2}+y_{2}^{2}\right)\right].
\end{eqnarray}
The one-loop $\beta$ functions of new parameters in the SDFDM model are as follows.
\begin{eqnarray}
\beta^{\mathrm{SDFDM}}\left(y_{1}\right)=&&\frac{1}{(4 \pi)^{2}}\left[\frac{5}{2} y_{1}^{3}+ y_{1} y_{2}^{2}-\frac{9}{20} g_{1}^{2} y_{1}-\frac{9}{4} g_{2}^{2} y_{1}+3 y_{t}^{2} y_{1}+3 y_{b}^{2} y_{1}+y_{\tau}^{2} y_{1}\right],\\
\beta^{\mathrm{SDFDM}}\left(y_{2}\right)=&&\frac{1}{(4 \pi)^{2}}\left[\frac{5}{2} y_{2}^{3}+ y_{1}^{2} y_{2}-\frac{9}{20} g_{1}^{2} y_{2}-\frac{9}{4} g_{2}^{2} y_{2}+3 y_{t}^{2} y_{2}+3 y_{b}^{2} y_{2}+y_{\tau}^{2} y_{2}\right]\label{betafunctionlambdaSDFDM}.
\end{eqnarray}

Note here that $g_{1} (g_{1}^2 = \frac{5}{3}g_{Y}^2)$, $g_{2}$, $g_{3}$ are the gauge couplings,
$y_t$, $y_b$, $y_{\tau}$, $y_1$, and $y_2$ are 
the Yukawa couplings, and $\lambda$ is the Higgs quartic coupling.
The one-loop anomalous dimension of the Higgs field is
\begin{equation}
\gamma^{\text {total }}=\gamma^{\mathrm{SM}}+\gamma^{\mathrm{SDFDM}}=\frac{1}{(4 \pi)^{2}}\left[\frac{9}{4} g_{2}^{2}+\frac{9}{20} g_{1}^{2}-3 y_{t}^{2}-3 y_{b}^{2}-y_{\tau}^{2}\right]+\frac{1}{(4 \pi)^{2}}(-y_{1}^2-y_{2}^2).
\end{equation}

\section{Renormalization of kinetic term in effective action}
\label{derivativecorr}

We compute effective action of an external field using derivative expansion. 
As long as the field varies slowly with respect to space and time, this is a valid approximation.
Keeping derivatives up to second order, the Euclidean effective action for a neutral scalar $\phi$ is written as
\begin{equation}
S_{\mathrm{eff}}[\phi]=\int d^{4} x\left[V_{\mathrm{eff}}(\phi)+\frac{1}{2}\left(\partial_{\mu} \phi\right)^{2} Z_{2}(\phi)\right],
\label{effaction}
\end{equation}
where $V_{\mathrm{eff}}$ is the effective potential. 
 The one-loop result of  $V_{\mathrm{eff}}$ in the SM in the background $R_{\xi}$ gauge is given in~\cite{DiLuzio:2014bua}. 
$Z_2$ can be obtained from the $p^2$ terms in self-energy Feynman diagrams.
We renormalize $Z_2$ to make $Z_2(\phi=0)=1$ which means that the
kinetic term goes back to the standard form when there is no external field.

\subsection{Feynman rules in background $R_{\xi}$ gauge}
\begin{figure}[H]
	\centering
	\includegraphics[width=0.8\linewidth]{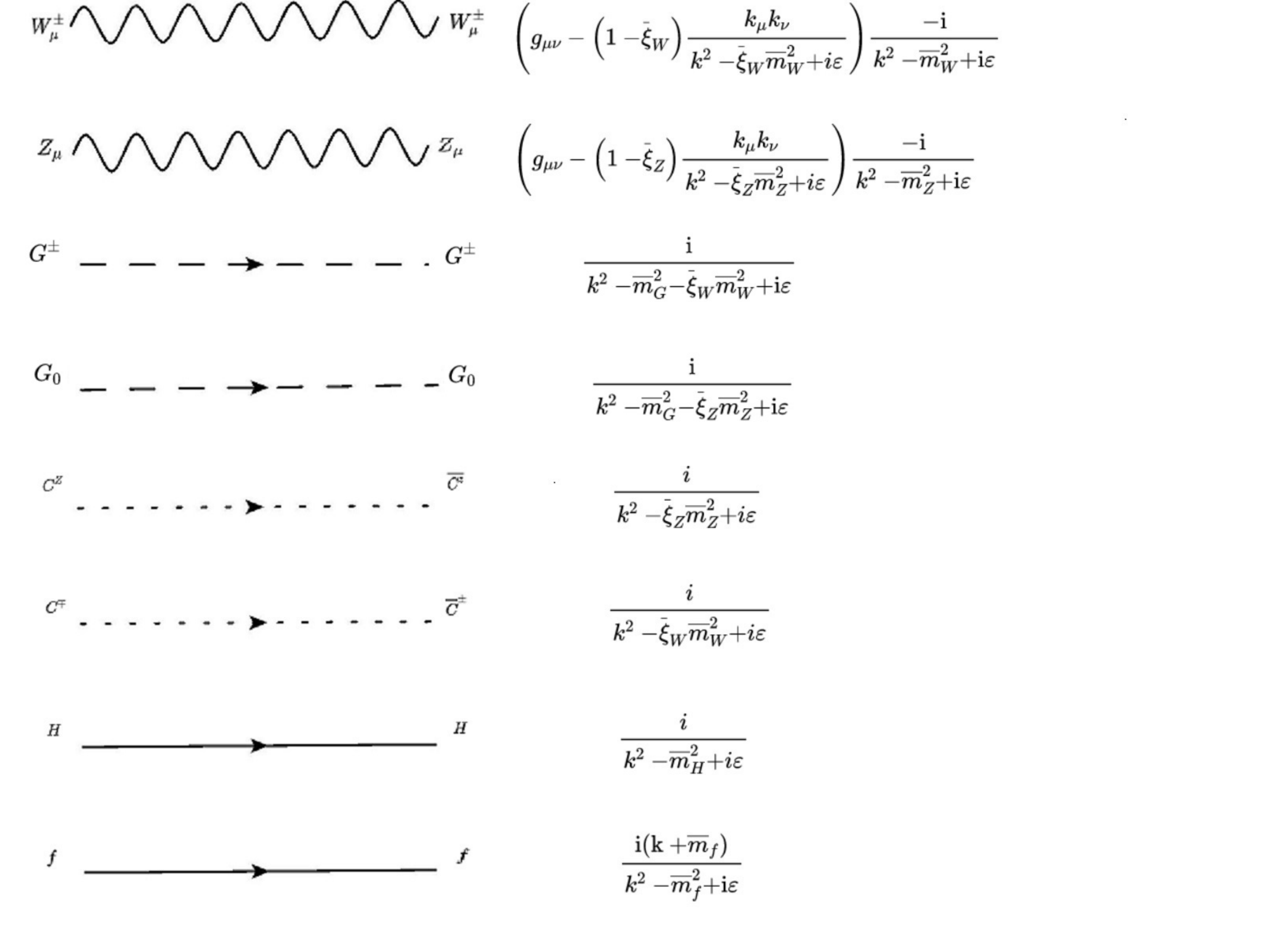}
	\caption{ Propagators for SM fields with external field $\phi$ in background $R_{\xi}$ gauge.
	 $\overline{m}^2_{h}$, $\overline{m}^2_{G}$, $\overline{m}_{W}$, $\overline{m}_{Z}$, $\overline{m}_{f}$ 
	 are the $\phi$-dependent masses, which can be defined as $\overline{m}^2_{G} = -m^{2}_\phi+\lambda{\phi}^2$ ,
	 $\overline{m}^2_{H} = -m^2_\phi+3\lambda\phi^2$ , $\overline{m}_{W} = \frac{1}{2} g \phi$ , 
	 $\overline{m}_{Z} = \frac{1}{2} \sqrt{g^2+g^{\prime2}} \phi$ ,  $\overline{m}_{f} = \frac{y_{f}}{\sqrt{2}} \phi$. 
     $m^2_\phi$ is the mass term in the Higgs potential,$y_{f}$ is the Yukawa coupling for alternative SM Fermion.
     Note here that $G^{\pm}$ and $G_{0}$ are the goldstone bosons,  $C_{Z}$ and $C^{\pm}$ are the ghost fields.
 }
  \label{fig:propagators}
\end{figure}

The Feynman rules with external field $\phi$ in the SM and in the SDFDM model are given in Fig.~\ref{fig:propagators}.
Here, we only list the vertices that we need in $Z_{2}$ calculation.
We have introduced
\begin{eqnarray}
\overline{m}^2_{G} = -m^{2}_\phi+\lambda{\phi}^2 , ~~\overline{m}^2_{H} = -m^2_\phi+3\lambda\phi^2
\end{eqnarray}
where $m^2_\phi$ is the mass term in the Higgs potential given in \eqref{treelevel-potential}.
The other $\phi$-dependent masses can be obtained by substituting the vacuum expectation value $v$ with $\phi$.

 We define the field-dependent masses of goldstone bosons and ghost particles as:
\begin{align}
& \overline{m}_{C^{\pm}}^{2}=\overline{\xi}_{W} \overline{m}_{W}^{2},\\
&\overline{m}_{C_{Z}}^{2}=\overline{\xi}_{Z} \overline{m}_{Z}^{2},\\
&\overline{m}_{G^{+}}^{2}=\overline{m}_{G}^{2}+\overline{\xi}_{W} \overline{m}_{W}^{2},\\
&\overline{m}_{G^{0}}^{2}=\overline{m}_{G}^{2}+\overline{\xi}_{Z} \overline{m}_{Z}^{2}.
\end{align}

\begin{figure}[H]
	\centering
	\includegraphics[width=0.7\linewidth]{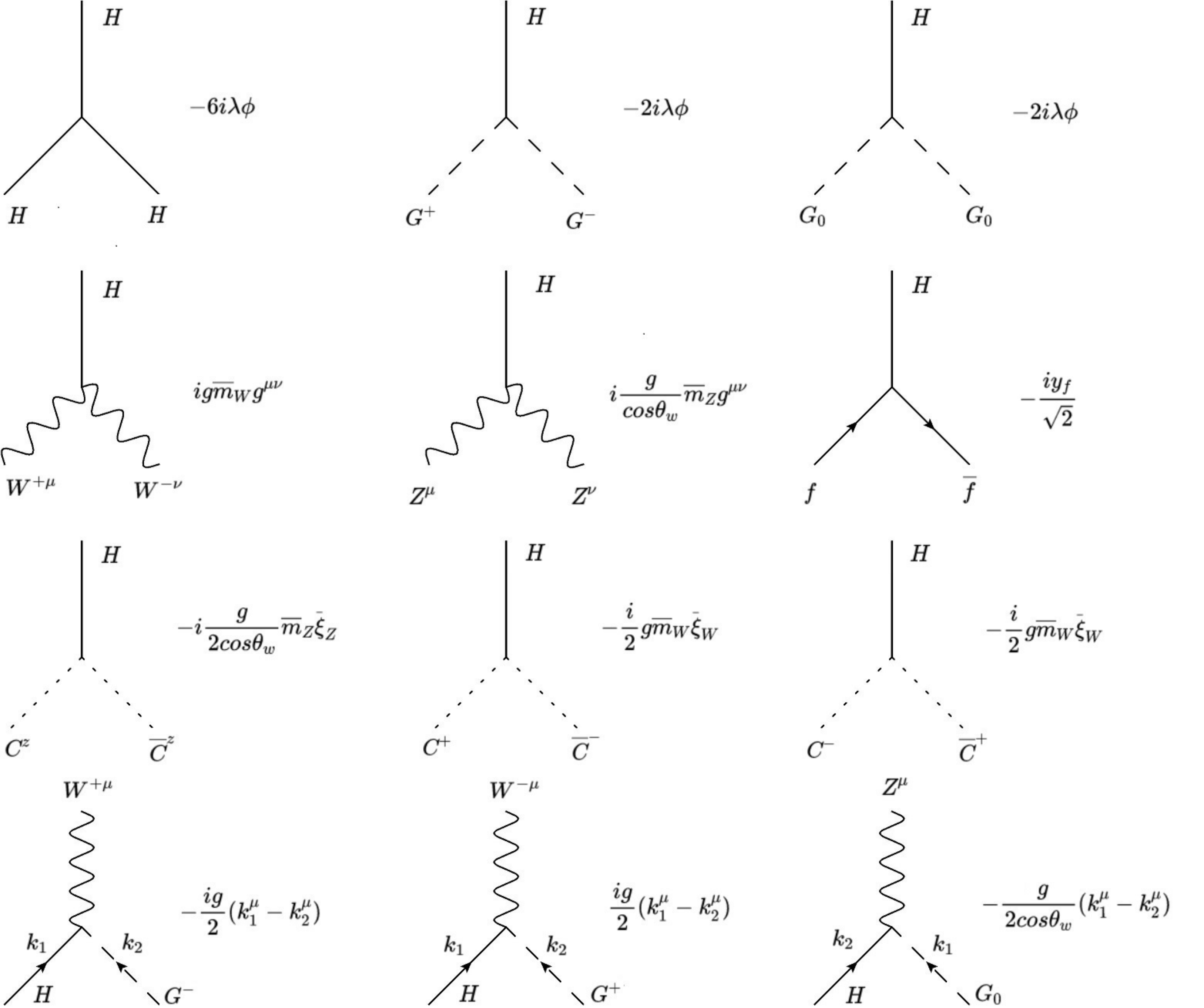}
	\caption{ Vertices with external field $\phi$ for the SM in background $R_{\xi}$ gauge.
	$\overline{m}_{W}$ and $\overline{m}_{Z}$ are the 
	 $\phi$-dependent masses as given in Fig. \ref{fig:propagators}.
	 $\overline{\xi}_{W}$ and $\overline{\xi}_{Z}$ are the gauge fixing parameters in background $R_{\xi}$ gauge.}
	 \label{fig:vertices}
\end{figure}

\subsection{$Z_{2}$ factor in the SM}
\begin{figure}[h]
	\centering
	\includegraphics[width=0.7\linewidth]{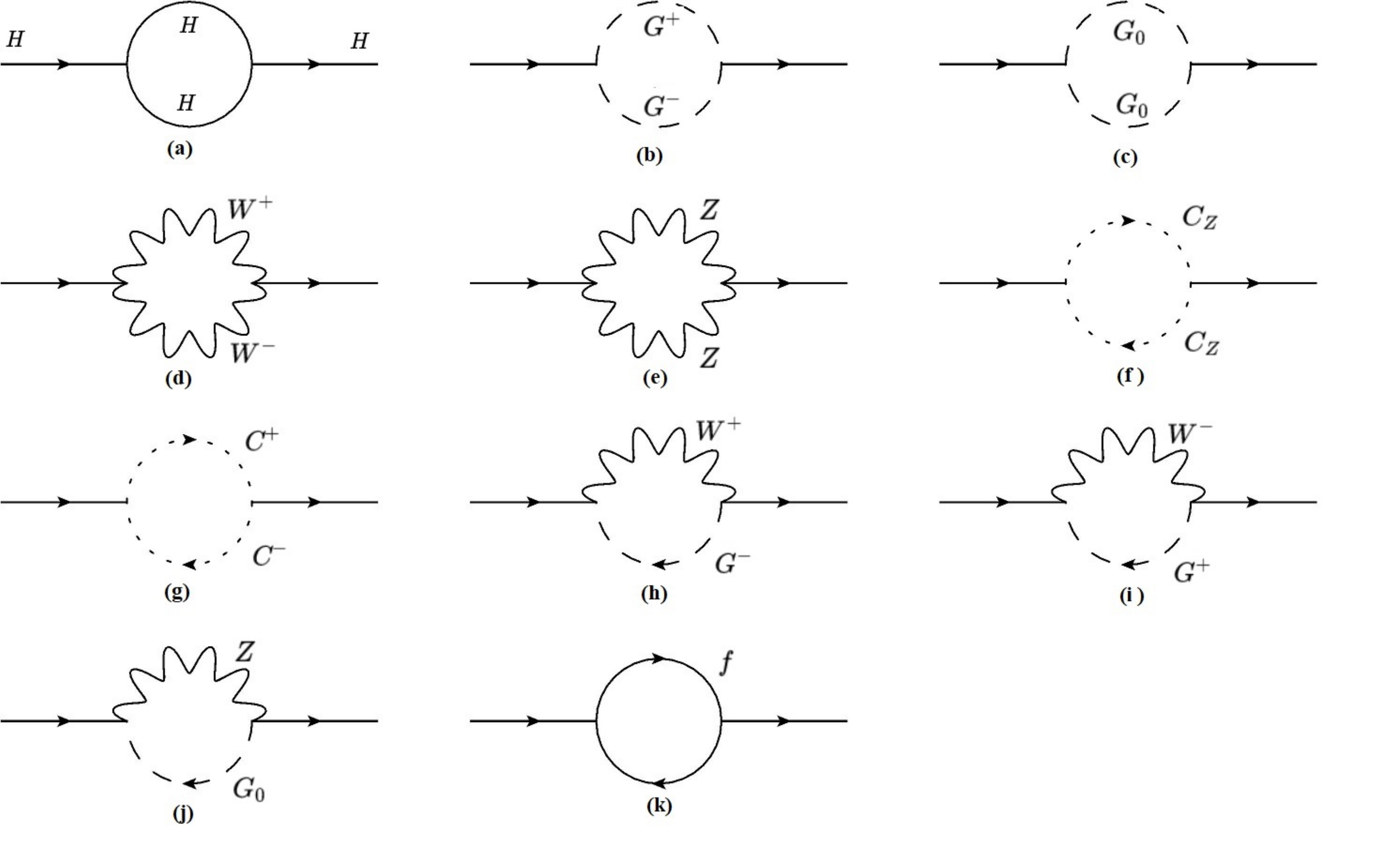}
	\caption{Self-energy diagrams contributing to $Z_{2}$ factor in the SM.
	           $G^{\pm}$ and $G_{0}$ are the goldstone bosons,
		$C_{Z}$ and $C^{\pm}$ are the ghost fields.} 
		\label{fig:selfenergydiagram}
\end{figure}

For simplicity, we calculate $Z_{2}$ in the 't Hooft-Feynman gauge with $\overline{\xi}_{W} = \overline{\xi}_{Z} = 1$. 
$Z_2$ comes from the $p^{2}$ term in the Higgs self-energy diagram in Fig.~\ref{fig:selfenergydiagram}.
Notations in~\cite{Tye:1996au} are used for the integrals calculated in the modified minimal
subtraction scheme:
\begin{eqnarray}
&& \frac{i}{16 \pi^{2}} B_{0}\left(m_{1}, m_{2}, p^{2}\right)  = \mu_{0}^{\epsilon} \int \frac{d^{d} k}{(2 \pi)^{d}} \frac{1}{\left(k^{2}-m_{1}^{2}\right)\left((k+p)^{2}-m_{2}^{2}\right)}, \\
&& \frac{i}{16 \pi^{2}} B_{0}\left(m_{1}, m_{2}, p^{2}\right) = \mu_{0}^{\epsilon} \int \frac{d^{d} k}{(2 \pi)^{d}} \frac{1}{\left(k^{2}-m_{1}^{2}\right)\left((k+p)^{2}-m_{2}^{2}\right)}, \\
&& B_{0}\left(m_{1}, m_{2}, p^{2}\right) =B_{0}^{0}\left(m_{1}, m_{2}\right)+B_{0}^{1}\left(m_{1}, m_{2}\right) \cdot p^{2}+O\left(p^{4}\right)+\dots
\end{eqnarray}
where $B_{0}^{0}\left(m_{1}, m_{2}\right)$ and $B_{0}^{1}\left(m_{1}, m_{2}\right)$ can be express as
\begin{eqnarray}
B_{0}^{0}(m_{1},m_{2})&&=1+\frac{m_{1}^{2} \ln \frac{m_{1}^{2}}{\mu^{2}}-m_{2}^{2} \ln \frac{m_{2}^{2}}{\mu^{2}}}{m_{2}^{2}-m_{1}^{2}},\\
B_{0}^{1}(m_{1},m_{2})&&=\frac{1}{2} \frac{m_{1}^{2}+m_{2}^{2}}{\left(m_{1}^{2}-m_{2}^{2}\right)^{2}}-\frac{m_{1}^{2} m_{2}^{2} \ln \frac{m_{1}^{2}}{m_{2}^{2}}}{\left(m_{1}^{2}-m_{2}^{2}\right)^{3}}.
\end{eqnarray}

 When $m_{1} = m_{2}=m$, $B_{0}^{0}(m_{1},m_{2})$ and $B_{0}^{1}(m_{1},m_{2})$ can be written
as $B_{0}^{0}\left(m\right)$ and $B_{0}^{1}\left(m\right)$. They are expressed as
\begin{eqnarray}
B_{0}^{0}(m)&& =-ln\frac{m^{2}}{\mu^{2}}, \\
B_{0}^{1}(m) &&=\frac{1}{6m^{2}}.
\end{eqnarray}

\begin{table}[h]
	\setlength{\tabcolsep}{.5em}
	\renewcommand{\arraystretch}{1.2}
	\begin{tabular}{|cc|}
		\hline 
		$a)$ &$\frac{1}{16\pi^2}(18\lambda^{2}\phi^{2}B_{0}^{1}(\overline{m}_{H}))$ \\ 
		$b)$ &$\frac{1}{16\pi^2}(4\lambda^{2}\phi^{2}B_{0}^{1}(\overline{m}_{G \pm}))$ \\  
		$c)$ &$\frac{1}{16\pi^2}(2\lambda^{2}\phi^{2}B_{0}^{1}(\overline{m}_{G_{0}}))$ \\ 
		\hline
		$d)$ &$\frac{1}{16\pi^2}(4g^{2}\overline{m}_{W}^{2}B_{0}^{1}(\overline{m}_{W}))$ \\ 
		$e)$ &$\frac{1}{16\pi^2}(\frac{4g^{2}}{cos^{2}(\theta_{w})}\overline{m}_{Z}^{2}B_{0}^{1}(\overline{m}_{Z}))$\\
		\hline
		$f)$ &$\frac{1}{16\pi^2}(-\frac{g^{2}}{4cos^{2}(\theta_{w})}\overline{m}_{C_{Z}}^{2}B_{0}^{1}(\overline{m}_{C_{Z}}))$\\
		$g)$ &$\frac{1}{16\pi^2}(-\frac{g^{2}}{4}\overline{m}_{C^\pm}^{2}B_{0}^{1}(\overline{m}_{C^\pm}))$\\
		\hline
		$h)$ &$\frac{1}{16\pi^2}(-\frac{g^{2}}{4})[(-2\overline{m}_{G^{-}}^{2}+\overline{m}_{W^{+}}^{2}) B_{0}^{1}(\overline{m}_{G^{-}},\overline{m}_{W_{+}})
		-2B_{0}^{0}(\overline{m}_{G^{-}},\overline{m}_{W_{+}})]$\\
		$i)$ &$\frac{1}{16\pi^2}(-\frac{g^{2}}{4})[(-2\overline{m}_{G^{+}}^{2}+\overline{m}_{W^{-}}^{2}) B_{0}^{1}(\overline{m}_{G^{+}},\overline{m}_{W_{-}})
		-2B_{0}^{0}(\overline{m}_{G^{+}},\overline{m}_{W_{-}})]$\\
		$j)$ &$\frac{1}{16\pi^2}(\frac{g^{2}+g^{\prime2}}{4})[(-2\overline{m}_{G_{0}}^{2}+\overline{m}_{Z}^{2}) B_{0}^{1}(\overline{m}_{G_{0}},\overline{m}_Z)
		-2B_{0}^{0}(\overline{m}_{G_{0}},\overline{m}_Z)]$\\
		\hline
		$k)$ &$\frac{1}{16\pi^2}(-g_{t}^{2})[-B_{0}^{0}(\overline{m}_t)+4\overline{m}_{t}^{2}B_{0}^{1}(\overline{m}_t)]$\\
		\hline
	\end{tabular}
    \caption{$p^{2}$ terms from the self-energy diagram in the SM which contribute to $Z_{2}$.
     Note that in these results we only list the fermion loop contribution from the top quark.}
	\label{selfenergy}
\end{table}

We list the $p^{2}$ terms of each self-diagram in Table.~\ref{selfenergy}.
Summing over all the $p^{2}$ term contributions, we obtain the $Z_{2}$ factor in the SM.
Since the RG equation for the kinetic term in the effective action can be solved in a way similar to the solution to $V_{eff}(\phi)$,
we can obtain the RG improved kinetic term by replacing $\phi$, $\mu$, $\lambda_{i}$ with $\phi(t)$, $\mu(t)$ and $\lambda(t)$.
Their expressions or equation are shown in Eqs.~\eqref{mutphit} and \eqref{runningeq}.
Taking $\mu(t) = \phi$ as mentioned above, we get the RG improved $Z_{2}$ factor in the SM for large $\phi$ field.
\begin{equation}
\label{Z2highphiSM}
  \begin{aligned}
  Z_{2}^{SM} = &1 + \frac{1}{16\pi^2}[\lambda+\frac{8\lambda^2}{12\lambda+3g^2}+\frac{4\lambda^2}{12\lambda+3(g^2+g^{\prime2})}+\frac{2}{3}(2g^2+g^{\prime2})-
  \frac{(2g^2+g^{\prime2})}{24}]\\
  &+\frac{1}{8\pi^2} \big\{ [\frac{-(4\lambda g^2+g^4)ln(\frac{4\lambda+g^2}{g^2})}{16\lambda^{3}}+\frac{2\lambda+g^2}{4\lambda^2}]
  \frac{8\lambda g^2 + g^4}{16}\\
  &-\frac{(4\lambda g^2 + g^4)ln(\lambda e^{2\Gamma}+\frac{1}{4} g^2 e^{2\Gamma})-g^4ln(\frac{1}{4}g^2 e^{2\Gamma})}{8\lambda}+\frac{g^2}{2} \big\} \\
  &-\frac{1}{16\pi^2}\big\{ [\frac{-(4\lambda (g^2+g^{\prime2})+(g^2+g^{\prime2})^2)ln(\frac{4\lambda+g^2+g^{\prime2}}{g^2+g^{\prime2}})}{16\lambda^{3}}+
  \frac{2\lambda+g^2+g^{\prime2}}{4\lambda^2}]
  \frac{8\lambda (g^2+g^{\prime2}) + (g^2+g^{\prime2})^2}{16}\\
  &-\frac{(4\lambda (g^2+g^{\prime2}) + (g^2+g^{\prime2})^2)ln(\lambda e^{2\Gamma}+\frac{1}{4} (g^2+g^{\prime2}) e^{2\Gamma})-(g^2+g^{\prime2})^2ln(\frac{1}{4}(g^2+g^{\prime2}) e^{2\Gamma})}{8\lambda}+\frac{g^2+g^{\prime2}}{2} \big\}\\
  &-\frac{3}{16\pi^2}[ln(\frac{y_{t}^2}{2} e^{2\Gamma}) y_{t}^2+\frac{2}{3}y_{t}^2]
  \end{aligned}
\end{equation}
with 
\begin{equation}
\Gamma(t)=-\int_{0}^{t} \gamma\left(\lambda\left(t^{\prime}\right)\right) d t^{\prime}
\end{equation}

\subsection{$Z_{2}$ factor in the SDFDM model}
\begin{figure}[h]
	\centering
	\includegraphics[width=0.7\linewidth]{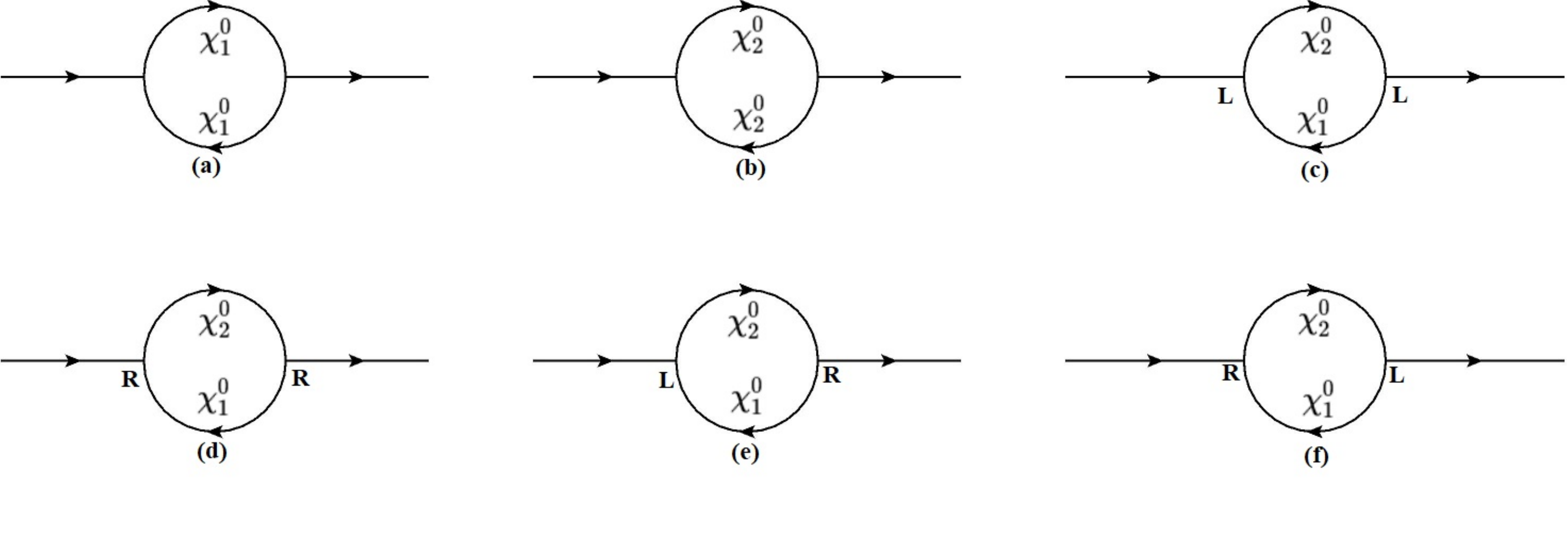}
	\caption{Self-energy diagrams contributing to $Z_{2}$ factor 
	by extra fermions in the SDFDM model.
	$\chi_1^{0}$ and $\chi_{2}^{0}$ are the new extra fermions in the SDFDM model.
	 }
	\label{fig:newselfenergydiagram}
\end{figure}

\begin{table}[h]
	\setlength{\tabcolsep}{.5em}
	\renewcommand{\arraystretch}{1.2}
	\begin{tabular}{|cc|}
		\hline 
		$a)$ &$\frac{1}{16\pi^2}(-A^2)(4\overline{m}_{\chi_{1}^{0}}^2B_{0}^{1}(\overline{m}_{\chi_{1}^{0}})-B_{0}^{0}(\overline{m}_{\chi_{1}^{0}})$ \\ 
		$b)$ &$\frac{1}{16\pi^2}(-B^2)(4\overline{m}_{\chi_{2}^{0}}^2B_{0}^{1}(\overline{m}_{\chi_{2}^{0}})-B_{0}^{0}(\overline{m}_{\chi_{2}^{0}})$ \\ 
		\hline 
		$c)$ &$\frac{1}{16\pi^2}(-\frac{CD}{2})(2(\overline{m}_{\chi_{1}^{0}} \overline{m}_{\chi_{2}^{0}} B_{0}^{1}(\overline{m}_{\chi_{1}^{0}}
		,\overline{m}_{\chi_{2}^{0}})$ \\ 
		$d)$ &$\frac{1}{16\pi^2}(-\frac{CD}{2})(2(\overline{m}_{\chi_{1}^{0}} \overline{m}_{\chi_{2}^{0}} B_{0}^{1}(\overline{m}_{\chi_{1}^{0}}
		,\overline{m}_{\chi_{2}^{0}})$ \\ 
		\hline 
		$e)$ &$\frac{1}{16\pi^2}(-\frac{D^2}{2})[(\overline{m}_{\chi_{1}^{0}}^2+\overline{m}_{\chi_{2}^{0}}^2)B_{0}^{1}(\overline{m}_{\chi_{1}^{0}}
		,\overline{m}_{\chi_{2}^{0}})-B_{0}^{0}(\overline{m}_{\chi_{1}^{0}},\overline{m}_{\chi_{2}^{0}})]$\\
 	   	$f)$ &$\frac{1}{16\pi^2}(-\frac{C^2}{2})[(\overline{m}_{\chi_{1}^{0}}^2+\overline{m}_{\chi_{2}^{0}}^2)B_{0}^{1}(\overline{m}_{\chi_{1}^{0}}
 	    ,\overline{m}_{\chi_{2}^{0}})-B_{0}^{0}(\overline{m}_{\chi_{1}^{0}},\overline{m}_{\chi_{2}^{0}})]$\\
		\hline
	\end{tabular}
	\caption{$p^{2}$ terms from the self-energy diagram contributed by extra fermions in the SDFDM model. 
	     Here we define 
		$A = \left(-y_{2} \cos \theta_{L} \sin \theta_{R}-y_{1} \sin  \theta_{L} \cos \theta_{R}\right)$,
	    $B = y_{2} \cos \theta_R \sin \theta_{L}+y_{1} \sin \theta_{R} \cos \theta_{L}$,
        $C = y_{2} \cos \theta_{L} \cos \theta_{R}-y_{1} \sin \theta_{L} \sin \theta_{R}$,
        $D = -y_{2} \sin \theta_{L} \sin \theta_{R}+y_{1} \cos\theta_{L} \cos \theta_{R}$.
        $\overline{m}_{\chi_{1}^{0}}$, $\overline{m}_{\chi_{2}^{0}}$ are the masses of new particles under the external field $\phi$.
        They are obtained by substituting $v$ in Eqs. \eqref{Mchi1} and \eqref{Mchi2} with $\phi$. }
        \label{newselfenergy}
	\end{table}
	The Feymann diagrams in the SDFDM model contributing to the Higgs self-energy are shown in Fig.~\ref{fig:newselfenergydiagram}.
$p^{2}$ term contributions to $Z_2$ in these diagrams are summarized in Table.~\ref{newselfenergy}.
Summing over all the $p^{2}$ term contributions in Table. \ref{selfenergy} and Table.~\ref{newselfenergy}, 
we obtain the $Z_{2}$ factor in the SDFDM model.
In the large $\phi$ limit, $Z_2$ factor in the SDFDM model can be expressed as
\begin{equation}
\label{Z2highphiSDFDM}
Z_{2}^{SDFDM} =Z_{2}^{SM} -\frac{y_{2}^2}{16\pi^2}[ln(\frac{y_{2}^2}{2}e^{2\Gamma})+\frac{2}{3}] 
                -\frac{y_{1}^2}{16\pi^2}[ln(\frac{y_{1}^2}{2}e^{2\Gamma})+\frac{2}{3}],
\end{equation}
where $Z_2^{SM}$ is given in Eq. \eqref{Z2highphiSM} .

\section{ Threshold effect, $Z_2$ and $\beta$ function in the general singlet-doublet fermion extension model}
\label{sec:thersholdeffectgeneral}
We summarize here the one-loop corrections to $\lambda$ from new particles in the general singlet-doublet extension model using Eq.~\eqref{90}.

The one-loop result is
\begin{equation}
\begin{aligned} 
\delta^{(1)} \lambda|_{fin} = &\frac{G_{\mu}}{\sqrt{2} (4\pi)^2} 
\sum_{i=1}^{Q+N}\sum_{j=1}^{Q+N}
\left\{ 2 Y_{ij}Y^{+}_{ji}[A_{0}(M_{\chi_i^0})+A_{0}(M_{\chi_j^0})-(M^2_h-M^2_{\chi_{i}^0}-M^2_{\chi_{j}^0})B_{0}(M_{\chi_i^0},M_{\chi_j^0},M_{h})]\right.\\
& \left. +2 (Y_{ij} Y_{ji}+Y^{+}_{ij} Y^{+}_{ji}) M_{\chi_i^0} M_{\chi_j^0} B_{0}(M_{\chi_i^0},M_{\chi_j^0},M_{h}) \right\} + \frac{G_{\mu}}{\sqrt{2} (4\pi)^2v}\sum_{i=1}^{N+Q}\left[(-2Y_{ii}-2Y^{+}_{ii}) M_{\chi_i^0} A_{0}(M_{\chi_i^0}) \right]\\
& + \frac{G_{\mu}}{\sqrt{2}}M_{h}^2 \left.\Delta r_{0}^{(1)}\right|_{\text {fin }} 
\end{aligned}
\label{one-loop-lambda-general}
\end{equation}

where $Y$ has been given in Eq.~\eqref{generalYmatrix} and the new contribution to $\Delta r_{0}^{(1)}|_{\text {fin }}$ is

\begin{equation}
\begin{aligned}
\left.\Delta r_{0}^{(1)}\right|_{\text {fin }}&= \frac{1}{(4 \pi v)^{2}} \left\{ 
4 \sum_{k=1}^{N} \left(\left(U_{L}^\dagger\right)_{i, k+Q}  \left(U_{R}\right)_{k+Q, i}+\left(U_{R}^\dagger\right)_{i, k+Q}  \left(U_{L}\right)_{k+Q, i}\right) \times \right.\\
&\left. \left[ \frac{M_{\chi_i^0} M_{\chi^{-}_{k}}}{M^2_{\chi^{-}_{k}}-M^2_{\chi_{i}^0}}\left( A_{0}(M_{\chi^{-}_{k}})-A_{0}(M_{\chi_i^0})\right) \right] \right.\\
&+\left. \sum_{k=1}^{N}\sum_{i=1}^{N+Q} \left( \left(U_{L}^\dagger\right)_{i, k+Q} \left(U_{L}\right)_{k+Q, i} +\left(U_{R}^\dagger\right)_{i, k+Q} \left(U_{R}\right)_{k+Q, i}\right) \times \right. \\
&\left.\left[ \frac{2M^2_{\chi_{i}^0}}{M^2_{\chi^{-}_{k}}-M^2_{\chi_{i}^0}}A_{0}(M_{\chi_i^0})-\frac{2M^2_{\chi^{-}_{k}}}{M^2_{\chi^{-}_{k}}-M^2_{\chi_{i}^0}}A_{0}(M_{\chi^{-}_{k}})+ M^2_{\chi^{-}_{k}}+M^2_{\chi_{i}^0} \right]  \right\}.\\
\end{aligned}
\end{equation}

Plugging Eq.  (\ref{one-loop-lambda-general}) into Eq. (\ref{one-loop-lambda-1}) we obtain $\lambda$ at one-loop order in the general singlet-doublet extension model. Contributions of extra fermions to 
$\delta^{(1)}\left.y_{t}\right|_{\mathrm{fin}}$, $\delta^{(1)}\left.g_{2}\right|_{\mathrm{fin}}$ and $\delta^{(1)}\left.g_{Y}\right|_{\mathrm{fin}}$ can
be similarly obtained.  Plugging them into Eqs. (\ref{one-loop-lambda-2}),  (\ref{one-loop-lambda-3}) and (\ref{one-loop-lambda-4})
we obtain relevant parameters at one-loop order in the general model.
Using these parameters in the $\MS$ scheme, we then carry out the calculation of the effective action in the $\MS$ scheme.
 
In the large $\phi$ limit, $Z_2$ factor in the Model \uppercase\expandafter{\romannumeral1} can be expressed as
\begin{equation}
\label{Z2highphiModel1}
Z_{2}^{\textrm{Model \uppercase\expandafter{\romannumeral1}}} =Z_{2}^{SM} - \frac{Ny_{2}^2 }{16\pi^2}[ln(\frac{y_{2}^2}{2}e^{2\Gamma})+\frac{2}{3}] 
-\frac{N y_{1}^2 }{16\pi^2}[ln(\frac{y_{1}^2}{2}e^{2\Gamma})+\frac{2}{3}] .
\end{equation}
$Z_2$ factor in the Model \uppercase\expandafter{\romannumeral2} and Model \uppercase\expandafter{\romannumeral3} can be expressed as
\begin{equation}
\label{Z2highphiModel23}
Z_{2}^{\textrm{Model \uppercase\expandafter{\romannumeral2}, \uppercase\expandafter{\romannumeral3}}} =Z_{2}^{SM} -\frac{N y_{2}^2}{16\pi^2}[ln(\frac{Ny_{2}^2}{2}e^{2\Gamma})+\frac{2}{3}] 
-\frac{N y_{1}^2}{16\pi^2}[ln(\frac{Ny_{1}^2}{2}e^{2\Gamma})+\frac{2}{3}] ,
\end{equation}

	The one-loop $\beta$ functions and the anomalous dimension of the three singlet-doublet extension models are as follows.
	\subsection{Model \uppercase\expandafter{\romannumeral1}}
	\label{betamodel1}
	\begin{eqnarray}
	\beta^{\mathrm{Model \  \uppercase\expandafter{\romannumeral1}}}\left(g_{1}\right)=&&\frac{1}{(4 \pi)^{2}}\left(\frac{2}{5}N\right) g_{1}^{3},\label{betafunctiong1model1}\\
	\beta^{\mathrm{Model \  \uppercase\expandafter{\romannumeral1}}}\left(g_{2}\right)=&&\frac{1}{(4 \pi)^{2}}\frac{2}{3}N g_{2}^{3},\label{betafunctiong2model1}\\
	\beta^{\mathrm{Model \  \uppercase\expandafter{\romannumeral1}}}\left(y_{\tau}\right)=&&\frac{1}{(4 \pi)^{2}}N\left(y_{1}^{2}+y_{2}^{2}\right) y_{\tau},\\
	\beta^{\mathrm{Model \  \uppercase\expandafter{\romannumeral1}}}\left(y_{b}\right)=&&\frac{1}{(4 \pi)^{2}}N\left(y_{1}^{2}+y_{2}^{2}\right) y_{b},\\
	\beta^{\mathrm{Model \  \uppercase\expandafter{\romannumeral1}}}\left(y_{t}\right)=&&\frac{1}{(4 \pi)^{2}}N\left(y_{1}^{2}+y_{2}^{2}\right) y_{t},\\
	\beta^{\mathrm{Model \  \uppercase\expandafter{\romannumeral1}}}(\lambda)=&&\frac{1}{(4 \pi)^{2}}\left[-N\left(2 y_{1}^{4} +2 y_{2}^{4}\right)+4N \lambda\left(y_{1}^{2}+y_{2}^{2}\right)\right]\label{betafunctionlambdamodel1}.
	\end{eqnarray}
	
	\begin{eqnarray}
	\beta^{\mathrm{Model \  \uppercase\expandafter{\romannumeral1}}}\left(y_{1}\right)=&&\frac{1}{(4 \pi)^{2}}\left[\frac{3}{2} y_{1}^{3}+Ny_{1}^{3}+ Ny_{1} y_{2}^{2}-\frac{9}{20} g_{1}^{2} y_{1}-\frac{9}{4} g_{2}^{2} y_{1}+3 y_{t}^{2} y_{1}+3 y_{b}^{2} y_{1}+y_{\tau}^{2} y_{1}\right],\\
	\beta^{\mathrm{Model \  \uppercase\expandafter{\romannumeral1}}}\left(y_{2}\right)=&&\frac{1}{(4 \pi)^{2}}\left[\frac{3}{2} y_{2}^{3}+N y_{2}^{3}+ Ny_{1}^{2} y_{2}-\frac{9}{20} g_{1}^{2} y_{2}-\frac{9}{4} g_{2}^{2} y_{2}+3 y_{t}^{2} y_{2}+3 y_{b}^{2} y_{2}+y_{\tau}^{2} y_{2}\right].
	\end{eqnarray}
	The one-loop contribution of new particles to the anomalous dimension is:
	\begin{equation}
	\gamma^{\mathrm{Model \  \uppercase\expandafter{\romannumeral1}}}=\frac{1}{(4 \pi)^{2}}N(-y_{1}^2-y_{2}^2).
	\end{equation}
	
	\subsection{{Model \uppercase\expandafter{\romannumeral2}}}
	\label{betamodel2}
	\begin{eqnarray}
	\beta^{\mathrm{Model \  \uppercase\expandafter{\romannumeral2}}}\left(g_{1}\right)=&&\frac{1}{(4 \pi)^{2}}\left(\frac{2}{5}N\right) g_{1}^{3}\label{betafunctiong1model2},\\
	\beta^{\mathrm{Model \  \uppercase\expandafter{\romannumeral2}}}\left(g_{2}\right)=&&\frac{1}{(4 \pi)^{2}}\frac{2}{3}N g_{2}^{3}
	\label{betafunctiong2model2},\\
	\beta^{\mathrm{Model \  \uppercase\expandafter{\romannumeral2}}}\left(y_{\tau}\right)=&&\frac{1}{(4 \pi)^{2}}N\left(y_{1}^{2}+y_{2}^{2}\right) y_{\tau},\\
	\beta^{\mathrm{Model \  \uppercase\expandafter{\romannumeral2}}}\left(y_{b}\right)=&&\frac{1}{(4 \pi)^{2}}N\left(y_{1}^{2}+y_{2}^{2}\right) y_{b},\\
	\beta^{\mathrm{Model \  \uppercase\expandafter{\romannumeral2}}}\left(y_{t}\right)=&&\frac{1}{(4 \pi)^{2}}N\left(y_{1}^{2}+y_{2}^{2}\right) y_{t},\\
	\beta^{\mathrm{Model \  \uppercase\expandafter{\romannumeral2}}}(\lambda)=&&\frac{1}{(4 \pi)^{2}}\left[-N^2\left(2 y_{1}^{4} +2 y_{2}^{4}\right)+4N \lambda\left(y_{1}^{2}+y_{2}^{2}\right)\right].
	\end{eqnarray}
	
	\begin{eqnarray}
	\beta^{\mathrm{Model \  \uppercase\expandafter{\romannumeral2}}}\left(y_{1}\right)=&&\frac{1}{(4 \pi)^{2}}\left[\frac{5}{2}N y_{1}^{3}+ Ny_{1} y_{2}^{2}-\frac{9}{20} g_{1}^{2} y_{1}-\frac{9}{4} g_{2}^{2} y_{1}+3 y_{t}^{2} y_{1}+3 y_{b}^{2} y_{1}+y_{\tau}^{2} y_{1}\right],\\
	\beta^{\mathrm{Model \  \uppercase\expandafter{\romannumeral2}}}\left(y_{2}\right)=&&\frac{1}{(4 \pi)^{2}}\left[\frac{5}{2}N y_{2}^{3}+ Ny_{1}^{2} y_{2}-\frac{9}{20} g_{1}^{2} y_{2}-\frac{9}{4} g_{2}^{2} y_{2}+3 y_{t}^{2} y_{2}+3 y_{b}^{2} y_{2}+y_{\tau}^{2} y_{2}\right].
	\end{eqnarray}
	The one-loop contribution of new particles to the anomalous dimension is:
	\begin{equation}
	\gamma^{\mathrm{Model \  \uppercase\expandafter{\romannumeral2}}}=\frac{1}{(4 \pi)^{2}}N(-y_{1}^2-y_{2}^2).
	\end{equation}
	
	\subsection{{Model \uppercase\expandafter{\romannumeral3}}}
	\label{betamodel3}
	\begin{eqnarray}
	\beta^{\mathrm{Model \  \uppercase\expandafter{\romannumeral3}}}\left(g_{1}\right)=&&\frac{1}{(4 \pi)^{2}}\left(\frac{2}{5}\right) g_{1}^{3},\\
	\beta^{\mathrm{Model \  \uppercase\expandafter{\romannumeral3}}}\left(g_{2}\right)=&&\frac{1}{(4 \pi)^{2}}\frac{2}{3} g_{2}^{3},\\
	\beta^{\mathrm{Model \  \uppercase\expandafter{\romannumeral3}}}\left(y_{\tau}\right)=&&\frac{1}{(4 \pi)^{2}}N\left(y_{1}^{2}+y_{2}^{2}\right) y_{\tau},\\
	\beta^{\mathrm{Model \  \uppercase\expandafter{\romannumeral3}}}\left(y_{b}\right)=&&\frac{1}{(4 \pi)^{2}}N\left(y_{1}^{2}+y_{2}^{2}\right) y_{b},\\
	\beta^{\mathrm{Model \  \uppercase\expandafter{\romannumeral3}}}\left(y_{t}\right)=&&\frac{1}{(4 \pi)^{2}}N\left(y_{1}^{2}+y_{2}^{2}\right) y_{t},\\
	\beta^{\mathrm{Model \  \uppercase\expandafter{\romannumeral3}}}(\lambda)=&&\frac{1}{(4 \pi)^{2}}\left[-N^2\left(2 y_{1}^{4} +2 y_{2}^{4}\right)+4N \lambda\left(y_{1}^{2}+y_{2}^{2}\right)\right].
	\end{eqnarray}
	
	\begin{eqnarray}
	\beta^{\mathrm{Model \  \uppercase\expandafter{\romannumeral3}}}\left(y_{1}\right)=&&\frac{1}{(4 \pi)^{2}}\left[\frac{5}{2}N y_{1}^{3}+ Ny_{1} y_{2}^{2}-\frac{9}{20} g_{1}^{2} y_{1}-\frac{9}{4} g_{2}^{2} y_{1}+3 y_{t}^{2} y_{1}+3 y_{b}^{2} y_{1}+y_{\tau}^{2} y_{1}\right],\\
	\beta^{\mathrm{Model \  \uppercase\expandafter{\romannumeral3}}}\left(y_{2}\right)=&&\frac{1}{(4 \pi)^{2}}\left[\frac{5}{2}N y_{2}^{3}+ Ny_{1}^{2} y_{2}-\frac{9}{2} g_{1}^{2} y_{2}-\frac{9}{4} g_{2}^{2} y_{2}+3 y_{t}^{2} y_{2}+3 y_{b}^{2} y_{2}+y_{\tau}^{2} y_{2}\right].
	\end{eqnarray}
	The one-loop contribution of new particles to the anomalous dimension is:
	\begin{equation}
	\gamma^{\mathrm{Model \  \uppercase\expandafter{\romannumeral3}}}=\frac{1}{(4 \pi)^{2}}N(-y_{1}^2-y_{2}^2).
	\end{equation}

\bibliographystyle{utphys}
\bibliography{SDFDM}

\providecommand{\href}[2]{#2}\begingroup\raggedright\endgroup

\end{document}